\documentclass[nonacm]{acm-book}
\usepackage[dvipsnames,table]{xcolor}

\usepackage{balance}

\usepackage{verbatim}
\usepackage{amsmath}
\usepackage{mathtools}
\usepackage[bb=boondox]{mathalfa}
\usepackage{comment}
\usepackage[normalem]{ulem}
\usepackage[inline]{enumitem}
\usepackage{bbm}
\usepackage{todonotes}
\DeclareMathOperator*{\argmax}{arg\,max}

\usepackage{hyperref}

\usepackage{booktabs}
\AtBeginDocument{%
  \providecommand\BibTeX{{%
    \normalfont B\kern-0.5em{\scshape i\kern-0.25em b}\kern-0.8em\TeX}}}

\presetkeys%
    {todonotes}%
    {inline,backgroundcolor=yellow}{}

\begin{document}

\title{Recommender Systems: A Primer}

\author{Pablo Castells$^a$ and Dietmar Jannach$^b$\\
$^a$Universidad Autónonoma de Madrid (UAM)\thanks{This manuscript results from work performed at UAM and is not associated with Amazon.}, Spain, \\$^b$University of Klagenfurt, Austria\\
pablo.castells@uam.es, dietmar.jannach@aau.at}

\date{February 2023\\ This manuscript is a preprint of a chapter to be published in ``Advanced Topics for Information Retrieval'' by R.~Baeza-Yates and O.~Alonso with ACM Press}


\newcommand{\prefacename}{Abstract}
\newenvironment{preface}{
    \vspace*{\stretch{2}}
    {\noindent \bfseries \Huge \prefacename}
    \begin{center}
        \thispagestyle{plain}
    \end{center}%
}
{\vspace*{\stretch{5}}}

\maketitle

\begin{preface}

Personalized recommendations have become a common feature of modern online services, including most major e-commerce sites, media platforms and social networks. Today, due to their high practical relevance, research in the area of recommender systems is flourishing more than ever. However, with the new application scenarios of recommender systems that we observe today, constantly new challenges arise as well, both in terms of algorithmic requirements and with respect to the evaluation of such systems.

In this paper, we first provide an overview of the traditional formulation of the recommendation problem. We then review the classical algorithmic paradigms for item retrieval and ranking and elaborate how such systems can be evaluated. Afterwards, we discuss a number of recent developments in recommender systems research, including research on session-based recommendation, biases in recommender systems, and questions regarding the impact and value of recommender systems in practice.
\end{preface}

\setcounter{tocdepth}{2}

\pagestyle{empty}
\tableofcontents
\newpage
\chapter{Basic Concepts}

\section{Introduction}
\label{sec:introduction}
Automated and often personalized recommendations are omnipresent in today's online world. Nowadays, whenever we go online, there is a good chance that we will very soon receive recommendations about more things to shop, trending apps to download, or new music or artists to discover. These personalized suggestions are provided to us by \emph{recommender systems}, which are software components that determine---based on statistics and machine learning models---the most suitable items that should be presented to an individual user. Due to their widespread use in practice and their demonstrated capabilities of creating value both for consumers and businesses, recommender systems can be seen as one of the most visible success stories of artificial intelligence. Today, research in this area is flourishing more than ever, both due to the practical relevance of such systems and due to the ongoing rapid developments in machine learning.

Historically, the area of recommender systems has various relationships to the field of information retrieval (IR). The problem of retrieving and ranking a set of items from a larger collection, for example, is central both in recommendation and search tasks. The two areas furthermore oftentimes rely on similar algorithms and machine learning models and they use the same or similar evaluation methodology to assess the performance of these models. Recommendation is however also different from IR in a number of ways. The selection of items in recommendation scenarios, for example, is typically based on individual user profiles and not on interactive queries.
From an IR perspective, one way to look at recommendation therefore is to see it as a problem setting where the item retrieval is based on an \emph{implicit} query---which may also depend on the user's context---rather than on \emph{explicit} queries.
Moreover, recommendation is more often a \emph{push} communication, whereas IR systems are typically reactive applications.

The roots of information retrieval can be traced to as far back as the 1950s \citep{sc2012pie}. The foundations of modern-day recommender systems, in contrast, were mainly laid in the 1990s. In 1992, \emph{Tapestry}, a prototype of a personalized email filtering system, was developed at Xerox PARC \citep{goldberg1992using}. One main innovative idea of this system was to leverage the opinion of others (\emph{ratings}) in the filtering process, and the term \emph{collaborative filtering} (CF) was popularized with this system. Soon later, several research groups worked on automating Tapestry's rule-based approach, and in 1994, the highly influential \emph{GroupLens} system was proposed \citep{ResnickGrouplens1994}. The main idea in this system was to automatically determine like-minded ``neighbors'' of a user and to recommend those news items to a user that these neighbors have rated highly. In the context of this early work, the recommendation problem was operationalized as a ``matrix filling'' problem, where the input to a recommendation algorithm is a user-item rating matrix and the goal is to predict the missing entries in this matrix. This research operationalization and the use of collaborative filtering techniques is still predominant in today's research and practice. Over the last almost three decades, however, many more elaborate machine learning algorithms were designed or applied for recommendation problems, including various matrix factorization techniques---which were also used first in the 1990s \citep{Billsus98Learning}---and, most recently, sophisticated deep learning techniques, e.g., \cite{liang2018}.

An alternative to using collaborative preference signals is to rank and filter documents in a personalized way by looking at their \emph{content}. The corresponding \emph{content-based} filtering techniques were also investigated in the 1990s, often under the terms \emph{information filtering} or \emph{personalized information filtering} \citep{Foltz1992PersonalizedInformationDelivery}. These systems are often largely based on known IR techniques like TF-IDF encodings\footnote{In \emph{Term-Frequency -- Inverse Document Frequency (TF-IDF)} encodings, documents are represented as vectors of real numbers, where each element stands for a term and its value represents the importance of a term in the document.} 
of the documents. In pure content-based recommender systems, the main idea is to suggest items to the users that are similar to those that they liked in the past, i.e., without considering the opinions of others. Nowadays, however, we very often observe that \emph{hybrid} systems are used in practice that combine different recommendation techniques and types of data. An early example of such a system is \emph{Fab} \citep{Fab1997}, a content-based collaborative system designed for web page recommendation at Stanford.

Most of the early systems discussed here were developed as research prototypes and mostly used in academic settings. However, already before the end of the 1990s a number of successful deployments of recommendation techniques
were reported for domains such as of e-commerce or music, see, e.g., \citep{Shafer1999ecommerce}. Later on, Amazon was probably one of the first organizations that relied on recommendation technology at scale \citep{linden2003}. Nowadays, as mentioned above, there are many online services where recommendations are central to the user experience, e.g., at Netflix \citep{Gomez-Uribe:2015:NRS:2869770.2843948}, Spotify \citep{10.1145/3298689.3346977}, or YouTube \citep{covington2016}. Over the years, also a number of success stories can be found in the literature, where the potential business value of recommender systems is documented, see
\citep{jannachjugovactmis2019} for an overview.

Overall, the field has reached a certain level of maturity and standardization, in particular with respect to the operationalization of the research problem and the evaluation methodology. However, the recommendation problem is far from being solved, as there is a constant stream of new application scenarios that have not been adequately addressed so far in the research community. This paper reflects this situation and consists of two main parts. In the first part, in Sections \ref{sec:recommendation-task} to \ref{sec:evaluation}, we provide an overview on the basic concepts of recommender systems and how recommender systems are commonly evaluated. Afterwards, from Section \ref{sec:sequential-session-based} on, we discuss a number of current and future topics in recommender systems research.

\section{The Recommendation Task}
\label{sec:recommendation-task}
The \emph{main computational task}\footnote{Other computational tasks of a recommender system could be to compute an explanation or to determine the next conversational move in case of an interactive system.}
of any recommender system is to determine which items to show to a user in a given situation. Therefore, in any individual application setting, before an algorithm is implemented and evaluated, the question has to be answered based on which criteria the items should be selected and ranked. At the most general level, any recommender system is designed to create a certain \emph{value} or \emph{utility} for one or more of the involved stakeholders \cite{JannachAdomavicius2016purpose}. In practice, there are many ways of how value can be created. From the perspective of a consumer, for example, a recommender system is usually assumed to reduce problems of information overload. From the perspective of the provider, on the other hand, providing personalized recommendations can help to increase business-related key-performance indicators (KPIs) such as sales numbers or customer retention. We will discuss related questions of the impact and the value of recommender systems in more depth later in Section \ref{sec:impact-and-value}.

Academic research often aims to abstract from the specifics of individual domains, particular applications, or business models of recommendation providers. The main computational task is therefore often framed in a correspondingly abstract way, usually \emph{to compute an estimate of the absolute or relative relevance of individual items for a given user}.
On a general level, this problem can be formalized as follows, see \cite{TowardsGedas2005}. Given
\begin{itemize}
\item a set of users $U$,
\item a set of items $I$ and
\item a \emph{utility function} $f: U \times I \rightarrow R$, which maps user and items to a utility value taken from a totally ordered set $R$ (e.g., of nonnegative real numbers),
\end{itemize}
recommend the item $i' \in I$, which maximizes the utility function, more formally:
\begin{equation}
\forall u \in U, i'_{u}=\argmax_{i \in I} f(u,i)
\end{equation}

Typically, we are interested in recommending more than one item. This can be done by returning those \emph{N} items that have the highest utility values. Using this definition, algorithmic research in recommender systems (see Section \ref{sec:algorithms}) amounts to \emph{defining} or \emph{learning} the utility function $f$, where $f$ can be based on different types of additional information.

In the collaborative-filtering GroupLens system from 1994 mentioned earlier and in countless subsequent works, for example, the utility function $f$ was designed to return a \emph{rating prediction}. The additional information that is used within $f$ was the user-item matrix of known ratings. Table~\ref{tab:user-item-matrix} shows an example of such a rating matrix, where the specific task that is highlighted in the table is to make a recommendation for user $u1$. This is accomplished by predicting the ratings of user $u1$ for the so far unrated items $i4$ and $i5$.

The specific implementation of $f$ in the GroupLens system was a nearest-neighbor method. In pure content-based recommendation approaches, in contrast, $f$ is not based on the user-item rating matrix, but on the observed past behavior of an individual user and additional information about the items in $I$. In hybrid approaches, finally, the implementation of $f$ might simultaneously consider the rating matrix, item meta-data, and  other sources of information including demographics, various other types of observed user behavior beyond past ratings, or the users' embedding in a social network, see also Section~\ref{subsec:cb-and-hybrid}.

\begin{table} [h!t]
\centering
\begin{tabular}{cccccc} \toprule
   & \textbf{i1} & \textbf{i2} & \textbf{i3} & \textbf{i4} & \textbf{i5} \\ \toprule
  \textbf{u1} &  3 & 4 & 3 &  \cellcolor{blue!25}\textbf{?} &  \cellcolor{blue!25}\textbf{?} \\
  \textbf{u2} &   & 4 & 3 & & 5 \\
  \textbf{u3} &   & 1 & 3 & 1 & \\
  \textbf{u4} & 4  &  &  3 & 2 & 3\\
  \textbf{u5} & 3  &  &  3 & 2 & \\ \bottomrule
  \end{tabular}
\caption{User-item Rating Matrix.}
\label{tab:user-item-matrix}
\end{table}

Collaborative filtering is the most frequently addressed problem setting in the literature and a large body of the literature aims at \emph{learning} $f$ from the usually \emph{noisy data} in the user-item matrix. Correspondingly, all sorts of machine learning algorithms were applied over the years to learn $f$.
Historically, the user-item rating matrix was considered to contain \emph{explicit}, user-specified item ratings. Nowadays, research is more focused on situations where only \emph{implicit} user feedback is available. In such cases, the entries in the rating matrix are not values ranging, e.g., from one to five stars, but unary. A \emph{positive} value, usually represented as a 1, in the matrix therefore expresses that a user has for example viewed or purchased an item in the past. Most commonly, implicit feedback datasets do not contain negative feedback values, i.e., a value is either 1 or it is missing. Clearly, a past purchase not necessarily means that a user has liked a certain item, and implicit feedback, i.e., observed past user behavior, can be more noisy than explicit item ratings.

In principle, all sorts of information can be incorporated in the utility function $f$. One piece of information that received special interest in the literature is that of \emph{context}, which refers to the particular situation in which a recommendation is made. The current context can have a major impact on the usefulness of a particular recommendation. One restaurant might, for example, be a good recommendation during summer but not in winter. Considering the time of the year as context may therefore be crucial. Overall, the difference to other types of side information is that the utility of an item $i$ can be different for a specific user $u$, depending on the current situation. One proposal therefore is to extend the signature of $f$ accordingly, to make this aspect explicit \citep{Adomavicius2022}. Correspondingly, we may have
\[f: U \times I \times C \rightarrow R\]
as an extended utility function, where $C$ denotes the context in which a recommendation should be made.

Generally, utility functions that make individual relevance predictions---with or without considering context---are widely used in the literature. Nowadays, however, predicting individual \emph{ratings} on an absolute scale, as was done in the Netflix Prize competition \cite{NetflixPrize2007}, is not considered the most relevant problem in practice anymore. Instead, more focus is put on creating ranked lists of suggestions, leading to the \emph{top-N} recommendation task. Technically, one can use the same algorithms that were designed for rating prediction problems, and rank the items based on the predicted rating. Alternatively, \emph{learning-to-rank} algorithms can be used, which do not consider the recommendable items individually, but directly aim to optimize the ranking.

As a result of this changed problem setting---creating a top-\emph{N} list of items instead of making point-wise predictions---an alternative utility function $f$ of the form
\[f: U \times L^* \rightarrow R\]
can be defined, where $L^*$ is the set of all permutations up to the length of $N$ of the powerset of $I$, see also \cite{QuadranaetalCSUR2018}. Given this function, we then recommend the list of items that maximizes the utility value $R$.
Correspondingly, the problem when designing an algorithm is to define or learn a function that predicts the utility of length-restricted ranked lists of items.

Regarding the concept of utility, note that we so far have not made any assumptions regarding how utility should be defined or measured. In the past, starting with the GroupLens system, ratings were considered as proxies for utility. The goal was to predict how a user would rate a yet unseen item and to then recommend the items with the highest predicted ratings, assuming that these items are the most useful ones. Note, however, that our definitions are not generally limited to \emph{consumer value}, and a utility function might consider the \emph{provider profit} as well, and thus consider the value perspective of more than one stakeholder, see also \citep{abdollahpouri2020,mmbld18cikm}.

In traditional settings, mostly utility functions of the form $f: U \times I \rightarrow R$ were considered, where each item's utility is considered independently from other items that might be shown to a user in a single recommendation list. Such an approach however does not allow us to assess the quality of top-$N$ item lists as a whole, e.g., in terms of their diversity. Therefore, utility functions---and corresponding evaluation metrics---are nowadays commonly used that consider more than point-wise utility estimates, see also Section \ref{sec:evaluation} on the evaluation of recommender systems.

\section{Recommendation Algorithms}
\label{sec:algorithms}

The development of recommendation algorithms has naturally mirrored the evolution of the task definition, hand in hand with the design of evaluation procedures and metrics (which we discuss later in Section \ref{sec:evaluation}) suited to the task. Recommendation can be addressed, in essence, as a supervised learning problem: given examples of observed user choices, we aim to predict present or future (yet unobserved) user interests. Variations in the task formulation give rise to different algorithmic approaches---and different metrics are appropriate to evaluate for different tasks.

As a machine learning problem, recommendation is quite unique. What makes recommendation singular in this field is, in essence, the human factor at the core of recommendation tasks. In these tasks, both the input signal and the prediction target consist of or involve user behavior at their core. This brings about a specific level of complexity compared to, for instance, recognizing shapes in an image or diagnosing a medical condition from medical tests. Furthermore, recommendation is often not just about predicting people's actions, but about enhancing (and hence changing) such actions by bringing awareness about potentially better choices. What makes a recommendation good (and therefore the algorithm that produces it) thus involves a great deal of subjectivity and is a challenging question, that we discuss later in Section \ref{sec:evaluation}.

Recommendation algorithms have been traditionally classified into \emph{collaborative} and \emph{content-based} \citep{Adomavicius2022}. The latter follow the principle that people's tastes are related to inherent item characteristics and tend to persist over time. The former build, in a myriad different ways, on the presumed existence of regularities and structure in the distribution of user preferences and choices over the user-item space. For instance, some users may have non-random similarities to other users in their personal interests. Again, such structures are assumed to persist over time.

Different algorithmic approaches have different strengths and weaknesses. Robust and effective recommender systems therefore combine elements or entire components from different such approaches. In an aim to make algorithm categorization exhaustive, such compound solutions have been often referred to as \emph{hybrid} systems. As with many machine learning or information retrieval systems, production recommender systems typically combine a core initial (commonly hybrid) algorithm that produces base item rankings, followed by post-processing algorithms, business rules (e.g. implementing marketing constraints) and fine enhancements to further improve the final recommendation ranking quality and optimizing for additional objectives \citep{ab15rshb,covington2016}.

\subsection{Recommendation as a Machine Learning Problem}

Modern recommendation algorithms tend to be formulated explicitly as a supervised machine learning task: a utility function $f$ (in some of the aforementioned forms in the previous section) is learnt that minimizes a cost function, which involves the utility function within its definition. Cost can be, for instance, the sum of squared prediction error over available training ratings (the latter playing the part of labels in supervised machine learning terminology), or the binary cross-entropy loss of a sigmoid function, or the amount (measured through some monotonic function) of item pairs incorrectly ranked by the utility function to be learned. In this formulation, recommendation algorithms are distinguished from each other in the form the utility function takes (a parameterized model), the cost function, and the procedure to minimize the latter.

An example utility function is the dot product of user and item \textit{embedding} vectors\footnote{Embeddings are low-dimensional (dense), learned representations of objects in the form of vectors of real numbers, see also Section~\ref{sec:mf}.}, where the coordinates of such vectors are the parameters of the model to be learned, as we shall see in Section \ref{sec:mf} below. The function therefore involves some arbitrary choice (of a family of functions often referred to as the \textit{hypothesis} space), and learning the function typically means learning its parameters (a ``model'') by fitting them to the training data (through minimization of the cost function on training labels). Some algorithms however do not express an explicit cost function, and the utility function is defined heuristically rather than learned. This is the case, for instance, of the common nearest-neighbors collaborative filtering formulation that we overview later in Section \ref{sec:knn}. In either case, the utility function has hyperparameters that need to be optimized, usually taking a separate validation subset out of the training data. The ultimate goal of all algorithms is to maximize some final metric(s) of interest, such as precision, recall, or revenue---this is why some algorithms take the final metric, or a more tractable approximation thereof, as the objective function to be optimized.

Next, we provide an overview of the main algorithmic approaches and principles in the recommendation field. We start by briefly discussing different types of input data for recommendation, which determine broad differences in the algorithmic approaches developed thereupon. Each algorithmic family may deserve a paper (or an entire book) by itself. Therefore, we will provide only a broad overview here and focus on some main highlights.

\subsection{Characterizing Approaches Based on their Input Data}
\label{sec:algorithms:cf}

Recommendation algorithms can be based on input data of different nature. Probably the most common data source for recommendation---and we might add, the one enabling the most powerful approaches---are records of logged user-item interactions: the rating matrix, as already discussed. When this is the only input of an algorithm we say the algorithm follows a collaborative filtering approach \citep{TowardsGedas2005}. Collaborative filtering is based on the abstract principle that people can benefit from the experience and discoveries of other people, and not just their own, in making future choices. The simplicity of this principle and the high potential of collective data as a source for prediction are also the Achilles heel of collaborative filtering as a pure approach: where the interaction matrix is sparse, the algorithm struggles to produce reliable predictions, due to a lack of sufficient input information. Collaborative techniques may fail to deliver proper recommendations in cold start situations where little interaction has yet been recorded, as is the case at the early stages of a new deployed system, or whenever a new user or a new item enter the system. However, when sufficient interaction records are available, collaborative filtering is one of the most effective and powerful approaches.

All other input that a recommendation algorithm can take beyond user-item interactions is often referred to as \emph{side information}, see Section~\ref{subsec:cb-and-hybrid}. Using such information is a good complement to collaborative filtering, particularly in sparse regions of the user-item matrix. A most common type of side information is any available data and metadata directly associated to the items, to which we may refer as item features: taxonomic classifications (e.g., movie genre, music style, online store section), tags, free text associated to the items (e.g., product descriptions and reviews), etc. Algorithms that solely use such data are referred to as being \emph{content-based} \citep{pb07aw}. Instead of turning to other users for hints on what a customer might like, pure content-based approaches just focus on one user at a time, considering the interaction records of the target user\footnote{The term \emph{target user} is commonly used to refer to the user for whom we make recommendations.}---in particular, the features of the items the user engaged with, and the features of the items to be recommended. Several other information sources can be effectively exploited by a recommender system to enhance the effectiveness and value of recommendation, such as social interactions between users \citep{g22rshb}, data about the user (e.g., demographic information), contextual information (e.g., user and/or item geolocation, session, time, device, etc.).

When substantial interaction data is available, collaborative filtering can be more effective than other approaches as a base algorithmic core, not just in terms of delivering accurate predictions, but also in providing users with rich options beyond their individual prior experience \citep{vc11recsys}. These algorithms are moreover fairly general and domain-agnostic, as they make no assumption about what the items are.

\subsection{Collaborative Filtering}
\subsubsection{Nearest Neighbors}
\label{sec:knn}

The earliest collaborative algorithms were inspired by a typical word-of-mouth human behavior where a person takes advice from trusted friends when making a decision---one of the main sources for guidance when people make decisions and choices.
Following this metaphor, so-called $k$-nearest-neighbor (kNN) algorithms suggest choices defined as the weighted average of the advice of the target user's friends.

Following this heuristic intuition, recommendation may be viewed as a regression problem, where a kNN algorithm predicts the level of preference of a user $u$ for an item $i$ as a linear combination of the observed preferences for $i$ in a subset of selected users: the target user's neighbors $N[u]$. Following the notation in the previous section, items are ranked by the following utility function:
\begin{equation}\label{eq:knn}
    f(u,i)=C_i \sum_{\substack{v\in N[u]\\r(v,i)\neq \emptyset}} w_{u,v}\ r(v,i)
\end{equation}
where $C_i$ is a normalizing term, and $r(v,i)\neq \emptyset$ means that interaction data (a `rating') is available involving $v$ and $i$. Typical definitions for the weights in the linear combination are based on pairwise user similarity $w_{u,v}=\textrm{sim}(u,s)$, where the similarity function can be, for instance, cosine, or Pearson correlation between the rating vectors of users $u$ and $v$. The standard definitions of those measures need to be further particularized to deal with missing values in the user vectors. For instance, when using cosine similarity, missing ratings may be assigned a value of zero.

The neighborhood $N[u]$ is usually picked as the top $k$ most similar users to $u$, where $k$ becomes a hyperparameter of the algorithm. Neighborhood selection is nonetheless a modular operation that has been the object of multiple variations and enhancements. When recommendation was addressed as rating prediction, the kNN score in Equation \ref{eq:knn} above was normalized by $C_i=1/\sum_{v\in N[u]:r(v,i)\neq \emptyset}|\textrm{sim}(u,v)|$, in order to produce values in the rating scale. When evaluated as a ranking task, omitting normalization (i.e. $C_i=1$) has often been found to produce better results \citep{ckt10recsys}.

Multiple further variations of the kNN scheme arise from here, such as the widely used item-based version (which roughly swaps the role of users and items), multiple similarity functions, different neighbor selection approaches, and so forth. Though kNN has been developed as a heuristic scheme, Cañamares and Castells \citep{cc17sigir} showed that kNN can be given a probabilistic formalization where the ranking function is derived from an objective function to be maximized: the expected number of recommended items the user will like. For a more comprehensive review of kNN variants see \citep{ndk22handbook}.

kNN is an old idea \citep{linden2003} yet it remains a competitive approach today \citep{ferraridacrema2020tois,ludewiglatifiumuai2020}. It is very easy to understand, simple to implement, and computationally inexpensive compared to other recent, more complex approaches. As such, it is an advisable reference baseline to include in experimental recommender systems research \citep{ckt10recsys,cc18sigir}.

\subsubsection{Matrix Factorization}
\label{sec:mf}

By the mid 2000s \textit{matrix factorization} became popular in the field and very soon became the preferred collaborative filtering approach, due to its empirical effectiveness, relative simplicity, and flexibility as a framework enabling multiple algorithmic developments and variations. Matrix factorization is based on the assumption that the interests of users can be described in a low-dimensional space of latent factors that synthesize the subjacent properties of items that determine why a person may like them \citep{kbv09computer}. In this perspective, users and items are assumed to be describable as vectors in a common latent factor vector space, in such a way that users with similar tastes would have similar factor vectors, and so would items that please similar users. These low-dimensional vectors have come to be referred to as \textit{embeddings} in the recent literature, making a connection to similar techniques in fields such as natural language processing and information retrieval \citep{mikolov2013, pennington2014}.

The latent factors do not necessarily correspond to actual properties of items or user traits as we would probably represent them in our human understanding. They are handled as abstract dimensions for which users and items are given values by an algorithmic approach, and are usually not interpretable by eye inspection or a clear natural intuition. Matrix factorization algorithms only require deciding how many factors we wish consider---which becomes a hyperparameter of the approach---but not what these factors are, other than axes of a multidimensional vector space.

Formally, for a $k$-dimensional factor space, users $u\in U$ and items $i\in I$ are represented by vectors $p_u\in\mathbb{R}^k$, $q_i\in\mathbb{R}^k$, where the coordinates $p_{u,z}\in\mathbb{R}$, $p_{i,z}\in\mathbb{R}$ represent how interested user $u$ is in factor $z$, and how much of factor $z$ is present in item $i$, respectively---the higher the $z$ coordinate value, the more the user interests are about this factor, and the more important the factor is in the nature of the item. Based on this representation, an additional assumption is made: that the interest of a user $u$ for an item $i$ can be captured by the dot product of their latent vectors---following our utility function notation, $f(u,i)=q_i^t p_u$.

The common approach to produce the latent vectors is by minimization of a cost function $\mathcal{R}$ (also referred to as risk or expected loss in the empirical risk minimization machine learning framework \citep{vapnik1998}), which generally involves the error in predicting the training data with some regularization. For instance, a basic formulation would be:
\begin{equation}\label{eq:mf}
    \mathcal{R}(p,q;\lambda) = \sum_{u,i}(r(u,i)-q_i^t p_u)^2 + \lambda(\|p\|+\|q\|)
\end{equation}
The term added to the squared difference is a common regularization factor to avoid overfitting, where $\|\cdot\|$ represents the L$_2$ norm, and $\lambda$ is a hyperparameter that is tuned by cross validation. The minimization is often solved for by stochastic gradient descent or alternating least squares, resulting in easy implementations \citep{kbv09computer}. Once the latent vectors are learned by the algorithm, the dot product $f(u,i) = q_i^t p_u$ is used as the ranking function for recommendation.

The summation in the cost function can be limited to the user-item pairs for which training data are available, but it is common to extend the sum to all or a sample subset of unobserved user-item preferences \citep{kbv09computer}. In the latter case, a value needs to be imputed to the missing preference observations, which can be done in different ways (such as a constant parameter value or a randomized value).
The error term in Equation \ref{eq:mf} can be weighted differently for every user-item pair, often representing confidence in the corresponding (observed or imputed) preference data point. Most usually, two different weight values are used for pairs having and not having training data \citep{hkv08icdm}. The combination of error weighting and missing value imputation reflect different nuances in the recommendation task and result in different algorithmic behavior (see e.g. \citep{s13recsys}).
Besides these configuration options, matrix factorization has been rich in a myriad of further variations, such as summing item, user and global bias terms as additional parameters in the rating representation $q_i^t p_u$ \citep{kbv09computer}, temporal awareness \citep{k09kdd}, probabilistic formulations \citep{h04tois,sm07nips}, and many other elaborations.
In particular, variations in the cost function have given rise to entire new approaches, which we summarize next.

\subsection{Learning to Rank}

With the realization that the effectiveness of recommendation in real scenarios relies more on item ranking than on point-wise rating prediction, a shift emerged in the field towards addressing recommendation as a learning to rank (LTR) problem, drawing from a similar earlier trend in the information retrieval field \citep{j01kdd}. Broadly speaking, LTR involves the introduction of loss functions that explicitly involve item ordering rather than rating error or interaction prediction error \citep{rfgs09uai}. In different possible ways, the cost functions typically represent the amount of ``contradiction'' between the rankings produced by the model to be learned, and how items should be ranked according to the training preference values. In other words, the model is trained to maximize order-wise agreement with the available observations. Most methods take a pairwise approach, where the ``ranking error'' is defined in terms of incorrectly ranked item pairs:
\begin{equation*}
    \mathcal{R}(\theta) = \sum_u\sum_i\sum_j \ell(u,i,j|\theta)
\end{equation*}
where $\theta$ are the parameters of the recommendation model, and the loss $\ell(u,i,j|\theta)$ quantifies how much in contradiction to the available observations the model $\theta$ ranks $i$ and $j$ for $u$ (and $\ell$ involves the utility function $f$---the model---to be learned). This is summed over all pairs of items $i,j$ that $u$ has available training data for.

A representative and effective example in this area is Bayesian Personalized Ranking (BPR) \citep{rfgs09uai}, which has become a common reference and baseline in the literature. Ranking inconsistency in BPR is expressed and developed in terms of the probability that the scores predicted by the learned model $\theta$ rank pairs of items in contradiction to the training rating data: essentially, and omitting details, $\ell(u,i,j) = \mathbb{1}_{r(u,i)>r(u,j)}P(j \textrm{ is ranked above } i \textrm{ for } u\, |\, \theta)\, P(\theta)$. The probability of ranking precedence is smoothed (for differentiability) as a logistic sigmoid of the ranking score difference.

For instance, for a matrix factorization model where $\theta$ consists of the latent vectors $p_u$, $q_i$ for $(u,i)\in U\times I$, and the utility scoring function is $f(u,i)=q_i^t p_u$, we get $P(j \textrm{ is ranked above } i \textrm{ for } u\, |\, p, q) \coloneqq 1/(1+e^{q_j^t p_u - q_i^t p_u})$, and under normality and factor independence assumptions, the prior $P(p,q)$ results in a typical L$_2$ regularization term. The resulting cost function can then be minimized by stochastic gradient descent.
BPR can thus be seen as a layer on top of matrix factorization and a means to train the latent vectors, alternative to point-wise (i.e., user-item-wise) cost minimization.

Many alternatives to such a scheme have been proposed on a similar principle. For instance, also in a pairwise approach, RankALS \citep{takacs2012} essentially takes $\ell(u,i,j) = ((r(u,i)-r(u,j)) - (q_i^t p_u - q_j^t p_u))^2$ as the core pairwise ordering error to be minimized, and alternating least squares instead of gradient descent as the minimization procedure. Other approaches introduce a target ranking evaluation metric---such as nDCG, Mean Average Precision (MAP) or Mean Reciprocal Rank (MRR)---in the objective function, i.e., the objective approximates how much is lost in the metric by a suboptimal item ordering. For instance, CLiMF \citep{sbloh12recsys} defines the objective function as a smooth lower-bound approximation of MRR. These methods are referred to as \emph{listwise} in the LTR literature \citep{l09now}, because even though the loss function is still often evolved into pairwise form, it reflects a ranking goodness function (the target evaluation metric) rather than a metric-agnostic item pair classification error in a binary order.

\subsection{Neural Recommendation}
While matrix factorization assumes the dot product of latent vectors as the scoring function, one may consider more complex, not necessarily linear\footnote{Technically, early matrix factorization models should be referred to as \textit{bilinear}: the scoring function is a polynomial of degree two on the model parameters (hence not really a linear function), but is linear, separately, with respect to the user parameters (latent factors) and the item parameters---this is what bilinear means.} utility scoring for recommendation. One such possibility is to use neural networks as a particular family of non-linear functions that have well-studied properties and training techniques \citep{hlznhc17www,zhang2022}. After revolutionizing machine learning application domains such as computer vision, speech recognition, or natural language processing, deep learning has gained blazing popularity in recent years as a basis for devising a wide variety of recommendation approaches. A comprehensive coverage of this area exceeds the scope of this paper---we provide instead a summarized overview; the reader is referred to e.g., \citep{zhang2022} for a wider survey.

Several technical and strategic reasons motivate an optimistic prospect for deep learning as an approach to recommendation: the ability to approximate any prediction function; efficient training (in comparison to other machine learning approaches at comparable complexity); smooth integration of heterogeneous data sources (including unstructured content), data dimensions and predictive signals; feature engineering effort savings; proved empirical effectiveness in other machine learning domains; and a growing availability of software resources and knowledge derived from a thriving activity in deep learning in many domains and fields. Furthermore, as universal function approximators, neural networks can be seen as a smooth generalization of virtually any simpler model (such as dot-product in matrix factorization as described earlier). From this perspective, a neural vs. non-neural dichotomy may become moot, strictly speaking. ``Depth'' in deep learning suggests however that the opportunity to grow and handle complexity is actually being leveraged by ``bigger'' layered models.

Neural networks provide arbitrarily high expressive power to capture complex relations in the user-item space, that
matrix factor models may fail to grasp. At the top layer (the output layer), the loss function can involve a pointwise (rating or binary prediction error) \citep{hlznhc17www} or pairwise (ranking error) element \citep{sycx19cikm} similar to other matrix factorization and LTR approaches discussed earlier. The generality of neural architectures further enables the smooth integration of side information in addition to user-item interaction. We discuss the use of side-information in recommendation more generally later in Section~\ref{subsec:cf-with-side-information}. The elaborations and combinations that can be developed in this area are as wide as imagination can afford. A huge variety of complex network architectures have been proposed in the literature using network structures such as autoencoders \citep{liang2018}, convolutional neural networks \citep{caser2018,yuan2019}, recurrent neural networks \citep{hidasi2018, wu2017}, and attention layers \citep{SASRec2018}, to name just a few. Particular architectures have been developed for specific tasks, such as session-based recommendation discussed later in Section \ref{sec:sequential-session-based}. Interestingly, the best performance in recommendation is sometimes
achieved with shallower neural models compared other domains \citep{steck2019,anellitop2022}.

Deep learning has proved to be empirically effective in recommendation, and steps have been taken towards its adoption in industry \citep{covington2016,wg20sigir,otot17kdd}. A certain hype in this domain may have also induced a degree of noise and haste in early empirical analyses that sometimes make it difficult to get a precise perception of the actual effectiveness of neural approaches, or any conditions for which they may be best suited \citep{ludewiglatifiumuai2020,kouki2020,ferraridacrema2020tois,rkza20recsys,Garg:2019,steck2019}. Clear effectiveness improvements have been found in specific tasks such as sequential recommendation \citep{slwploj19cikm},
and for specific conditions when, for instance, side information or massive data are available. Deep learning can often also simplify and reduce feature engineering efforts. See \cite{steck2021} for a discussion of deep learning for recommendations at Netflix.

Deep learning faces similar challenges in recommendation as it does in other domains: a large number of hyperparameters to tune, a heavy training cost tradeoff to reach the effectiveness potential, the need for massive data availability, and a black box nature involving interpretability and explainability challenges \citep{ah20recsys}. As a result, the cost and complexity of neural approaches may not necessarily always pay off in enhanced effectiveness as a universal solution for all recommendation settings and problems \citep{ferraridacrema2020tois,rkza20recsys}. On the other hand, the data richness and volume often required to make the best of deep recommendation architectures is not always within reach of research outside corporate boundaries. Neural recommendation is nonetheless a thriving area where a profuse stream of work is being published in top tier research outlets as we speak,
and industry is striving to leverage its potential \cite{steck2021,covington2016,cen2020,feng2020}.
We suggest the reader to follow through recent literature and surveys \citep{zhang2022} and to stay tuned to ongoing developments and future potential breakthroughs.

\subsection{Content-based and Hybrid Recommender Systems}
\label{subsec:cb-and-hybrid}
Collaborative filtering (CF) methods, as discussed in the previous section, are highly effective in many application scenarios due to their ability to detect and utilize preference or behavior patterns in a user community. Furthermore, they are particularly good at helping users to discover types of content that were previously unknown to them, e.g., by considering preferences of like-minded users. One intriguing aspect of CF methods is that they are able to accomplish this even without knowing anything about the items themselves, for example, an item's category in an e-commerce store.

However, in cases where we dispose of such knowledge about the items and a user's preference towards such individual item properties, it may of course make sense to consider it in the recommendation processes. For example, if we know that a user usually prefers certain movie genres or types of news, it is only intuitive to recommend more content of the same type. Moreover, in case we do not have logged transaction or rating data yet, i.e., if the user has not rated many items yet (\emph{user cold-start}) or if there are new items in the catalog without purchase history (\emph{item cold-start}), collaborative filtering methods may not work well.

\subsubsection{Pure Content-based Systems}
\label{subsec:cb-systems}
In such cases, when there is little or no collaborative information available, one option is to match \emph{user-individual} past preferences with knowledge about item features. Historically, this process is called \emph{content-based filtering}\footnote{A prominent use case in early recommender systems was the personalized filtering of incoming news messages, and the filtering in these applications was based on the document contents, i.e., on text documents. The term \emph{content-based} was later on also used for systems which rely on item meta-data, e.g., genres. See \cite{pb07aw} for an early survey on the topic.}.
Technically, many early content-based filtering methods were based on ideas and techniques from information retrieval. Both in recommendation and retrieval scenarios, the goal is to identify and rank a set of relevant documents. While in the IR case the starting point is a (search) query, the retrieval process in a recommender system is based on a content-based user profile.

On a general level, content-based recommendation can be characterized as follows \cite{TowardsGedas2005}:
\begin{equation}
f(u,i)=\text{\emph{score(ContentBasedProfile(u,Content(i))}}
\end{equation}

\noindent where the predicted relevance $f$ for a given user $u$ and item $i$ is based on a scoring function \emph{score} which matches a content-based profile of $u$ with the content features of an item $i$.

Various ways of implementing the details of such a content-based approach are possible. In case where textual descriptions of an item $i$ are available, a traditional way would be to represent the items using a TF-IDF (Term-Frequency Inverse Document Frequency) encoding. In such an implementation, the function \emph{Content(i)} would return a vector of real values, where each element of the vector corresponds to a term (word), and the value indicates the importance of the term. The importance values in a TF-IDF encoding are determined by multiplying two relatively simple counting statistics, TF and IDF. The term frequency TF corresponds to the number of times a word appears in the given document $i$, usually normalized to account for different document lengths. The Inverse Document Frequency IDF in contrast indicates how ``informative'' a term is in the given document collection. The underlying logic is that if there is a term that appears in almost all documents, it carries little information to distinguish one document from each other.

The next question is how to represent the interest profile of a user $u$, i.e., how to implement the function \emph{ContentBasedProfile}. A common goal is to use a representation that can be easily matched with the content representation of individual items. In the case of text documents, one could for example first encode all items which the user liked in the past with TF-IDF. Then, the \emph{ContentBasedProfile} could be defined as being the average of the vectors of the liked items. Finally, now that the user profile and the items are represented in a compatible way, the function \emph{score} can for example be implemented through the \emph{cosine similarity} function.

Various alternative implementations of the different functions are of course possible. The options range from even more simple approaches, e.g., by counting overlapping features such as actors in a movie recommender systems, to more elaborate embeddings, which aim to better capture the semantics of the documents.\footnote{While TF-IDF encodings often lead to good results in certain applications, a clear limitation of this simple (``shallow'') encoding is that there is no meaning attached to the words and that the positions of the terms in a document are not taken into account.} The prediction function can of course be learned in a supervised approach, with input features based on item content information, and any sort of suitable shallow or deep model. Furthermore, a variety of additional pieces of information beyond the document itself were considered in the past decades to better understand the similarity or relations between different items. Besides item meta-data, various forms of \emph{exogenous} information about the items, e.g., from Wikipedia, have been incorporated into content-based (or: semantics-aware) recommender systems, see \cite{Musto2022} for an in-depth discussion.

In terms of application areas, pure content-based systems are often used when there is not sufficient collaborative information available. A typical use case is the recommendation of news, where we have to deal with a constant stream of new items, see \cite{KirshenbaumForbes} for a case study from industry. Content-based systems have however also been applied in practice in other domains such as movies or mobile apps \cite{Bambini2011,JannachHegelich2009}, where the user preferences can be relatively stable, e.g., in terms of the preferred genres. A special use case of content-based methods are ``similar item'' recommendations, where the reference point to making a recommendation is not a user profile, but a specific item. Similar item recommendations are common on video streaming platforms, e.g., to recommend movies that are similar to the user just has watched, see \cite{Yao2018judging,TrattnerJannach2019}.

\subsubsection{Hybrid Recommender Systems}
\label{subsec:hybrids}
Pure content-based systems can have certain limitations. An intrinsic feature of these systems is that they recommend ``more of the same'' by design, thus limiting the support for discovery of new types of content for users. In addition, content-based systems might surface content that is too niche. A movie recommender system that for example only considers the genre or plot descriptions may miss important quality aspects of a movie. It may thus return movies which are content-wise related to those that the user liked in the past, but in the end are not recommendations that the user will like.

A common solution to deal with such problems is to build \emph{hybrid} systems, i.e., systems that not only rely on one single paradigm, e.g., collaborative filtering, but combine different approaches. This way, the shortcomings of individual methods should be avoided while the advantages are combined. In our example of content-based movie recommendations, one could apply some quality filters before the recommendations are returned, e.g., by removing all movies that have a low average community rating.

In the literature, a variety of ways of building hybrid recommender systems have been identified. In \cite{Burke02umuai}, Burke identified seven such ways, which were later organized in three larger categories in \cite{JannachZankerEtAl2010} as follows:

\begin{itemize}
  \item \emph{Monolithic}: In such a design, aspects of different recommendation strategies are implemented in one algorithm. A very common realization of such an approach is to design a machine learning model that includes both collaborative signals and content-based features (``side information'').
  \item \emph{Parallelized}: In parallelized designs, the outcomes of two or more algorithms are first determined independently and then combined in some way. The combination could for example be done on the user interface level, where users are presented a list that contains personalized recommendations as well as recently trending items. Or, some weighted approach can be applied where each item receives a score from each recommender. An extreme case of such a weighting approach would be a switching hybrid, where the recommendations are always taken from one particular algorithm, depending on the situation.\footnote{In the mobile game recommender system in \cite{JannachHegelich2009}, for example, a content-based technique was used until the user had purchased a certain minimum number of items.}
  \item \emph{Pipelined}: In this design, one algorithm uses the outputs from another one as an input. In practice, one could for example retrieve a number of recommendation candidates based on popularity, and then only rank these items according to the assumed user preferences in a personalized way.\footnote{Such staged approaches are common in industrial settings with huge item catalogs, e.g., at YouTube~\cite{covington2016}.}
\end{itemize}

\subsubsection{Collaborative Filtering with Side Information}
\label{subsec:cf-with-side-information}

\begin{figure*}[h!t]
    \centering
    \includegraphics[width=0.75\textwidth,clip]{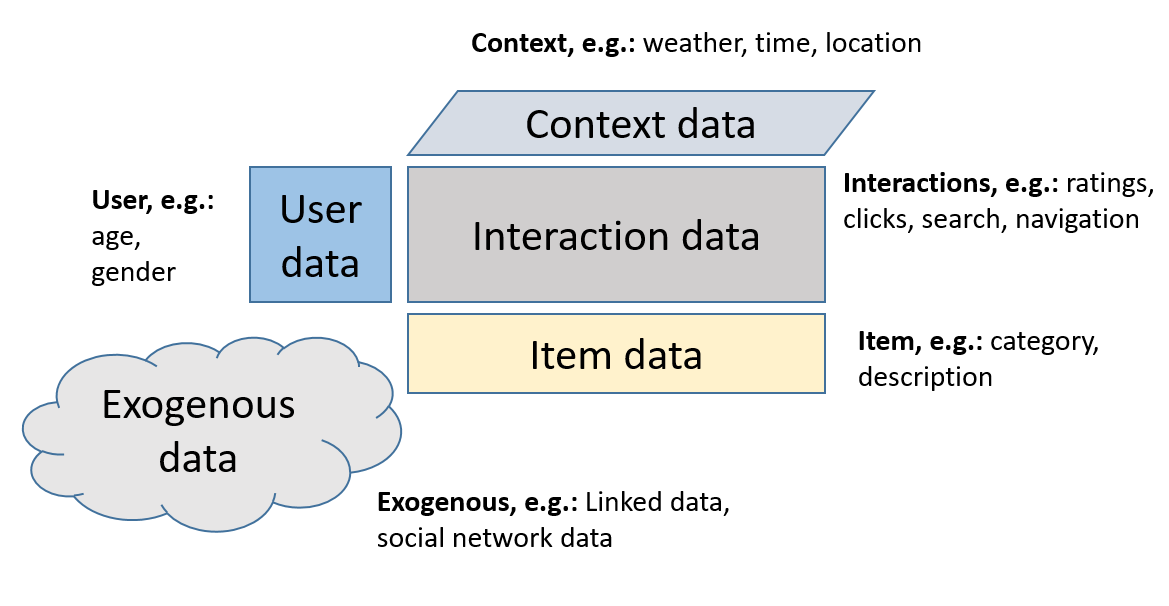}
    \caption{Types of data (extended from \cite{Wu2022}).}
    \label{fig:types-of-data}
\end{figure*}

Probably the most common hybrid approach is to enhance the power of collaborative filtering with various forms of side information. Such side information is however not limited to item-related information as in content-based filtering approaches. Wu et al.~\cite{Wu2022} categorize side information into the following categories of information that a recommender system can use in addition to the observed user-item interaction data. Figure~\ref{fig:types-of-data} 
visualizes the different forms of information.

\begin{itemize}
  \item \emph{User} data: These are static or slowly changing features of the user, such as age, gender, or nationality. Historically, systems that leverage such information were called \emph{demographic} recommender systems. In addition, researchers have also explored the consideration of personality traits in the recommendation process, see~\cite{Dhelim2022}.
  \item \emph{Item} data: These are aspects that are tied to specific items. Various forms of domain specific item-related features has been considered in the literature, including pre-structured ones like categories, unstructured ones like textual item descriptions, reviews or images, or community-provided semi-structured information in the form of tags.
  \item \emph{Context} data: Various forms of comparably frequently changing context information have been explored in the literature, most importantly the geographical location of users, the time of the day, weather conditions or the end user device, see \cite{Adomavicius2022}.
\end{itemize}

In terms of the user-item interaction data, research has historically focused on ratings as the only form of interaction. In more recent years, recommending based on implicit feedback signals has been become predominant, see \cite{JannachLercheZanker2018} for a survey. In most published research works, only one type of implicit feedback is assumed, indicating, e.g., if a user has interacted with an item or not. In reality, multiple types of interactions are available and could be used. In an e-commerce shop, the available signals may included item views, add-to-cart-events, purchases, category navigation events, search actions and even later item returns etc. Research on leveraging combinations of such types of interaction data is somewhat limited today, probably due to the application specific nature of the interaction events. Moreover, in real-world interaction logs, information is often stored about the point in time when an interaction happened. With that information, time-aware recommender systems \cite{Campos2014} and sequence-aware based approaches can be implemented, see also Section~\ref{sec:sequential-session-based}.

Finally, as mentioned above, various approaches exist in the literature that aim to incorporate \emph{exogenous} information into the recommendation process. Examples of such information (including ``world knowledge'') are Wikipedia articles, publicly accessible domain ontologies and knowledge graphs, or data that can be accessed or queried through Linked Open Data endpoints, see \cite{Musto2022} for a discussion. Besides such types of information that refer mostly to the items, exogenous information that is tied to individual users has been explored in the literature as well, for instance, the users' social network, their trust relationships, or reviews written by them, see \cite{DONG2022108954,He2010,Chen2015reviews}.


Recent surveys on technical approaches to build ``information-rich'' collaborative filtering systems can be found in \cite{Wu2022}. and \cite{SUN2019100879}. In particular the latter work provides an overview of how various types of deep learning models incorporate side information of different types, including flat features, hierarchical features, knowledge graphs, image features and so forth.

\subsection{Discussion}
We have reviewed a selection of important algorithmic approaches to recommendation in this section. We emphasize that this is only a high-level rundown and the reader can find a myriad of other algorithms and variations in the literature. For instance, \citet{rendle2010,rendle2012} developed the \textit{factorization machines} framework, generalizing several factor models including matrix factorization. Like neural approaches, factorization machines can smoothly leverage user and item side information in the framework for an integrated hybrid recommendation approach. Around that time and in the scope of linear models, \citet{nk11icdm} proposed SLIM, a generalization of item-item collaborative filtering that, in a way, learns the similarity matrix as a regularized optimization problem akin to Equation \ref{eq:mf} for matrix factorization in Section \ref{sec:mf}. This algorithm has become a common baseline in the recommender systems literature for its empirical effectiveness. Later on, \citet{steck2019} derived a closed form solution for computing a very similar model to SLIM. By removing the need for a costly iterative optimization procedure, the method is considerably more efficient.

Other algorithmic developments are targeted at specific recommendation scenarios such as sequential recommendation \citep{QuadranaetalCSUR2018}, discussed later in Section \ref{sec:sequential-session-based}, and/or revised perspectives of the recommendation problem such as reinforcement learning \citep{Li2016,xin2022}, which we consider in Section \ref{sec:bias}.

Generally, despite the success of in particular of collaborative filtering approaches in practice, recommendation is far from a solved problem: we suggest the reader to retrospect on their own experience as a recommender system user, and form their own opinion. Many elements are involved in bringing the recommendation experience anywhere near, for instance, the effectiveness of modern search engines. Strictly speaking, this is not exactly possible---such a comparison is unfair, as search systems receive an explicit expression of intent from the user: the query. Competing against search would anyway misrepresent the purpose of recommendation: recommendation brings value in areas that search cannot solve, or for which search should not be needed. Be that as it may, room for improvement of current recommendation technology certainly remains, and different angles need to be addressed to achieve it, as we cover in the rest of this paper. The development of better algorithms is certainly one of them, and a major focus of attention and efforts in the field.

\section{Evaluation of Recommender Systems}
\label{sec:evaluation}

As discussed in the previous section, there is a myriad of choices available when one wants to deploy a recommender system, including a multitude of algorithms, their variants and specific configurations. Making an informed choice requires suitable and reliable evaluation methodologies. The design and selection of such reliable evaluation methodologies is therefore a central piece in recommendation technology development and research. At the same time, these evaluation methodologies drive algorithmic evolution in the field---as a fitness function, they shape the state of the art.

Evaluation generally involves a comparison between two or more recommender systems or variants. The most direct approach to compare systems in a production setting is online A/B testing, see for example \citep{gfdabh14recsys}. Online experiments are a finite resource though, they take time and involve a potential cost in user experience degradation by exposition to suboptimal system variants. Offline evaluation is therefore used as a complementary instrument to filter out which variants and change proposals are brought to more expensive online testing \citep{gccmhtc19wsdm,sg22handbook}. Offline evaluation enables, on the other hand, fast and safe hyperparameter exploration, and remains the main empirical resort widely available to academic research.

In the sections that follow we summarize the main lines of approach to recommender system evaluation. We briefly address online approaches (a perspective that will find further elaboration later in Section \ref{sec:impact-and-value}) to then focus on offline evaluation. More specific evaluation methodologies in the context of session-based recommendation are also discussed in Section \ref{sec:sequential-session-based:evaluation}.

\subsection{Online Evaluation}
\label{sec:evaluation:online}
Recommendation technologies as we experience them on Spotify, Netflix, Amazon, YouTube, Booking, Twitter, Facebook---you name it---have undergone a filtering funnel from the early conception of an algorithmic idea (perhaps in an academic research context) to the final production system \citep{ab15rshb}. This pipeline involves a combination of offline and online experimentation where new ideas are compared to each other, to established baselines, and finally, to the recommendation algorithms currently operating in a live application. By and large, offline experimentation precedes online evaluation, given the cost and bandwidth constraints of live testing \citep{gccmhtc19wsdm}. The final test in validating a new idea consists in launching it in the production platform alongside the current version, driving a fraction of user traffic to it, and comparing its performance to that of the present system---whichever wins takes over or remains the working system version.

This is called an A/B test, since two systems A and B are compared, where by A one usually means the current system (often referred to as \emph{control}, borrowing from clinical trial terminology), and B is the challenger (\emph{treatment}). In practice, it is common to run so-called multivariate tests comparing more than two systems simultaneously. Popular applications such as the ones mentioned at the beginning of this section are running hundreds of simultaneous A/B tests as we speak \citep{ab15rshb}. One genuine characteristic of A/B tests is that users do not know which version of the system they are facing, just like subjects in drug research do not know whether they are dispensed the actual treatment under trial (system B) or a placebo (system A). This aims to maximize the objectivity of the comparison by avoiding the introduction of confounding distortions in the experiment.

In addition to testing ideas in the most realistic possible conditions, A/B tests allow measuring the effects of algorithms on ultimate business performance and goals, beyond theoretical, more speculative effectiveness metrics. Effects can be measured on engagement (clicks, playcount, dwelling time), sales (revenue, profit, number of purchases), customer retention, and any other metric or key performance indicator the business relies on. We further elaborate on the business value of recommendation from wider perspectives in Section \ref{sec:impact-and-value}, also considering other possible stakeholders---beyond the customer and the system---involved in and impacted by recommendations.

A/B tests are commonly run until a statistically significant difference is found between system A and B, or until an affordable time limit has elapsed without a clear conclusion. A typical duration for an A/B test is in the order of weeks---the time needed to collect sufficient evidence to make data-driven decisions about system changes. This, and the risk involved in exposing customers to new untested changes, limits the number of experiments than can be run simultaneously on a production platform. For this reason, offline evaluation is used to shortlist a small selection of change proposals to be brought to live evaluation---we discuss offline experimentation methodology in some detail in the next section.

One open challenge to this respect is the weak or missing correlation often observed between the outcome of offline and online comparisons \citep{Gomez-Uribe:2015:NRS:2869770.2843948,jannachjugovactmis2019,gfdabh14recsys,kouki2020}.
While this remains a major open research question in the field \citep{jannach2021mcnamara,gcnad18wsdm,rsz16recsys}, offline evaluation is routinely practiced as a selection yardstick before online testing in industry, and is by far the main resource for empirical observation in academic research.

Besides A/B testing, other forms of online experimentation and user studies  can be designed with specific purposes \citep{kw15rshb}, not necessarily involving a production system. We discuss such studies with humans in the loop in Section \ref{subsec:impact-of-rs}.

\subsection{Offline Evaluation}
\label{sec:evaluation:offline}

An offline experiment can be seen as a simulation of the system interacting with users, where different proxies---i.e., offline metrics---for system effectiveness are measured \citep{castells2022}. The most informative offline experiment is the one that best simulates and represents the production setting. In industrial developments, very specific production settings may be required. Therefore, in general research, a number of abstractions are usually made.

The difference between online and offline evaluation can be neatly defined as follows: whereas online evaluation acquires user feedback data (for metric computation) \emph{after} recommendations are delivered to users---from these same users---in offline evaluation the test data is collected \emph{before} recommendations are produced---and users are in fact never delivered the evaluated recommendations.

Setting up a basic offline experiment for evaluating a recommendation algorithm may seem a straightforward endeavor at first sight, given the wide body of well established methodologies and experimental design principles, in fields such as machine learning and information retrieval, to draw upon. The recommendation task has however peculiarities of its own that result in a fair degree of hidden complexity and pitfalls, that can produce inconsistent evaluation outcomes more easily than one might think \citep{ccm2020irj}. The experimenter is advised to carefully consider, understand and report the detailed design choices made in an experiment and their implications. We overview some of them in the following sections.

\subsection{Offline Data}
\label{sec:evaluation:data}

Offline evaluation generally distinguishes \emph{training} data and \emph{test} data, which should be strictly disjoint \citep{g22rshb}. Training data is supplied as input to the recommender system being evaluated, whereas test data is hidden from the system and is used to compute metrics on the returned recommendations. Training data in an offline experiment can be similar to the data that a production system may have available (the more similar the better), whereas test data is used in the experiment to simulate user feedback in reaction to the delivered recommendations. The training and test data for recommendation usually reflect user-item interactions as one of their major components---that is, they can be seen as disjoint subsets of the rating matrix. Training data can however also include additional side-information that specific recommendation algorithms may consume.

Training and test data can be acquired in many different ways. A good approach for the training set is to export a subset of the input data that a real recommender system is using at a certain point in time.
Over the years, a number of such datasets were made publicly available, most prominently the datasets from the non-commercial MovieLens system\footnote{https://www.movielens.org}. Various other datasets are nowadays used by researchers containing items rating, purchase information or listening events, see also \cite{MovieLens2015}\footnote{A collection of additional datasets can be found at \url{https://github.com/RUCAIBox/RecSysDatasets}.}. Most recently, several datasets were also published that do not contain a matrix of user-item interactions, but sequential logs of recorded interactions, which can be used to evaluate session-based recommendation algorithms, see Section~\ref{sec:sequential-session-based}. Unfortunately, for many of the datasets it is not clear under which particular circumstances they were collected. For example, when an e-commerce or  music streaming platform already has a recommender in place at the point in time of data collection, what we observe in the logs may be to certain extent be biased by the existing recommendation algorithm. In any case, given such a dataset of logged interactions, a trivial means to obtain test data is to subsample from the training set. This is usually referred to as \emph{splitting} the data, and can be carried out in different ways, which we discuss in the next subsection.

The data typically collected for recommender system experimentation displays heavy sampling biases, originated by the working system (its algorithms, user interface and business rules) through which it was collected, external biases (marketing, fashion, social influence, etc.) and behavioral biases in user engagement. This raises issues in evaluation that we discuss later in Section \ref{sec:bias}. Test data can however also be obtained from a separate source from the training set. For instance, the Yahoo! R3 dataset \citep{mz09recsys} includes a test set consisting of randomly sampled ratings from users for music---which means users were required to rate music that was uniformly sampled for them---whereas the training set was collected from free user interaction with music. The Coat dataset \citep{skadljz17nips} was built in a similar way in the clothing domain. The CM100k dataset \citep{cc18sigir} collected user ratings for music entirely at random, and provides a complementary label for user familiarity with the music, which is suggested as a proxy for non-uniformly distributed input training data.

\subsubsection{Data Splitting}
\label{sec:evaluation:offline:split}

When the data collected for offline evaluation does not include a separate test subset, the latter is usually subsampled from the available dataset. The sampling---referred to as splitting---procedure can be carried out in different ways. A first parameter of the procedure is the \emph{split ratio}, i.e. the proportion of train-to-test data, typically expressed as a percentage or a ratio in $[0,1]$. It is common to allocate a larger data share for training (e.g. $\geq 80\%$) than test, given the usual data sparsity challenges when training recommendation models.

A second dimension of choice in the sampling procedure is random sampling vs.~temporal splitting. For a chosen point in time, a temporal split places all the data produced prior to that point in the training subset, and the rest in the test subset \citep{k09kdd}. The time point is sometimes chosen in terms of meaningful time units (a number of weeks, months, etc., worth of data), or in such a way as to obtain a specific split ratio (a proportion between the amount of training and test data). More elaborate variants involving segmentation into multiple time windows have also been explored \citep{l10phd}. A temporal split generally adds to the soundness of the evaluation methodology, as it better represents a real recommendation task: predicting future (or present) usefulness based on past observed user behavior. Furthermore, it provides a cleaner data separation, since mixing past and future records can be seen as a form of data leakage. To this respect, a global time split point can be cleaner than different points for different users (as per e.g., a leave-last-out approach). Otherwise we might be predicting that a user will like some item in the ``future'' based on a leakage of foresight information that, for instance, the item would become popular or trendy by then (according to the training records of users with a later split time point), when in fact the item might not have even been created yet during the training period for that user.

Temporal splitting can however not always be possible, or not strictly required. For instance, meaningful data timestamps might not be available, or user needs might be relatively stable, or the experimenter might aim to focus on a specific problem where time is not a priority variable. The common alternative is to sample test and training data randomly, based on the split ratio. Random sampling is more flexible than a temporal split and enables, for instance, $n$-fold cross-validation, where $n=5$ is a typical number ($n=10$, for instance, is also usual).

A natural implementation of a random split is by i.i.d. Bernoulli sampling (coin flip) $\textrm{B}(1,p)$ of data records with $p=$ the split ratio. Other implementations have been documented that sample an exact number of data records uniformly at random without replacement, sometimes separately for each user or each item. Other authors have explored sampling an equal amount (rather than an equal ratio) of test data per item or per user, as a way to reduce evaluation biases \citep{bcc17irj}. Research on specific problems can also deploy orthogonal variations of the split procedure to simulate specific conditions such as cold-start or long-tail by placing the items or users of interest in the test set \citep{hz11toit}.

\subsubsection{Candidate Item Sampling}
\label{sec:evaluation:offline:target}

A somewhat hidden option in the design of offline experiments is selecting the set of candidate items (sometimes referred to as \textit{target} items \citep{sarwar2010,cc20recsys}) that the evaluated system should rank for each target user. A natural setting for this option might consider selecting all the items in the catalog. This is not the case however when recommendation is viewed as a matrix completion problem: the known matrix cells need not be completed. In other words, an item is not included in the recommendations to users who have the item in their training records when discovery is part of the aim of recommendation, as is the case of most of the literature.

Experimenters may consider restricting candidate items to an even smaller set. \citet{k08sigkdd} was the first to suggest a fixed number of target items per user. The idea caught up and is still common nowadays in the research literature~\citep{kr20kdd,cc20recsys}. When small candidate sets are used, some authors refer to this option as computing \textit{sampled metrics} \citep{kr20kdd}. An extreme, often called \emph{condensed rankings} \citep{bv04sigir}, consists in taking only the items with test ratings in the candidate set---this is the option when the error metrics are used (see Section \ref{sec:evaluation:metrics:error} above), as it is not possible to compute a rating error where no rating is available.

Recent studies show that candidate sampling can have a deep impact on the outcome of offline evaluation and should be better understood \citep{cc20recsys,s13recsys,kr20kdd}. Authors unanimously report disagreements in system comparisons arising when all vs. no test-unrated items are included in the candidate set. \citet{cc20recsys} further show that the extremes in this settings have each their own shortcomings, and suggest that some point in between condensed and full rankings might optimize the informativeness of offline evaluation.

\subsection{Recommendation Task and Metrics}
\label{sec:evaluation:metrics}
In practice, every recommender system is designed to fulfil a certain purpose in order to create value for users and organizations, see also Section~\ref{sec:impact-and-value}. Depending on the purpose, different computational tasks have to be implemented in the system \cite{JannachAdomavicius2016purpose}. In their seminal work on recommender systems evaluation, \citet{herlocker2004} identified tasks such as ``find good items'', ``annotation in context'' (predict ratings), or ``recommend sequence''. When designing new algorithms, the main question is to understand how effective this new algorithm is in terms of fulfilling these tasks. In offline experiments, this assessment is done with the help of computational metrics, which serve as \emph{proxies} for the effectiveness of the system in its context of use.

Traditionally, the most important general aim of recommendation has been understood to involve an accurate grasp of user needs. Metrics devised to assess this have therefore been broadly referred to as \emph{accuracy} metrics. Recommendation accuracy is nowadays understood as a synonym for ranking quality---metrics and evaluation protocols have therefore been borrowed and adapted from the information retrieval field \citep{bcc17irj}, as we discuss next in Section \ref{sec:evaluation:metrics:ranking}. We nonetheless first discuss earlier perspectives based on rating prediction, for historical interest.

\subsubsection{Rating prediction}
\label{sec:evaluation:metrics:error}

As mentioned in Section \ref{sec:recommendation-task}, the recommendation task was initially understood as a rating prediction problem---that is, a regression task where a function $f:U\times I\rightarrow R$ is to be learned from examples. As such, it seemed natural to evaluate recommendation by error metrics such as Mean Average Error (MAE) and Root Mean Squared Error (RMSE). The error was measured across the test data in the rating matrix:
\begin{equation*}
    \textrm{MAE} = \frac{1}{|T|}\sum_{(u,i)\in T}|f(u,i)-r(u,i)|\hspace{8mm}
    \textrm{RMSE} = \sqrt{\frac{1}{|T|}\sum_{(u,i)\in T}\left(f(u,i)-r(u,i)\right)^2}
\end{equation*}
where $T\subset U\times I$ denotes the subset of test data records, and the lower the value of these metrics, the better the recommender system is considered.

Such metrics were used for almost two decades in the recommender systems literature, and were the basis for such an important initiative in the growth of the field as the Netflix Prize \citep{Liu2007}. The community has moved beyond rating prediction nonetheless, and nowadays tends to see and address recommendation as a ranking task, also in line with business models that are prominent in industry \citep{ckt10recsys}.
Many recommendation algorithms, on the other hand, output scores for user-item pairs that do not have a meaningful interpretation as ratings, but are highly effective to select lists of top-scored items to recommend---error metrics are not meaningful when applied to such scores.

\subsubsection{Ranking Quality: Recommendation as an IR Task}
\label{sec:evaluation:metrics:ranking}

During the 2000's the view that delivering recommendations has many commonalities with delivering search results grew stronger in the community \citep{herlocker2004}.
Both recommendation and search systems assist users in accessing relevant information or products from a large collection. One main difference lies in the absence of an explicit query in the recommendation task---still the problem can be understood to involve a user need to be satisfied, even if it is not explicitly conveyed by the user. In fact, search and recommendation often work together and complement each other in many applications. This view was recognized time before \citep{bc92comacm} but did not seem to catch up community-wide as---to some degree---a paradigm change in evaluation until the 2010's \citep{ckt10recsys,bcc17irj}.

In this realization, the key for recommendation effectiveness is in the returned ranking: effective recommendations should return as many relevant items as early as possible in the ranking, and as few non-relevant items as possible in the top positions. The ranking determines what items the user will see, and how soon, when browsing recommendations. The interpretation of the system scores, by which items are ranked, as accurate rating predictions becomes irrelevant.
The notion of ranking can be naturally generalized to other, not necessarily linear displays of recommendations (e.g. a ``shelve'' matrix), where some regions in the screen layout---equivalent to the top rank notion---are more likely than others to be examined by users.

Ranking evaluation naturally motivated researchers to borrow concepts and methodologies from the information retrieval field, where offline evaluation procedures and metrics had been researched and developed for half a century. The adaptation is however not straightforward and involves subtleties that need to be handled with care. We discuss this adaptation through the main elements involved in offline evaluation practice in the information retrieval field \citep{s10now}.

\paragraph{Collection.} The set of items $I$ can be considered an equivalent of the set of all ``documents'' in the IR literature. Item collections---often referred to as ``item catalogs''---are in fact often
the retrieval space for complementary search and recommendation functionalities in most recommender system applications.

\paragraph{Query and Information Need.} Search and recommendation are both motivated by a need on the user side that the system aims to help satisfy. Whereas users actively describe their need with explicit queries to a search engine, recommender systems derive user needs from observed user activity and interactions with the items. The information need representation in a recommender system is therefore considerably vague and incomplete compared to a search engine, and calls for different retrieval techniques. Search queries can have different degrees of vagueness too, and recommendation can be seen as the endpoint in a continuous spectrum of query specificity for a retrieval task, where the query is just empty.

\paragraph{Relevance.} The notion of relevance, central to IR, is just as meaningful in recommendation, and can be taken in quite the same sense: a recommended item is relevant to a user if they like, enjoy, are pleased by, are interested in, etc., the item. Relevance can be considered as a necessary condition for a recommendation to bring any value to the user. If a recommended item is not relevant the user will ignore it and no gain will be derived.

A major difference should be noted though as to how relevance is handled in offline experiments: while relevance is often assumed to be objectifiable---i.e. user-independent---in face of a query as a reasonable simplification, this assumption is not reasonable in recommendation, where relevance is acknowledged to be highly user-dependent, and this is typically an intrinsic characteristic of the recommendation task. This has important consequences in offline evaluation when it comes to eliciting relevance judgments:
delegating item labeling as relevant or non-relevant to assessors on behalf of users is too far a stretch to enable any kind of meaningful evaluation of personalized recommendations. Offline evaluation is therefore not separable from actual end users in the way search evaluation can be.

In offline recommendation experiments, test data obtained from target users play the role of relevance judgments in IR evaluation. Compared to judgment pooling in IR \citep{s10now}, test data can introduce extremely sharp selection biases in evaluation when data are collected in the wild, and/or by mediation of a recommender system, as is typically the case. Test data is very unevenly distributed over items, displaying a so-called popularity bias which can heavily distort evaluation outcomes \citep{jlkj15umuai,cc18sigir,bcc17irj}, as we discuss further in Section \ref{sec:bias}.

\paragraph {Metrics.} Once an equivalent of relevance judgments is defined and obtained, any IR metric can be applied to the output of a recommender system: precision, recall, mean average precision (MAP), mean reciprocal rank (MRR), normalized discounted cumulative gain (nDCG), are commonly used to assess the ranking quality of recommendations. Since precision and recall are rank-insensitive metrics, they are usually measured on a subset of top-$n$ recommended items---a ranking \emph{cutoff}---as $\textrm{P}@n$ and $\textrm{Recall}@n$. Cutoffs can also be taken in rank-sensitive metrics such as nDCG, MAP and MRR, to further focus measurements on the ranking top.

The Area Under the ROC Curve (AUC) \citep{b75jmp} can also be considered a ranking metric inasmuch as it provides a perspective of the relevant vs.~non-relevant recommended items tradeoff across the ranking. AUC is commonly used as a metric for classifier evaluation, but then so are precision and recall: they view recommender systems as binary classifiers into the relevant and non-relevant classes. Focusing the measurements on top $n$ rank positions does the trick for such measures, turning them into informative ranking metrics.

Compared to search experiments, bias and user-dependence in relevance judgments for recommendation exacerbate the sparsity of relevance labels in the ranking tops to be evaluated  in offline  recommendation experiments. Deciding how to handle the missing judgments becomes a key issue when importing IR ranking metrics, as discussed earlier in Section \ref{sec:evaluation:offline:target}. An additional challenge is the non-random nature of the missing data, and the resulting bias in evaluation results. We discuss this further in Section \ref{sec:bias}.

\subsection{Beyond Accuracy}
\label{sec:evaluation:beyond}

While accuracy was the primary perspective on algorithmic evaluation for a long time, it eventually became apparent that this is an incomplete view of what makes a recommendation useful and profitable for consumers and providers \citep{mrk06chi,Ge2010}. For example, a fan of the Beatles would very likely love the song `Yesterday'---the song is highly relevant in the IR sense---but how useful would this be as a recommendation? The song is a most widely known music piece worldwide, more so for a Beatles fan. The user might as well search for it when they fancy. The song might make sense as a convenience recommendation in some scenarios, e.g., in an automatic playlist while driving on a trip, but the added-value of such a recommendation (e.g., with respect to searching and browsing) is clearly more restricted than might be, for instance, the discovery of new music or a new band or a new taste the user was not aware of and can now enjoy.

Likewise, when a streaming platform lists movies and series the user might enjoy, it may be wise to include movies from a variety of genres and directors, to better represent the diversity of interests people have---and the varying mood. While people's tastes can be stable, we may feel like watching comedy one day and a documentary the next day, and those changes are very difficult to guess. Rather than risking a double-or-nothing bet on all-comedy or all-documentary recommendation, a diversified offering of potential favorites would seem a more sensible approach considering this. Moreover, by diversifying recommendations the system takes opportunities to explore unseen user interests and learn about them to keep improving in the future.

Novelty and diversity are just two dimensions in the objectives beyond accuracy that can maximize the value of recommendation, and we focus on them here. It is useful to distinguish between novelty and diversity as highly related but different perspectives, a distinction we reflect in the summary that follows. Broader perspectives on the direct key measures of the value, user satisfaction, business performance and the impact at large that recommendation has on the involved stakeholders, are discussed in Section \ref{sec:impact-and-value}. A recommender system finds several motivations to enhance novelty and diversity, and different angles to these general notions, that can be operationalized into metrics and algorithms that enhance and measure them, as we discuss next. For a more extensive review of novelty and diversity in recommender systems, the reader is referred to \citep{chv22handbook}.

\subsubsection{Novelty}
\label{sec:evaluation:beyond:novelty}

At an abstract level, recommendation novelty can defined as the difference between the recommended items and a certain context of reference \citep{vc11recsys}. The context can typically be the past experience of the person to whom a recommendation is delivered or, considering the sparse and partial system knowledge about that, the aggregated experience of all users in the system. In the latter case, novelty equates to rarity: a recommended item is novel if it is in the \emph{long-tail} of the interaction frequency distribution \citep{ch08recsys}. Personalized novelty, often referred to as unexpectedness \citep{at14tist},
would reflect how different a recommendation is from the items the target user was observed interacting with in the past. Long-tail novelty and unexpectedness can be quantified in different ways, which we summarize next.

\paragraph{Long-tail novelty.} This dimension can be measured, for instance, as the average of a monotonically decreasing function $\phi$ on the amount of past engagement $L_i$ with each recommended item:
\begin{equation*}
    \textrm{LT}(R) = \frac{1}{|R|} \sum_{i\in R} \phi(|L_i|)
\end{equation*}
where $R$ is the returned recommendation and $L_i$ denotes the set of observed interactions involving item $i$. The $\phi$ function has been defined in the literature, for instance, as the complement or the negative log of the ratio of observed interactions involving the item  \citep{vc11recsys}:
\begin{equation*}
\phi(x)=1-\frac{x}{|L|}\hspace{20mm}
\phi(x)=-\log \frac{x}{|L|} =\log \frac{|L|}{x}
\end{equation*}
where $L$ represents the set of all observed interactions. With the first definition, novelty can be read as the probability that a random user has never interacted in the past with a recommended item in $R$. When repeated interaction is ignored (i.e., each user-item pair is counted at most once in the above equations), the second definition is equivalent to inverse document frequency as defined in IR models for document retrieval, users being here the equivalent of documents---hence sometimes referred to as \emph{inverse user frequency} \citep{bhk98uai}. When measured this way, novelty can be seen as a condition of coldness (lack of data) or unpopularity.

\paragraph{Unexpectedness.} This notion is typically quantified as:
\begin{equation*}
\textrm{Unex}(R)=d(R,E_u)
\end{equation*}
where $d$ denotes a set dissimilarity operator and $E_u\in I$ is a reference set of items representing ``the expected'' for each user $u\in U$. The expected set can be, for instance, the items that the target user has interacted with, or the recommendations delivered by a supposedly conventional algorithm, or any other proxy for an unsurprising experience \citep{at14tist}. Set dissimilarity can be any conventional measure such as Jaccard, Hausdorff, average item-wise distance, or other meaningful elaborations to quantify how different the two sets are. When pairwise item distance is used, feature-based dissimilarity (measured e.g., by the cosine or Jaccard distance between feature vectors/sets) is typically a good option. Other set-based dissimilarity operators can have a probabilistic interpretation, e.g. $d(R,E_u)=|R\setminus E_u|/|R|$ can be read as the probability that a recommended item is not expected.

\paragraph{Serendipity.} An additional important notion in the scope of novelty is \emph{serendipity}. While the definitions slightly vary in the literature, the dominant convention considers a recommendation as serendipitous if it is surprising (i.e. novel) and valuable \citep{Chen2019}. If we equate value, in this sense, to relevance, a straightforward way to measure serendipity is to restrict the computation of the above novelty metrics to the recommended items that are relevant to the target user (according to the available test data), and ignore the rest.

\paragraph{Further notions.} The above novelty notions are probably the most common in the literature but they are not meant to be exhaustive \citep{chv22handbook}. \emph{Freshness} (how recently a recommended item was created), for instance, is often an important property by itself, and typically a signal that correlates with the above notions. Lathia and Amatriain \citep{lhca10sigir} explored finer temporal notions of novelty and diversity, involving the history of past recommendations---capturing, broadly speaking, how much a recommender system is repeating itself.

\subsubsection{Diversity}

Related to but different from novelty notions, diversity is generally defined as the amount of variety covered within recommendations. For instance, `Orinoco Flow' by Enya and `Highway to Hell' by AC/DC can hardly be considered novelties (anyone has listened to this music sometime), but together they would make for a highly diverse music recommendation because they are very different to each other. Likewise, two very similar music rarities would not be diverse while they may be highly novel.

\paragraph{Intra-list dissimilarity.}
A common and straightforward way to measure internal diversity is as the average pairwise dissimilarity of recommended items \citep{smyth2001,Ziegler2005}:\footnote{\citet{Ziegler2005} and other authors define intra-list \textit{similarity} instead of \textit{dissimilarity}, as a metric to minimize rather than maximize. The two views can be seen as (explicitly or conceptually) complementary to the same extent that similarity and distance can be.}
\begin{equation*}
\textrm{Div}(R)=\frac{2}{|R|(|R|-1)}\sum_{i\in R}\sum_{j\in R}d(i,j)
\end{equation*}
where $d:I^2\rightarrow\mathbb{R}$ is a dissimilarity function typically based on item features. In this definition, diversity can be seen as a particular case of unexpectedness, where the expected set is the recommendation itself: $E_u\leftarrow R$.

\paragraph{Aspect-based diversity.} Alternatively, the assortment within recommendations can be defined in terms of a single particularly meaningful feature, referred to as an \emph{aspect} space, such as movie genre, music style, nationality, etc. So-called aspect-based metrics such as aspect recall \citep{zcl03sigir} or the intent-aware metric framework \citep{aghi09wsdm} defined in the context of search systems, can be readily applied to evaluate recommendation diversity \cite{wh16recsys,vcv11sigir}.

\paragraph{Coverage.} Another very relevant dimension of diversity in recommendation is the spread of recommendations over the item space, viewed at the aggregate level across users. This angle of diversity is commonly referred to as \emph{coverage}, \emph{aggregate diversity}, or \emph{diversity of sales}. Metrics that capture this notion are borrowed (intentionally or not) from the ecology domain \citep{pt82jasa}, and go from very simple formulations (number of items that are recommended at least to one user) to finer measures such as the Gini index \citep{sg22handbook}. This angle on diversity is gaining increasing importance as it directly relates to issues of fairness and equal opportunity for all items (and their providers, vendors or creators) to have a chance to be exposed to consumers.

\subsubsection{Enhancing Novelty and Diversity}

Enhancing novelty and diversity can be addressed as the maximization of the measure of interest, taking metrics such as the ones discussed above as the target for maximization. This finds two main problems. The first and perhaps most important is that maximizing for novelty or diversity is not aligned with maximizing relevance, in the IR sense. In fact, both objectives sometimes display a tradeoff---for example, random recommendations typically score extremely high on novelty and diversity, and extremely low on relevance. Rather than a single optimum the problem displays a Pareto front. As a multi-objective problem, diversity and novelty enhancement can be addressed by common strategies: scalarization, evolutionary algorithms, etc. \citep{vrlmhz14tist}. Since state of the art recommendation algorithms do not necessarily lie on the Pareto front, some works report improvements in both directions \citep{jannach2015,Vargas2014}.

In doing so, a top-level distinction can be made between diversification strategies: intrinsic or extrinsic. Extrinsic approaches re-rank an initial relevance-oriented recommendation, seeking to improve the relevance-diversity (and/or novelty) tradeoff. A basic, typical approach in this line is re-ranking the initial recommendation by greedy maximization of a linear combination of relevance and diversity. In intrinsic approaches relevance and diversity are addressed internally at the same time as built-in targets in the algorithm \citep{Vargas2014}. For instance, novelty and/or diversity can be injected along with accuracy in a multi-objective function to be maximized \citep{vrlmhz14tist} in the perspective discussed earlier in Section~\ref{sec:algorithms}.

A second difficulty arises when the diversity metric is not defined separately for each user and aggregated or averaged afterwards, but is defined on the set of all recommendations---as is the case of coverage, explained earlier. In those cases the relevance-diversity tradeoff cannot be optimized on each user independently, and the problem space becomes even larger: recommendations become interdependent in their contribution to the global diversity, and should ideally be optimized at the same time. Different levels of greediness in the approach are still possible, though an additional tradeoff needs then to be withstood between cost and optimality \citep{sc18recsys}.

In most recent years, diversification approaches that operate on the user-individual level have become more popular. In an earlier work, \citet{Oh2011} proposed to adapt the novelty of the recommendations according the past popularity tendencies of individual users. A similar and more generic re-ranking approach was later proposed in \cite{JugovacJannachLerche2017eswa}, which also supports the consideration of multiple optimization objectives per user in parallel. Today, such approaches are known as \emph{calibrated} recommendation \cite{steck2018calibration}. An important aspect of calibrated recommendations that is currently not covered in depth so far is that a user's diversity of novelty preferences may change over time and depending on the context, see \cite{Kapoor2015ILike} for an analysis of such phenomena in the music domain.

\subsubsection{User Perceptions of Diversity and Novelty}
Over the last decade, a multitude of algorithmic approaches were proposed to create more novel and diverse recommendations by optimizing the list of suggestions according to corresponding computational metrics. As with any computational metric, it is however important to ensure that these metrics are actually valid proxies of the users' perceptions, e.g., of the novelty and diversity of the recommendations. Moreover, it is crucial to understand how these perceptions then impact the users' satisfaction with the system.

Intra-list dissimilarity, as mentioned above, is widely used in the literature to express the diversity of a recommendation list. In \cite{Ziegler2005}, the impact of \emph{topic diversification} on user satisfaction was explored with the help of a user study\footnote{See Section~\ref{subsec:impact-of-rs} for a deeper discussion of user studies in recommender systems.}. Different levels of diversifications were tried and the obtained peak value in satisfaction indicated that the participants preferred a certain degree of diversification in their book recommendations. Similar observations were made in \cite{Castagnos2013} in the context of movie recommendations. In their study, the authors also presented study participants with recommendation lists that had different degrees of diversity. The authors found that (i) participants notice the different diversity levels and (ii) that diversity may positively influence user satisfaction. However, in case of diversified lists it was observed that providing additional explanations may be advisable so that users can better link the recommendations with their preferences. Positive effects of diversity on user satisfaction in the movie domain was also reported in \cite{Ekstrand2014user}; however, in this study it also turned out that increased levels of novelty---e.g., when too many unfamiliar items are recommended---may negatively impact user satisfaction.

In the mentioned studies in the book and movie domains, the used similarity functions were based on topic categories and movie attributes and proved effective in the experiments. Such a validation of the particular similarity function is however often missing in other works. More work is therefore required to understand which features of an item are determining the similarity perceptions of users, see \cite{Yao2018judging,TrattnerJannach2019} for studies in the movies and food domain. Moreover, in the traditional intra-list dissimilarity measure, the position of the diverse elements does not matter when assessing the overall diversity of a given list. In reality, indications exist that the ordering of the items may impact user perceptions \cite{GeJannachEtAl2012}. In this direction, drawing from earlier work on IR metrics built upon user models (e.g., \citep{Moffat2008}), \citet{vc11recsys} propose a probabilistic metric framework where a wide variety of novelty and diversity metrics---including the ones mentioned in this section---can be endowed with rank sensitivity.

Instead of varying the positions of the items in one recommendation list, as done in \cite{GeJannachEtAl2012}, \citet{Hu2011b} suggested to organize the items in groups. In a user study, in which also the participants' eye movements were tracked, they then found that the ``organization'' interface helped to increase the diversity perception of users. Finally, in a more recent work, \citet{Chen2019} report the outcome of a large-scale user study on the quality perception of recommendations on an e-commerce platform. The study revealed that \emph{serendipity} plays a decisive role for user satisfaction and is more important than diversity and novelty alone.

\chapter{Recent Topics}

\section{Sequential and Session-based Recommendation}
\label{sec:sequential-session-based}
The traditional problem formulation of recommender systems, as described in Section \ref{sec:recommendation-task}, is designed for making non-contextual item suggestions based on stable long-term user preferences. Such an approach is for example meaningful to present landing-page recommendations to a frequent user of a media streaming site. However, there are many other important application scenarios where we cannot assume to have long-term preference information available. Think, for example, of visitors of an e-commerce shop who are not logged in or who are first time users.  Moreover, on an e-commerce site, a user's interests and needs might change from shopping session to shopping session. Therefore, even if we would know the identity of the user it is important to consider the user's short-term history and current intents when we determine what we should recommend to the user next.

\subsection{Problem Definition and Terminology}

\begin{figure*}[h!t]
    \centering
    \includegraphics[width=0.75\textwidth,clip]{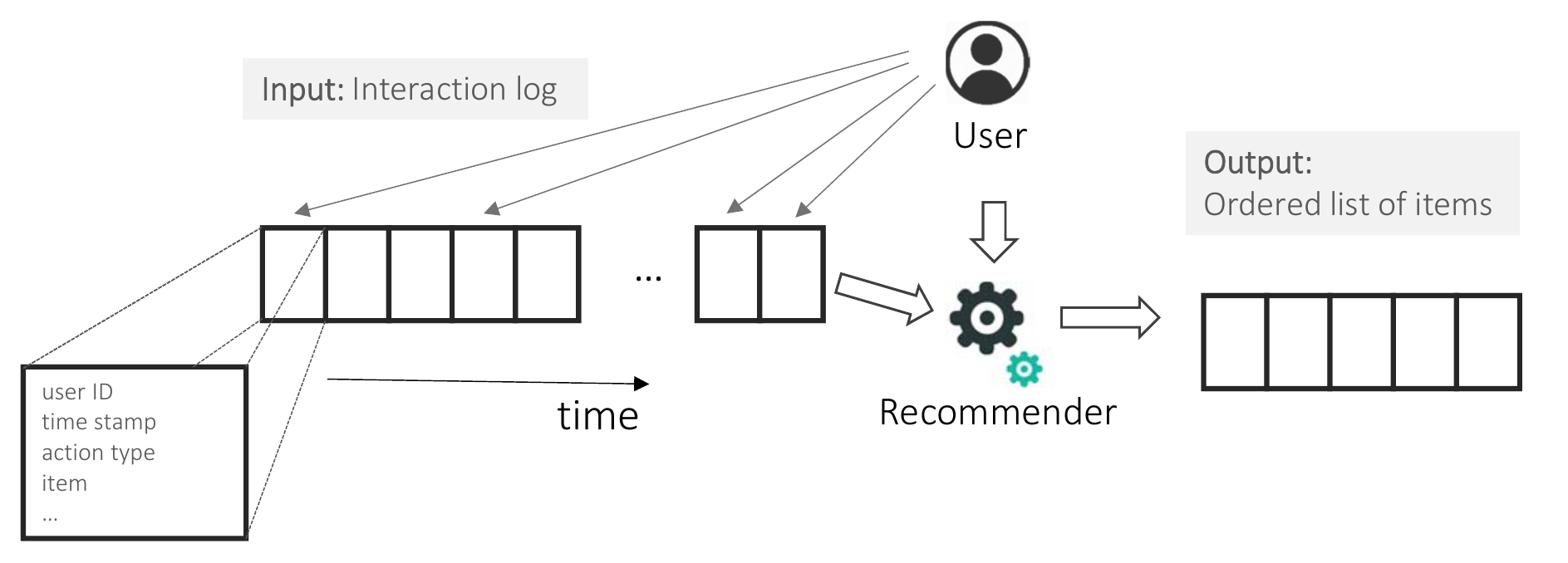}
    \caption{Sequential Recommendation: Overview (adapted from \cite{QuadranaetalCSUR2018}).}
    \label{fig:sequential-recommendation}
\end{figure*}

These requirements, which cannot be easily addressed with the traditional matrix completion abstraction, led to the development of different types of \emph{sequence-aware} recommender systems \citep{QuadranaetalCSUR2018}. Instead of a user-item rating matrix, the main input to this type of recommenders consists of a sequence of logged interactions. The entries of these logs are typically collected through the application, e.g., a web store, and they can have different types. A web shop might for example record item views, add-to-cart events, or purchases. Moreover, each log entry is usually associated with an individual user. This can be a known user, who is logged in to the web shop, or an anonymous one which is only identified through a cookie or IP address. Finally, a last difference to the user-item rating matrix is that the interaction log may contain multiple user-item interaction pairs, for example, when the user viewed an item multiple times before a purchase. In addition, a log entry may have additional types of meta-data attached such as a time stamp. Overall, while the \emph{inputs} are different from the matrix-completion problem, the \emph{output} of a sequence-aware recommender is usually a ranked list of items. Figure \ref{fig:sequential-recommendation} shows an overview of the problem setting.

Within the family of sequence-aware recommender systems, two related terms
can be identified in the literature: \begin{enumerate*}[label=\emph{(\roman*)}]
\item \emph{sequential} recommendation and
\item  \emph{session-based} recommendation.
\end{enumerate*} Sequential recommendation refers to the more general problem of making \emph{next-item} recommendations. A typical example is the problem of recommending the next Point-Of-Interest (POI) to a user in a tourism scenario or to recommend an entire next shopping basket in an e-commerce scenario. 
In session-based recommendation scenarios, the problem is also to make next-item predictions, but the underlying assumptions are that \emph{(a)} the past interaction log is organized in usage sessions and that \emph{(b)} recommendations are made in the context of an ongoing session.  Another underlying assumption of session-based approaches typically is that the user's interests and needs can change from session to session. Furthermore, in many practical problem settings, no long-term preference information is available at all for the current user, which means that the recommendations must be based on a small set of interactions---e.g., a few item view events---that are observed in an ongoing session.

\begin{figure*}[h!t]
    \centering
    \includegraphics[width=0.75\textwidth,clip]{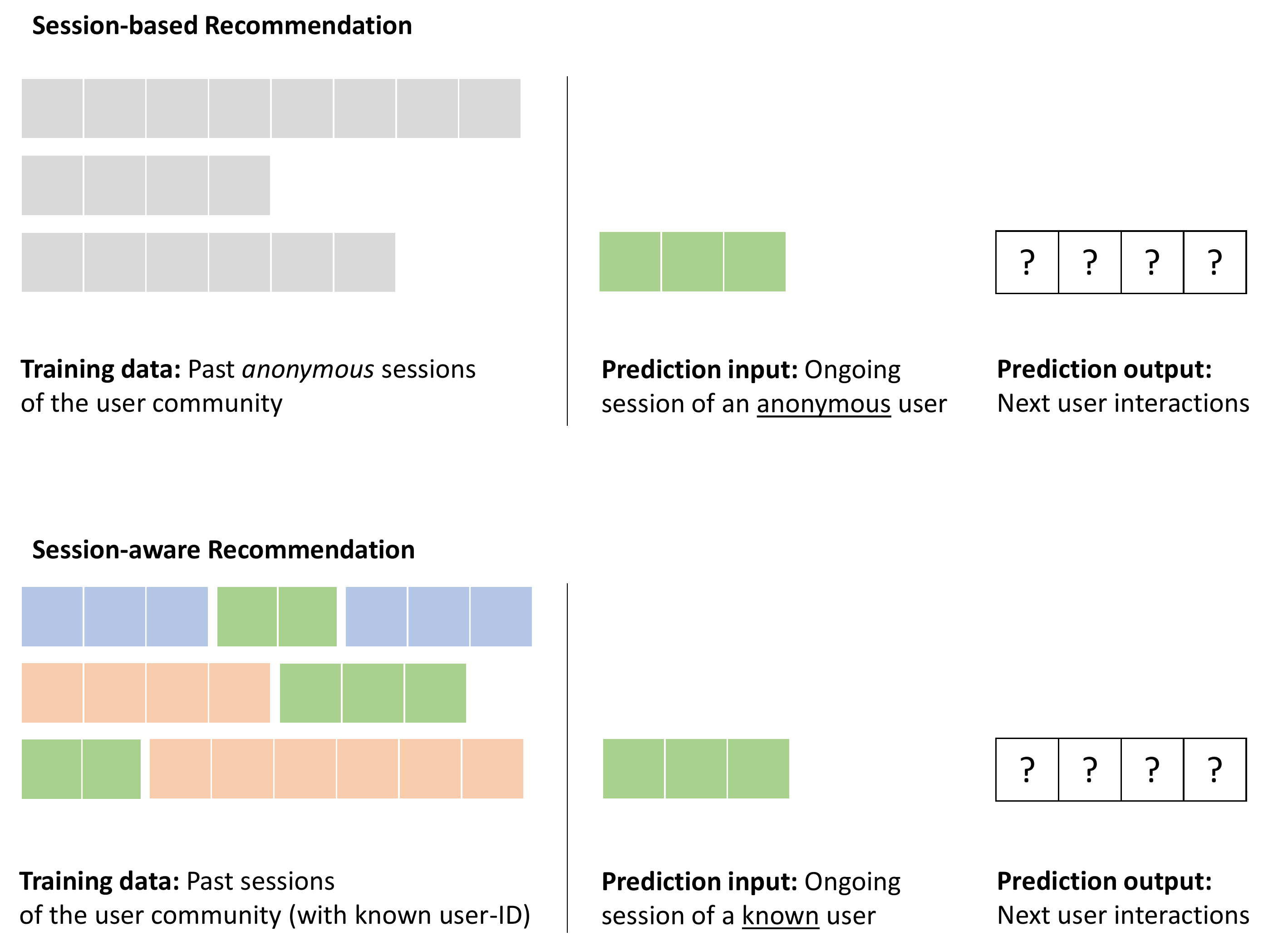}
    \caption{Session-based and Session-Aware Recommendation: Overview (adapted from \cite{latifi2020sessionaware}).}
    \label{fig:sequential-session-based}
\end{figure*}

In the literature, two alternative situations are considered when recommendations have to be made in the context of an ongoing session. In one case, nothing is known about the current user but the interactions that are observed in the session. In the other case, also information about past sessions of the same user is available. This second scenario is often referred to as \emph{personalized session-based recommendation} or \emph{session-aware} recommendation. Figure \ref{fig:sequential-session-based} visualizes the differences between (pure) session-based and session-aware recommendation problems.

\subsection{Algorithms for Sequential and Session-based Recommendation}
Various algorithmic approaches were explored for sequence-aware recommendation problems. The following main types of technical approaches were identified in \cite{QuadranaetalCSUR2018}:
\begin{enumerate*}[label=\emph{(\roman*)}]
  \item Sequence Learning Approaches
  \item Sequence-Aware Matrix Factorization
  \item Hybrid Approaches, and
  \item Nearest Neighbors and Other Methods.
\end{enumerate*}

\paragraph{Sequence Learning Approaches.} Approaches in this category can be subdivided into several main paradigms.
\begin{itemize}
\item \emph{Frequent Pattern Mining} techniques were historically often applied for the problem of detecting (unexpected) patterns in shopping baskets,  e.g., in the form of \emph{association rules}. Later, these techniques were extended to consider \emph{sequential patterns}, in which the order of the elements is relevant.
\item \emph{Sequence Modeling} approaches try to go beyond mined patterns and aim to learn models from past time-ordered data to predict future events. A large variety of corresponding machine learning models were proposed over the years. The predominant approaches in the literature are based on Markov Models, Reinforcement Learning, or Deep Learning, in particular in the form of Recurrent Neural Networks (RNNs). Often, modern sequential models are based on ideas from Natural Language Processing to model the sequences in the interaction data, see e.g., \cite{BERT4REC2019}.
\item \emph{Distributed Item Representations} are another form of encoding sequential information. These representations (embeddings) are projections from sequences of item-related events (e.g., views) into a lower-dimensional representation in which the transition between items is preserved.
\item \emph{Supervised Learning with Sliding Windows}, finally, is a rather specific approach, where the idea is to reframe the next-item prediction problem to a supervised learning problem by moving a sliding window over the sequence of events and to derive the corresponding feature values from this sliding window to predict the next event.
\end{itemize}

\paragraph{Sequence-Aware Matrix Factorization.} A few examples exist in the literature that extend matrix factorization approaches to consider sequential information, usually derived from timestamps that are available for the entries in the user-item rating matrix. These approaches are conceptually related to \emph{time-aware} recommender systems \cite{Campos2014}, which aim to identify changes in user preferences over larger time spans.

\paragraph{Hybrid Approaches.} A more common approach in the literature is to combine sequence modeling with the power of latent factor models. One of the earlier approaches of that type is the \emph{Factorized
Personalized Markov Chain (FPMC)} method \cite{rendle10FPMC}. This method jointly factorizes a first-order Markov Chain, which models user-specific item transitions, and a traditional user-item rating matrix to make next-item predictions. Similar approaches were proposed later on, e.g., in \cite{Fossil2016}. More recently, methods that combine neural techniques (e.g., RNNs, Convolutional Neural Networks, or Attention) with matrix factorization layers, have become more popular, e.g., \cite{SASRec2018,caser2018}.

\paragraph{Nearest-Neighbors and Other Methods.} In particular in the context of session-based recommendation, a simple yet effective method is often to apply neighborhood-based techniques. In \cite{JannachLudewig2017RecSys} and subsequent works, the idea was explored to use those past sessions in the data as a basis for predicting the next event, which are most similar to the ongoing one. Despite the conceptual simplicity of the approach, it turned out to be competitive or even superior to much more complex models. Finally, a number of alternative proposals were identified in \cite{QuadranaetalCSUR2018}, including methods that rely on (non-neural) graph-based models or techniques from discrete optimization, where the latter class of problems may however suffer from scalability problems given the usually high number of recommendable items.

\subsection{Evaluation of Sequential and Session-based Recommender Systems}
\label{sec:sequential-session-based:evaluation}

In practical applications, recommender systems are evaluated with respect to the organizational goals and purpose they should fulfill (e.g., increase customer retention) and the specific computational tasks they implement that serve this purpose (e.g., help users find novel content) \cite{JannachAdomavicius2016purpose}. While the general goals of sequence-aware recommender systems are usually not very different from traditional ones, sequential and session-based recommenders often implement a number of computational tasks based on the particular nature of their input data, as identified in \cite{QuadranaetalCSUR2018}:
\begin{itemize}
\item \emph{Context Adaptation} is the fundamental problem of session-based algorithms, i.e., to guess the user's situation and intents from the interactions observed in an ongoing session. The observed user behavior thus represents a form of \emph{interactional} context \cite{Adomavicius2022}.
\item \emph{Trend Detection} is a still under-explored topic in academic research and relates both to individual trends (e.g., interest drift) and community trends that might, for example, depend on seasonal patterns or popularity peaks of items.
\item \emph{Repeated Recommendations}, which are usually not considered in matrix-completion problems, can both be valuable in the context of consumable items (e.g., printer ink) and for the sake of reminding users of items they liked in the past (e.g., a past favorite artist).
\item \emph{Consideration of Order Constraints and Patterns:} In a number of domains, there is a strict order in which items should or must be recommended (e.g., learning course recommendation); in other domains, there might be a ``natural'' order in which things should be recommended (e.g., in the case of movie sequels). Sequence-aware recommenders can learn and/or enforce such patterns as a computational task.
\end{itemize}

In principle, all evaluation approaches outlined in Section \ref{sec:evaluation} can be applied for the evaluation of sequential and session-based recommender systems, including offline experiments, user studies, and field tests (A/B) tests.   Offline experiments dominate the research landscape and academic research also for sequence-aware recommenders and, as usual, such experiments aims to abstract both from the value perspective and the particular computational tasks.

\subsubsection{Offline Evaluation}
Remember from above that the output of a sequence-aware recommender system as usual is a ranked list of item suggestions. Following the usual ``hide-and-predict'' evaluation approach in offline experiments, common evaluation measures from information retrieval for ranking accuracy such as precision and recall can be applied. The evaluation procedure of sequence-aware recommender systems therefore involves the hiding of a number of interactions in a given sequence of events and the subsequent measurement of how good an algorithm is to predict the hidden interaction(s).

In evaluations of traditional matrix completion problem settings, the selection of the hidden elements (i.e. the test set) is often done in a randomized way, e.g., by randomly selecting 20\,\% of the data for testing. Moreover, cross-validation is in many cases applied to increase the confidence in the observed results. Such an entirely randomized approach is not meaningful for sequence-aware recommendation problems, and alternative procedures have to be applied.

\paragraph{Data Splitting.} When splitting the data into training, validation, and test data sets, the sequential order must be considered, and the held-out interactions to be predicted must happen after those that are used for training. Different strategies are possible in the data splitting process. The following are relatively common.
\begin{itemize}
\item In \emph{sequential} problem settings, often the splitting of the data is done \emph{per user}. Commonly, the very last interaction of each user is hidden and put in the test set.\footnote{While this is a commonly used approach in the literature, it may violate the principle that interactions events of other users are used for training which actually happened after the interaction that is predicted for a particular user.}
\item In \emph{session-based} scenarios, it is, in contrast, more common to apply a \emph{time-based} split. Commonly, research datasets cover an extended period of time, e.g., a few weeks. Often, the sessions of the very last or the last few days are used for testing (and one or more preceding days for validation).
\end{itemize}

\paragraph{Making the Measurement.}
In sequential recommendation problems, when the last interaction of each user is hidden, the making the measurement amounts to determining if and at which position an algorithm was ranking the hidden interaction within a top-n list. Common evaluation measures therefore include the Hit Rate, the NDCG or the MRR at a certain cut-off length.

For session-based problems, these measures can be applied as well, but the evaluation procedure is slightly different. Remember that the test dataset typically contains entire user sessions and that the specific problem in session-based recommendation is to predict the next interaction(s) given the session beginning. The options reported in the literature are to ``reveal'' all but the last interaction in a session, to reveal the first $k$ elements, or to incrementally reveal one interaction after the other when measuring. Different variants also exist how the accuracy measures are applied. One can, for example, measure how good an algorithm is to predict the immediate next item in a session, using the Hit Rate, NDCG, and MRR as typical measures for a given cut-off threshold. Alternatively, and probably more importantly, one can measure how many of the subsequent (and hidden) interactions in the session are contained in the top-n list returned by a recommender. In that case, measures like precision and recall are often used.

Note that like in traditional recommendation scenarios, quality factors other than accuracy can be considered. Beyond-accuracy aspects considered in the literature in particular include the diversity of the recommendation lists, catalog coverage, popularity biases, or scalability.

Generally, observe that while we hide \emph{interactions} in the data, we are usually making predictions for \emph{items}. In case there is only one type of interactions in the logs, e.g., listening events on a music streaming site, an item prediction corresponds to predicting a listening event. However, real-world datasets often consist of interactions of different types, e.g., item views, add-to-cart actions, or purchases on an e-commerce site. In many academic research works, only the most frequent type of interactions is considered, for example item views. In reality, however, it might be much more important to predict item purchases, and to use all available types of interactions in the training and prediction process.

\paragraph{Cross-Validation.} Given the sequential nature of the data, performing cross-validation based on random splits is not meaningful. In the literature, often only one single train-test split is applied, which, however, bears a certain risk that the obtained results are specific to that split. An alternative approach therefore is to create multiple overlapping or non-overlapping time-based splits of the data, and to repeat the measurement on these splits, see \cite{ludewiglatifiumuai2020}.

\subsubsection{User-Centric Evaluation}
User-centric research, while not uncommon in the recommender systems literature in general, is still very rare in the context of sequential and session-based recommendation. In one study in the music domain \citep{KamehkhoshJannach2017}, researchers compared different techniques for next-track music recommendations. In this study, participants were presented with a number of alternative continuations for a given playlist. In another study \cite{ludewigjannach2019radio}, an online interactive radio station was created for the purpose of the experiment. Different algorithms were used to create playlist based on the a start track provided by the user and based on the feedback during the radio listening session. Both mentioned studies led to interesting insights regarding the quality perceptions of users and investigated to what extent offline accuracy measures align with user perceptions.  Besides such specific research setups, the use of general frameworks for the user-centric evaluation of recommender systems can be applied, e.g., \cite{Pu:2011:UEF:2043932.2043962} or \cite{umuai2012knijnenburg}.

\subsubsection{Real-World Evaluation}
For more traditional recommendation problems, a number of success stories from real-world deployments of recommender systems exist, see \cite{jannachjugovactmis2019}. In the area of sequential and session-based recommendations, such reports are rare. An exception is the work in \cite{kouki2020}, which reports on the use of session-based recommendation technology for an online shop in the home improvement domain. Ultimately, the use of a modern neural approach led to a significant increase in relevant business KPIs when compared to an existing commercial software for retail recommendation.\footnote{Unfortunately, the inner workings of this previous system are not known, i.e., it is for example unclear if it implements a session-based approach or a traditional one.}

Another interesting aspect of this work lies in how the decision was made which of several existing session-based algorithms should be put in practice. First of all, the authors benchmarked a number of techniques, including conceptually simple and more complex ones, through an offline evaluation. Reproducing previous offline experiments \cite{ludewiglatifiumuai2020}, they found that simple methods in many cases outperform the latest neural techniques in terms of typical accuracy measures. A subsequent evaluation with human experts however revealed that the well-performing nearest-neighbor technique might not be the best choice in the given application domain. While it is often able to correctly predict a single hidden element, other algorithms turned out to be more successful in recommending more than one possibly relevant item. In particular, other algorithms often returned items that are similar to the hidden one, which could therefore represent purchase alternatives for the customers. As a result, this study emphasizes not only the possible limitations of offline evaluations, it also stresses that in practice it is important to understand the specifics of a particular application and how a recommender is expected to create value.

\subsection{Discussion and Outlook}
In particular session-based and session-aware are highly relevant practical problems for which we have seen a number of technical proposals in recent years.\footnote{For a recent, more detailed discussion, see also \cite{JannachHBSessions2022}.}
However, despite the many complex algorithms that are developed, it is surprising to see that in many cases very simple techniques based on nearest-neighbors are competitive or even outperform the latest neural techniques in offline evaluations \cite{ludewiglatifiumuai2020,Garg:2019,kouki2020}. Such phenomena are however not specific to session-based recommendation problems, see \cite{ferraridacrema2020tois} for a discussion of similar problems that were observed for traditional top-n recommendation scenarios. Overall, this points to certain methodological issues that may hamper progress in the field.

On the other hand, these observations also indicate that there is huge potential for the development of even better algorithms for session-based and sequential recommendation. In particular, the use of side information, e.g., in the form of item meta-data, the use of contextual information, and the consideration of multiple types of interactions in parallel seem to be promising areas for future work.

\section{Popularity, Bias and Recurrence in Recommendation}
\label{sec:bias}

As the view of recommendation shifted from rating prediction to item ranking, an new phenomenon was observed: offline evaluation tends to reward the recommendation of popular items, that is, items involved in frequent interaction with users \cite{ckt10recsys}. This becomes a strong bias to the same extent that the data distribution over items---the so-called popularity distribution---is heavily skewed in a long-tail shape. With a random data split, the popularity effect in evaluation is a clear self-fulfilling prophecy: the items with most training data---the popular items---have the most test data, which are used as relevance judgments, as discussed in Section \ref{sec:evaluation:data}. Hence, by ranking popular items highly in recommendations, we get fewer missing ratings in the ranking top, resulting in improved chances of producing high relevance metric values \citep{bcc17irj,cc18sigir}. Even with temporal splits, the popularity distributions on the respective sides of the split point are usually correlated (to the extent that item popularity persists over time), whereby the popularity bias remains \citep{bcc17irj,mccrs21tois}.

Given this effect, it seems not surprising that state of the art collaborative filtering algorithms are also heavily biased to recommend popular items, as researchers have found systematically \citep{jlkj15umuai,cc17sigir}. This realization has fueled research in two directions: a) avoiding or countering the biases in both evaluation and algorithms, under an implicit view of bias as a potential distortion of learning and evaluation \citep{s10kdd,s11recsys,bcc17irj,jmo2021tois}; and b) better understanding where the bias may come from, and wondering whether it might be harmless or even useful to some extent \citep{cc18sigir}. An additional fundamental perspective has recently gained momentum that considers the dual role of recommendation in satisfying users, on one side, and learning from their interactions with the delivered recommendations, on the other. We briefly discuss these perspectives in the following paragraphs.

\subsection{Countering Bias}

From a certain angle, popularity can be considered intrinsically undesirable from an added-value perspective that, as such, should be avoided. This is the viewpoint of novelty as a complementary goal of recommendation as discussed in Section \ref{sec:evaluation:beyond:novelty}. Even from this perspective, novelty should still be combined with relevance, and relevance needs to be evaluated; the relevance-seeking component, and its evaluation, are likely to be affected by popularity biases.
Specific evaluation procedures have been researched aiming to remove the popularity bias from the metrics \citep{s10kdd,s11recsys,bcc17irj,ssscj16icml} and the algorithms \citep{lcdhpm20sigir,hhg14jlmr,JannachLudewigLerche2017umuai}. Among them, counterfactual reasoning approaches for off-policy learning and evaluation are being actively researched, using techniques such as inverse propensity scoring (IPS) \citep{skadljz17nips,ssscj16icml,jmo2021tois}. IPS produces unbiased estimators for metrics and objective functions by modeling the distribution of observations over the user-item space, and compensating for it in the evaluation metrics and the learning algorithms, by downweighting observed preferences by the probability of observing them. Challenges involved in the application of IPS in this context include modeling the propensity bias from biased samples, and the boundless variance of IPS weights when observation probability can be indefinitely small.

\subsection{Understanding Bias}

Beyond the capability to remove or harness biases, some authors argued for a better understanding of popularity distributions and their potential harm or innocuity as a distorting factor in evaluation \citep{cc18sigir}. After all, recommender system applications make deliberate use of popularity signals \citep{ab15rshb}, where popularity is not necessarily an entirely bad property---whereas naive popularity countering in evaluation might assess this signal as equivalent to random recommendation. \citet{cc18sigir} theorized about the formation of popularity distributions, and found out specific conditions that determine whether or not system comparisons are preserved in evaluation under popularity biases. The key to the answer lies in the conditional dependencies between observation, relevance and items as random variables, which explain why majority choices and majority tastes can be useful in delivering effective suggestions. These findings have been extended later in experiments showing a high degree of agreement in evaluation with biased and unbiased data \citep{mccrs21tois}. In other words, the general findings in this line are that feedback loops and Matthew effects \citep{Merton56} can sometimes have a virtuous component \cite{nematzadeh2017}. Acting upon this awareness, \citet{zhang2021} explore a finer handling of popularity at training and inference time, seeking to mitigate distortion while leveraging the useful signal.

\subsection{The Feedback Loop}

A major source of bias in offline data is the production system through which the interaction data---the input for recommendation---are collected. Recommender systems are commonly integrated in multi-component applications where items are exposed to user feedback through different pathways: search, browsing, recommendation, front page, etc. Each of such components introduces bias of its own. To the extent that a recommendation component feeds on user interaction with its own output, we have a feedback loop \citep{fh09mansci,cse18recsys}. This loop, in combination with offline evaluation, creates a magnification effect (also known as snowball, Matthew effect, rich-gets-richer), where the data collected for evaluation reward and reinforce the hypotheses ingrained in the production system, making popular items ever more popular, and penalizing change \citep{chaney2018}. As a result the system can get stuck in its own assumptions and miss out on value and opportunities that remain hidden in unobserved choices.

Furthermore, perverse incentives can arise from such loops, where user engagement, sometimes tightly associated to short-term revenue, is blindly targeted by the recommendation strategy---recommendation might then be inducing compulsive or toxic behavior, rather than (or mixed with) serving a constructive purpose \citep{whittaker2021}. Such rabbit hole traps may result in suboptimal long-term business performance at best, or unwillingly contribute to toxic phenomena such as misinformation spreading, polarization, and radicalization, at worst. Preventing such risks and harmful effects is not all that simple, and to a great extent requires continual monitoring and direct specialized intervention, a revision of the business model and incentives, and dealing with non-trivial ethical questions \citep{stray2020,zuckerberg2018}.

From a more generic and algorithmic viewpoint, a promising approach to breaking feedback loops is the reinforcement learning perspective, where the algorithm is understood to be an agent that aims to please and learn simultaneously \citep{sb21rlintro}. A recommendation is seen as an opportunity to match user interests to the best of the current system knowledge (immediate optimization), but also to extend this knowledge by exploring yet unknown or uncertain areas, so as to produce better recommendations in the future. Optimal solutions care for both the long and short term (exploration and exploitation), striving to achieve a subtle balance. After all, the mission of a recommender system is not circumscribed to a single recommendation for every user, but hopefully a recurrent, cyclic, mutually beneficial relationship with the customer over an extended time span. Much of the literature in the field builds however implicitly on the view of a once-in-a-lifetime interaction with the user.

Developments in this area have adapted so-called multi-armed bandit (MAB) approaches \citep{sb21rlintro}, where items are seen as actions to be selected (choices to be recommended) \citep{hill2020}. When chosen, actions return a reward (a utility or value for the user), drawn from an unknown distribution, and the goal of the system is to maximize the cumulative rewards over time. MAB algorithms break the feedback loop by injecting a controlled degree of guided stochastic exploration beyond local maxima (a controlled bit of informed randomness, to put it simply), seeking a better explore vs. exploit balance than traditional, pure exploitation approaches (which include all the algorithms discussed in Section \ref{sec:algorithms}). In order to define personalized solutions, so-called contextual variants of MAB are developed, where the context fundamentally includes the user, and the reward depends on the context \citep{Li2016}.

Further generalization beyond contextual MAB considers the effect that recommendations have on the user state, thus representing the interaction of users with the system as a Markov Decision Process (MDP) \citep{xin2020}. MDP are also applied as a natural fit to represent sequential recommendation \citep{xin2022}, that we discussed earlier in Section \ref{sec:sequential-session-based}. MDP-oriented solutions typically require large amounts of online interaction in order to be effective, which may take more time to start producing good recommendations than is affordable \citep{afsar2022}. This can be mitigated by complementary offline training using more abundant logged interaction data, which then introduces bias from the logging policy \citep{xin2022b}, that needs to be dealt with as discussed earlier. Reinforcement learning in recommendation is a rapidly growing area, with tight connections to counterfactual learning and evaluation; the reader may find a good starting point towards more in-depth readings in the survey by \citet{afsar2022}.

As a final note from a broader abstract perspective, popularity biases can be seen as arising from a compound feedback loop involving manifold channels of user-item discovery and feedback, interacting with and mutually reinforcing each other. Direct novelty enhancement, bias-countering techniques, and reinforcement learning can be seen as multiple solutions in a common direction seeking a degree of healthy exploration in recommendations.

\section{Impact and Value of Recommender Systems}
\label{sec:impact-and-value}
Over the last thirty years, recommender systems have been successfully deployed in many application domains, and their value both for consumers and businesses is undisputed~\cite{jannachjugovactmis2019}.
However, like many other areas of applied machine learning---including various subfields of information retrieval---academic \emph{research} in recommender systems is
mostly not based on studies with deployed systems, but
largely relies on a data-centric approach. Moreover, differently from observational and analytical types of data-based research, where the goal for example is to understand certain phenomena from recorded data, recommender systems research almost exclusively focuses on developing novel prediction models. The research question in such experiments is therefore mostly not ``What made a recommendation successful?'' \cite{JannachLudewigLerche2017umuai} or ``Are popular items, when recommended, more likely to be purchased by users?'', but almost exclusively if a new prediction model is able to more accurately \emph{predict} the held-out \emph{past} data.

The holy grail in recommender systems research today therefore is to improve the prediction accuracy of algorithms. The underlying and generally plausible assumption is that better accuracy leads to the effect that more relevant items appear at higher positions in the recommendation lists. As a result, the more relevant items are easier to discover by the users, thereby reducing information overload.
In this context, one central feature of almost any recommender system lies in the \emph{personalization} of the item suggestions.\footnote{
While the personalization of search results has been extensively studied in IR settings as well, non-personalized recommendations, e.g., of generally popular items, are commonly not considered to be competitive with personalized ones.
} As such, recommender systems can seen as strategies to implement digital one-to-one marketing and hyper-segmentation.

Unfortunately, the assumed correspondence of higher prediction accuracy on past data with better overall \emph{value} for the user or the provider of the recommendations is less than certain. A number of academic research works for example report that higher prediction accuracy only sometimes lead to a better quality \emph{perception} by users, as obtained through user studies. Also various works from industry, e.g., by Netflix, report that the offline experimentation is often not predictive of \emph{online success} \cite{Gomez-Uribe:2015:NRS:2869770.2843948}. The limitations of relying solely on accuracy measures can be easily illustrated. Imagine a user who liked the first two ``Avengers'' movies and rated them highly on an online video rental platform. An algorithm that then recommends three additional Avengers films along with related superhero movies, might even reach an accuracy of 100\,\%. While being 100\,\% accurate, the recommendation might be of little value both for the user---as the recommendation list is obvious and maybe too monotone---and the provider, because the user would have rented these movies anyway. Recommending these movies can therefore even represent a missed opportunity to promote other items.

Given these observations, it is pivotal that the field of recommender systems moves beyond research that is solely based on offline experiments and abstract accuracy measures. Such a change is highly important, in particular because it is  not clear if small improvements on a specific accuracy measure for a given dataset---such things are commonly reported in published research---would even matter in practice.

In this section, we therefore first emphasize the importance of considering the human in the loop when proposing new algorithms or user interfaces for recommender systems (Section \ref{subsec:impact-of-rs}). Then, we discuss the various ways in which recommender systems can create value, both for consumers and recommendation providers, and why it is important to evaluate recommender systems with respect to their intended purpose (Section \ref{subsec:value-of-rs}).

\subsection{Understanding the Impact of Recommendations with the Human in the Loop}
\label{subsec:impact-of-rs}
There are a number of relevant questions one might have to address when building a recommender system in practice, for example:
\begin{itemize}
\item Would users find the recommendations provided by a system \emph{useful}?
\item Would they find them novel, diverse, or entertaining, and would they ultimately be \emph{satisfied} with the recommendations?
\item Would they believe that the system's recommendations are \emph{trustworthy}, would they like to see an explanation?
\item Will they consider the system's recommendations in their decisions in the future?
\end{itemize}

All these aspects might have an impact on the ultimate success of a recommender system, which depends on whether or not users adopt the system's recommendations. Unfortunately, none of these questions can be answered with typical offline experiments.
Ultimately, recommending online---to a large extent---is a problem of human-computer interaction, where a computerized, interactive system is designed to support its users in an information-finding or decision-making task. Thus, studies that involve users are a necessary means to explore questions like those mentioned above.

The most common form of such studies are \emph{experiments} in the form of a \emph{randomized controlled trial}, where the study participants are randomly assigned to either one or more \emph{treatment} groups or to a \emph{control} group. In many such studies, the participants interact with different versions of a prototype system in a controlled environment, often in a lab. For example, in an experiment on effects of providing explanations to users, one participant group would interact with a system that provides explanations, and the other one with one that does not. The existence of the explanations would therefore be the manipulated \emph{independent} variable in the experiment. Depending on the specific research question, various \emph{dependent} variables can be measured. These dependent variables can either be objectively measured, e.g., by recording the participants' behavior like mouse clicks or interaction time, or they can be obtained by asking participants to self-report their perceptions in a questionnaire. If the research question, for example, is if explanations help reducing the user's \emph{decision effort}, one can both measure the time needed to make a choice from the given recommendations, and ask participants about the \emph{perceived} effort of the task. After the experiment, various statistical analyses can be made to assess if the existence of explanations may have caused or is at least correlated with the observations for decision effort.

Differently from typical offline experiments, such user studies are guided by explicit research questions or hypotheses, which are commonly derived from theoretical considerations. As a consequence, each user study requires a specific experimental design that is suited to answer these research questions. This fact and the need to recruit participants for the study, which are representative for at least a subset of the target population, makes user studies typically more effortful than offline experiments, which are often not based on theoretical considerations or explicit research questions.

\begin{figure*}[h!t]
    \centering
    \includegraphics[width=0.75\textwidth,clip]{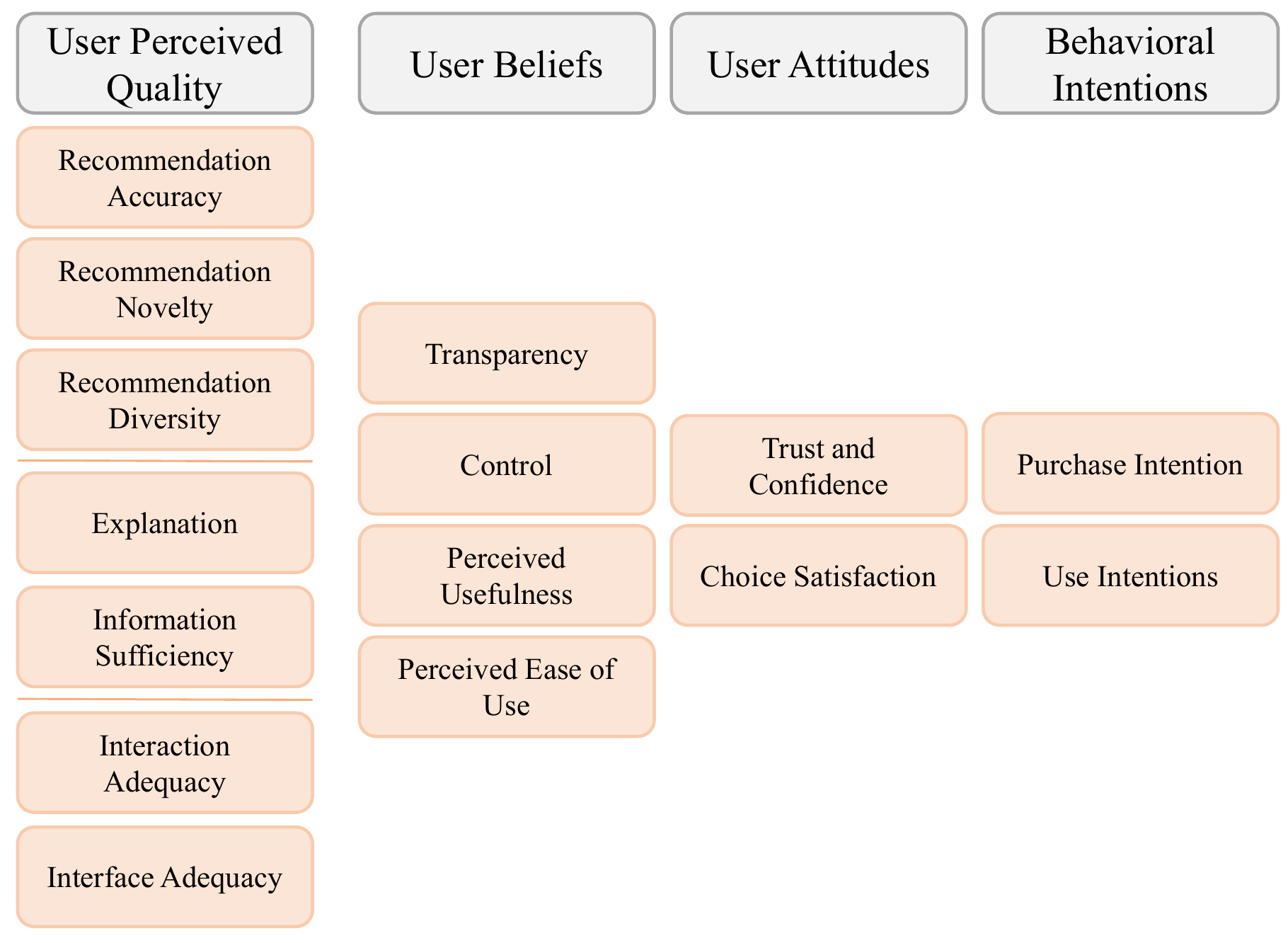}
    \caption{Overview of the ResQue framework for User-Centric Evaluation of Recommender Systems (see \cite{Pu:2011:UEF:2043932.2043962}).}
    \label{fig:resque}
\end{figure*}

User-centric evaluation frameworks, as proposed in \cite{Pu:2011:UEF:2043932.2043962} or \cite{umuai2012knijnenburg} are one way of standardizing such user experiments at least to some extent. These frameworks include a number of general quality factors and the potential effects of recommender systems on user behavior. Figure \ref{fig:resque} provides an overview of the ResQue framework \cite{Pu:2011:UEF:2043932.2043962}, which---partly inspired by the Technology Acceptance Model---proposes to assess the potential impact of user quality perceptions on their beliefs, attitudes and future behavioral intentions. Besides the identification of the relevant constructs (variables) in the model, the framework furthermore provides a set of questionnaire items that can be used to measure the participants' subjective perceptions in an experiment.

Generally, controlled experiments with users are much more common in the Information Systems literature than in the Computer Science literature on recommender systems. Given that many important questions---in particular also regarding the user interface of recommenders \cite{JugovacJannachTiis2017}---cannot be answered with the predominant but narrow focus on improving prediction models, it is important that research in recommender systems more often considers evaluations with users in the loop. Clearly, user studies also have their limitations, e.g., that the study participants are not interacting with a real system or are typically not making real purchase decisions. Nonetheless, such studies can provide us with relevant indications of what might matter for users in the real world.

Finally, note that controlled experiments are not the only form of evaluations with the human in the loop. Various other types of \emph{observational} studies as well as \emph{qualitative} research methods, such as focus groups or interviews, are commonly applied in academia and in industry.

\subsection{Consumer and Business Value of Recommender Systems} 
\label{subsec:value-of-rs}
Ultimately, any recommender system is designed to create utility or value for one or more of the involved stakeholders. The academic literature historically focuses on the \emph{consumer value}, at least implicitly. One underlying assumption often is that if a recommender system is able to create value for consumers, e.g., by helping them to discover novel items, this at least indirectly leads to value for the business or organization that provides the recommendations. Such considerations seem plausible, for example, for music or video streaming services that are based on flat-rate subscription rates. The more the consumers are engaged with the service---due to the constant discovery of new content---the more likely they might be to renew their subscription at the end of the month. However, the goals of the provider might not always be fully aligned with the value perspective of consumers, leading to \emph{multi-stakeholder} recommendation problems.

\subsubsection{Recommendation as a Multi-Stakeholder Optimization Problem}
Different types of entities can be involved in the recommendation process \cite{abdollahpouri2020,jannach2021mcnamara}.
\begin{itemize}
\item \emph{Consumers} or \emph{End Users} are the recipients of the recommendations. This can be individual users or groups of users. In the latter \emph{group recommendation} setting, see \cite{Masthoff2022}, typically the needs and preferences of different group members must be balanced, i.e., there can be multiple stakeholders in the group.
\item \emph{Recommendation Service Providers} are the organizations that are in the control of the recommender system, which they run as part of their business. Typical examples include large online retailers or media streaming providers.
\item \emph{Suppliers} are the entities that create or provide the items that are offered through the recommendation services. In the e-commerce example, this could be the manufacturers; in the media domain, it could be artists or labels, or other content providers. Note that the recommendation service providers can also be the suppliers of the items they offer.
\item \emph{Society}: In case a recommendation service affects a large portion of a society, e.g., feed recommendations on a social media site, recommender systems can also have a societal impact, e.g., when the recommender system reinforces certain political opinions in an unbalanced way.
\end{itemize}

Consider the following example, which illustrates the potential challenges of considering multiple stakeholders interests \cite{jannach2021mcnamara}. Imagine an online hotel search service which charges a commission fee to hotel owners for each booking that is made through the site. From the consumer's perspective, the best recommendation is one that satisfies their needs, usually given certain budget constraints. The recommendation service provider, on the other hand, might try to maximize its short-term profit (commission). A recommender system might therefore select items that represent a reasonable match for the consumer, but lead to higher profit for the platform. At the same time, the hotel platform might also try to ensure that all hotel suppliers are recommended from time to time, so that the platform remains attractive for the suppliers. The hotel chains, finally, might have an interest that the recommender system pushes a certain subset of the offerings through recommendations, for example, the more expensive ones.

Today, research in multi-stakeholder recommender systems is still limited. Works exist, for example, for the specific problems of reciprocal recommendation \cite{Reciprocal2021}, price and profit aware recommender systems \cite{JannachAdomaviciusVAMS2017}, or in the area of algorithmic fairness and biases, e.g., \cite{DBLP:conf/recsys/AbdollahpouriMB20,mmbld18cikm}. Generally, a particular challenge in the context of multi-stakeholder recommendation lies in the fact that the most common evaluation approach---accuracy evaluation on historical datasets---are typically not sufficient to investigate these problems. Instead, alternative methodological approaches are needed, e.g., based on simulation techniques \cite{Gedas2020Longitudinal,ghanem2022balancing}. This is particularly important because balancing multiple stakeholder goals often requires longitudinal analyses that are able to uncover the long-term effects on the involved stakeholders.

\subsubsection{Impact and Value Oriented Recommender Systems Evaluation}
In practical deployments, the performance of recommender systems is determined in terms of Key Performance Indicators (KPIs) that are selected or defined by responsible organizational units. What is actually being measured primarily depends on the purpose that is intended to achieve with the recommender system. Therefore, the particular choice or design of the KPI also depends on the specific business model of the provider.

A survey of real-world success stories of recommender systems \cite{jannachjugovactmis2019} identified the following main types of measurements that are taken in practice:
\begin{itemize}
  \item \emph{Click-Through-Rates (CTR)}: The CTR is traditionally most suitable in ad-based business models, i.e., where page impressions matter. In recommendation scenarios, optimizing for click-rates may however be too short-sighted in many applications.
  \item \emph{Adoption and Conversion Measures}: These measures go beyond counting click events. They, for example, only count a recommendation as successful, if there are signs that it was truly useful for consumers, e.g., when they watched a recommended movie.
  \item \emph{Sales and Revenue}: This is the most direct measurement and can, for example, consist of determining how much additional sales was stimulated by a recommender.
  \item \emph{Effects on Sales Distributions}: Measures of that type are useful to assess the impact of the recommendations on consumer behavior, e.g., if they could be persuaded to select different options compared to a situation with no recommender.
  \item \emph{User Engagement}: This measure is often considered to be related with long-term customer loyalty. When consumers interact more with a service, this is considered a sign that they will come back for purchases in the future or that they will continue or renew their subscription.
\end{itemize}

The specific choice of these KPIs, as discussed, can be strongly tied to the specific market situation and individual business models of the recommendation providers. Academic research typically does not target at the idiosyncrasies of such specific situations but aims at providing generalizable insights. Nonetheless, it is important---also in academia--which kinds of value recommender systems can generate, both for consumers and providers.

Today's research, as outlined above, is largely (implicitly) focused on the value for consumers, which is often circumscribed as helping users to find items of interest (or: ``Find good items'' \cite{herlocker2004}). Commonly, prediction accuracy metrics are used as proxies to measure this value by determining how good an algorithm is at predicting previously recorded and held-out positive interactions between a user and an item (see Section \ref{sec:evaluation}). While this abstraction is generally useful and provides as with a standardized way of operationalizing the research problem, it may represent an oversimplification of the problem and even misguide our research efforts, as discussed above. In reality, there are many different ways in which recommender systems can create value, see \cite{jannach2021mcnamara}.

To overcome this somewhat narrow research operationalization, researchers have proposed a purpose- and value-oriented framework for the evaluation of recommender systems \cite{JannachAdomavicius2016purpose}. The main goal of this conceptual framework is to guide researchers---both in academia and industry---in terms of how they measure the success of a recommender system. Specifically, the framework aims to ensure that what is measured, i.e., the KPIs or metrics that are used, is aligned with the intended purpose of the system.

The proposed framework, which considers both the consumer and provider perspective, has four layers:
\begin{enumerate*}[label=\emph{(\roman*)}]
  \item Overarching Goals,
  \item Purpose of the Recommender,
  \item Computational Task,
  \item Evaluation Approach.
\end{enumerate*}
To understand how the framework works, consider the case of an online media streaming service. The \emph{overarching goal} from a business perspective---which can be entirely independent of any recommendation technology---might be to ensure the long-term profitability of the service. The specific \emph{purpose} of a recommender in that context that serves this goal might be to increase user engagement. One way to achieve this increased engagement can be to provide consumers with a mix of familiar and novel recommendations to support discovery. The \emph{computational task} of the recommender system therefore is to determine such a mix. Finally, the \emph{evaluation approach} must be chosen that it is aligned with the original overarching goals In the example, it could be a mixed methods approach, where \emph{(a)} offline evaluations is used to gauge the accuracy and novelty levels of alternative algorithms, where \emph{(b)} user studies are conducted to assess the intention of potential consumers to use the service in the future and where \emph{(c)} field studies determine the impact of the recommendations on the user's media streaming activity.

The most important point in this context---which should become obvious when using the proposed framework---is that relying solely on computational measures (such as prediction accuracy), can be too limiting. In practice, there are probably not too many cases where measures like accuracy strictly correlate with aspects that matter more in practice, such as customer retention or customer satisfaction. Therefore, as discussed in more depth in \cite{jannach2021mcnamara}, a paradigm shift seems needed in terms of how we evaluate recommender systems. Such a paradigm shift should ultimately lead to more impactful research in this area. To achieve this progress researchers should, for example:
\begin{itemize}
  \item more often adopt a purpose-oriented evaluation approach, as described, where goals and evaluation measures are ensured to be better aligned;
  \item consider a multi-method research approach, which also includes human-in-the-loop evaluations both in terms of controlled experiments and qualitative research;
  \item expand their methodological repertoire also in terms of improved data-based evaluations, considering for example simulation-based approaches to investigate longitudinal effects;
  \item consider domain- and application specifics more often where appropriate, acknowledging that there is no ``best model'' and that the same set of recommendations might be good in one situation and sub-optimal in another one.
\end{itemize}

Following such advice might ultimately help to avoid a \emph{dead end} in algorithms research, which is not only heavily focused on abstract metrics, but is also suffering from certain methodological problems \cite{ferraridacrema2020tois}, which may limit the impact of current recommender research in practice.

\section{Summary and Future Directions} %
\label{sec:summary-future-directions}
We conclude this paper with a brief summary, pointers to further readings, and an outlook on important future directions in this area.

\subsection{Summary}
Today, personalized recommendations are a central component of many online services, and various success stories are reported in the literature how such systems create value both for users and organizations. Due to their wide success in practice, research interest has correspondingly been continuously growing over the past twenty-five years. Ultimately, this led to the development an own recommender systems research field, which is historically strongly rooted in information retrieval,  machine learning, and human-computer interaction. In this paper, we have first characterized the recommendation problem on an abstract and application-independent level and then reviewed the most common way of operationalizing the research problem as a matrix completion problem. After discussing the most common types of recommendation algorithms, we elaborated on how to evaluate recommender systems ``in the lab'' and ``in the wild''. Finally, we reviewed recent technical developments in the area of session-based recommendation and approaches based on reinforcement learning.

\subsection{Further Readings and Future Directions}
In this paper, we largely focused on algorithmic aspects of recommender systems, which traditionally is also the most active area of research. However, building a recommender systems in practice requires much more than just a clever algorithm \cite{Xiao:2007:EPR:2017327.2017335,JannachResnickEtAl2016}. Questions of the user experience of recommender systems---the HCI perspective---and the impact of recommenders on users---the Information Systems perspective---seem to be underexplored compared to the large amount of algorithmic research that we observe today. More research often seems also required regarding the idiosyncrasies of particular application domains of recommender systems, acknowledging that there is  no ``best'' application-independent algorithm and that the success of a recommender can only be evaluated relative to its intended purpose. As a result, an algorithm that works well for predicting movie relevance scores based on long-term user preferences might not be suited well for the recommendation of possible short-lived news articles or the context-dependent recommendation of music or points-of-interest in the tourism domain. Various applications of recommender systems are discussed in more depth in the Recommender Systems Handbook \cite{RSHandbook2022}.

In the same handbook, a variety of other timely research topics in recommender systems are discussed, which were not considered here, including the recent algorithmic developments, context-awareness, attacks on recommender systems, privacy aspects, or human factors in the evaluation process. In the remainder of this paper, we would like to discuss a small set of additional promising topics of which we believe that they require more research in the future.

\paragraph{Conversational Recommender Systems.}
Most of today's online recommender system have a rather simple user interaction model. At defined places in the application, the system presents users with recommendations, and often the only available user action is to inspect or accept these recommendations.\footnote{Only sometimes further mechanisms are provided to let users give feedback on the recommendations or exert some control on what is shown \citep{JannachNaveedEtAl2016}.} With the recent developments in natural language processing, the current progress in machine learning in general, and the increased spread of voice-enabled devices, more interactive forms of information retrieval and recommendation \cite{RadlinskiConvSearch2017} have received renewed interest in recent years. One vision of such \emph{conversational recommender systems} (CRS) is that they are able to act like a human and engage in a ``natural'' conversation with their users, see \citep{jannach2021crscsur} for a survey. Correspondingly, such future systems will be able to support a variety of user intents \citep{cai2019department}, thus supporting interactive preference elicitation and revision, explanation, and also chit-chat. While historically such CRS were often built based on engineered knowledge, e.g., about possible dialog states, many of today's approaches also consider ``end-to-end learning'' as their main approach, where a machine learning model is trained on larger collections of dialogues between humans. Considerable progress was made in that context in recent years. Still, pure learning-based approaches still have their limitations, and it is expected that future CRS will be often based on a mix of explicit domain and inference knowledge and of learning components.

\paragraph{Fairness in Recommender Systems.}
Fairness and ethical concerns have grown around the exponential development of artificial intelligence and technology, permeating and transforming all realms of modern societies---and recommendation technologies are no exception to these matters. Recommendation functionalities can determine the success of a music artist on Spotify, a seller on Amazon, a job candidate or a recruiter through LinkedIn, a research author on Google Scholar, a hotel owner on Booking.com, an Internet celebrity on YouTube, Instagram or TikTok, a political party or a world view or a value system through online news and online media. As mediators between providers and consumers, recommender systems, willingly or not, get laded with a burden of responsibility they can hardly let go of.

Fairness, and more generally, responsible AI, has indeed become a pressing issue in recommender systems research that finds reflection in the recent research literature \citep{Mehrotra2018}, keynote talks \citep{b20recsys}, new dedicated research outlets \citep{elish2021}, and public debate. Metrics are being proposed to measure fairness, and algorithms and enhancements are being researched to avoid unfair treatment of groups \citep{Yao2017}, minorities or individuals, as both providers and consumers of recommended choices \citep{Patro2020}. A first realization with the onset of this new research area is the complexity of fairness as a concept and the difficulty of achieving universal solutions. Studies are unanimous in stressing the importance of bias awareness and understanding as a key step (and possibly a considerable part of the solution) to prevent or mitigate unfairness. An ample strand of reflection and research efforts can be expected to continue in the times to come.

\paragraph{Offline / Online Misalignment in Evaluation.}
As mentioned in Section \ref{sec:evaluation:online}, offline evaluation results are often weakly aligned with the outcomes of online A/B testing \citep{Gomez-Uribe:2015:NRS:2869770.2843948,jannachjugovactmis2019,gfdabh14recsys}. Known causes may partially account for this, such as a metric mismatch, bias in the offline data, or interference from external components when the evaluated techniques are integrated in complex, multi-step recommendation pipelines usually involved in industrial-scale applications \citep{ab15rshb}. But a more complete explanation seems not to have been reached, and the problem may not have been fully understood yet. Important efforts are being made in this direction from different angles, as discussed in earlier sections \citep{jannach2021mcnamara,gcnad18wsdm}. Improvements in bringing offline evaluation closer to a measure of the final effect of recommendation would have a great positive impact in the future development of recommendation technologies.



\begin{thebibliography}{190}

\providecommand{\natexlab}[1]{#1}
\providecommand{\url}[1]{#1}
\providecommand{\doi}[1]{DOI: #1}

\bibitem[Abdollahpouri et~al.(2020{\natexlab{a}})Abdollahpouri, Adomavicius,
  Burke, Guy, Jannach, Kamishima, Krasnodebski, and Pizzato]{abdollahpouri2020}
H.~Abdollahpouri, G.~Adomavicius, R.~Burke, I.~Guy, D.~Jannach, T.~Kamishima,
  J.~Krasnodebski, and L.~Pizzato.
\newblock 2020{\natexlab{a}}.
\newblock Multistakeholder recommendation: Survey and research directions.
\newblock \emph{User Modeling and User-Adapted Interaction}, 30: 127--158.

\bibitem[Abdollahpouri et~al.(2020{\natexlab{b}})Abdollahpouri, Mansoury,
  Burke, and Mobasher]{DBLP:conf/recsys/AbdollahpouriMB20}
H.~Abdollahpouri, M.~Mansoury, R.~Burke, and B.~Mobasher.
\newblock 2020{\natexlab{b}}.
\newblock The connection between popularity bias, calibration, and fairness in
  recommendation.
\newblock In \emph{Fourteenth {ACM} Conference on Recommender Systems (RecSys
  2020)}, pp. 726--731. {ACM}.

\bibitem[Adamopoulos and Tuzhilin(2014)]{at14tist}
P.~Adamopoulos and A.~Tuzhilin.
\newblock 2014.
\newblock On unexpectedness in recommender systems: Or how to expect the
  unexpected.
\newblock \emph{Special Issue on Novelty and Diversity in Recommender Systems,
  ACM Transactions on Intelligent Systems and Technology}, 5(4).

\bibitem[Adomavicius and Tuzhilin(2005)]{TowardsGedas2005}
G.~Adomavicius and A.~Tuzhilin.
\newblock 2005.
\newblock Toward the next generation of recommender systems: A survey of the
  state-of-the-art and possible extensions.
\newblock \emph{IEEE Transactions on Knowledge and Data Engineering}, 17(6):
  734--749.

\bibitem[Adomavicius et~al.(2022)Adomavicius, Bauman, Tuzhilin, and
  Unger]{Adomavicius2022}
G.~Adomavicius, K.~Bauman, A.~Tuzhilin, and M.~Unger.
\newblock 2022.
\newblock Context-aware recommender systems: From foundations to recent
  developments.
\newblock In F.~Ricci, L.~Rokach, and B.~Shapira, eds., \emph{Recommender
  Systems Handbook, $3^\textrm{rd}$ edition}, pp. 211--250. Springer, New York,
  NY, USA.

\bibitem[Afchar and Hennequin(2020)]{ah20recsys}
D.~Afchar and R.~Hennequin.
\newblock 2020.
\newblock Making neural networks interpretable with attribution: Application to
  implicit signals prediction.
\newblock In \emph{Proceedings of the $14^\textrm{th}$ ACM Conference on
  Recommender Systems}, RecSys 2020, pp. 220--229. ACM, New York, NY, USA.

\bibitem[Afsar et~al.(2022)Afsar, Crump, and Far]{afsar2022}
M.~M. Afsar, T.~Crump, and B.~Far.
\newblock June 2022.
\newblock Reinforcement learning based recommender systems: A survey.
\newblock \emph{ACM Computing Surveys}, in press.

\bibitem[Agrawal et~al.(2009)Agrawal, Gollapudi, Halverson, and
  Ieong]{aghi09wsdm}
R.~Agrawal, S.~Gollapudi, A.~Halverson, and S.~Ieong.
\newblock 2009.
\newblock Diversifying search results.
\newblock In \emph{Proceedings of the 2$^\textrm{nd}$ ACM Conference on Web
  Search and Data Mining}, WSDM 2009, pp. 5--14. ACM, New York, NY, USA.

\bibitem[Amatriain and Basilico(2015)]{ab15rshb}
X.~Amatriain and J.~Basilico.
\newblock 2015.
\newblock Recommender systems in industry: {A} netflix case study.
\newblock In F.~Ricci, L.~Rokach, and B.~Shapira, eds., \emph{Recommender
  Systems Handbook, $2^\textrm{nd}$ edition}, pp. 385--419. Springer, New York,
  NY, USA.

\bibitem[Anelli et~al.(2022)Anelli, Bellogin, Noia, Jannach, and
  Pomo]{anellitop2022}
V.~W. Anelli, A.~Bellogin, T.~D. Noia, D.~Jannach, and C.~Pomo.
\newblock 2022.
\newblock Top-n recommendation algorithms: A quest for the state-of-the-art.
\newblock In \emph{30th ACM Conference on User Modeling, Adaptation and
  Personalization (UMAP 2022)}.

\bibitem[Baeza{-}Yates(2020)]{b20recsys}
R.~Baeza{-}Yates.
\newblock 2020.
\newblock Bias in search and recommender systems.
\newblock In R.~L.~T. Santos, L.~B. Marinho, E.~M. Daly, L.~Chen, K.~Falk,
  N.~Koenigstein, and E.~S. de~Moura, eds., \emph{$14^\textrm{th}$ {ACM}
  Conference on Recommender Systems (RecSys 2020)}, p.~2. {ACM}.

\bibitem[Balabanovi\'{c} and Shoham(1997)]{Fab1997}
M.~Balabanovi\'{c} and Y.~Shoham.
\newblock Mar. 1997.
\newblock Fab: Content-based, collaborative recommendation.
\newblock \emph{Communications of the ACM}, 40(3): 66--72.

\bibitem[Bamber(1975)]{b75jmp}
D.~Bamber.
\newblock 1975.
\newblock The area above the ordinal dominance graph and the area below the
  receiver operating characteristic graph.
\newblock \emph{Journal of Mathematical Psychology}, 12(4): 387--415.

\bibitem[Bambini et~al.(2011)Bambini, Cremonesi, and Turrin]{Bambini2011}
R.~Bambini, P.~Cremonesi, and R.~Turrin.
\newblock 2011.
\newblock \emph{Recommender Systems Handbook}, chapter {A Recommender System
  for an IPTV Service Provider: a Real Large-Scale Production Environment}, pp.
  299--331.
\newblock Springer.
\newblock Eds. Francesco Ricci, Lior Rokach, Bracha Shapira, and Paul B.
  Kantor.

\bibitem[Belkin and Croft(1992)]{bc92comacm}
N.~J. Belkin and W.~B. Croft.
\newblock 1992.
\newblock Information filtering and information retrieval: Two sides of the
  same coin?
\newblock \emph{Communications of the ACM}, 35(12): 29--38.

\bibitem[Bellog{\'i}n et~al.(2017)Bellog{\'i}n, Castells, and
  Cantador]{bcc17irj}
A.~Bellog{\'i}n, P.~Castells, and I.~Cantador.
\newblock 2017.
\newblock Statistical biases in information retrieval metrics for recommender
  systems.
\newblock \emph{Information Retrieval}, 20(6): 606--634.

\bibitem[Bennett et~al.(2007)Bennett, Elkan, Liu, Smyth, and
  Tikk]{NetflixPrize2007}
J.~Bennett, C.~Elkan, B.~Liu, P.~Smyth, and D.~Tikk.
\newblock 2007.
\newblock {KDD} cup and workshop 2007.
\newblock \emph{{SIGKDD} Explor.}, 9(2): 51--52.

\bibitem[Billsus and Pazzani(1998)]{Billsus98Learning}
D.~Billsus and M.~J. Pazzani.
\newblock 1998.
\newblock Learning collaborative information filters.
\newblock In \emph{Proceedings of the Fifteenth International Conference on
  Machine Learning}, ICML 1998, pp. 46--54. Morgan Kaufmann Publishers Inc.

\bibitem[Breese et~al.(1998)Breese, Heckerman, and Kadie]{bhk98uai}
J.~S. Breese, D.~Heckerman, and C.~Kadie.
\newblock 1998.
\newblock Empirical analysis of predictive algorithms for collaborative
  filtering.
\newblock In \emph{Proceedings of the 14$^\textrm{th}$ Conference on
  Uncertainty in Artificial Intelligence}, UAI 1998, pp. 43--52. Morgan
  Kaufmann Publishers Inc.

\bibitem[Buckley and Voorhees.(2004)]{bv04sigir}
C.~Buckley and E.~M. Voorhees.
\newblock 2004.
\newblock Retrieval evaluation with incomplete information.
\newblock In \emph{Proceedings of the $27^\textrm{th}$ Annual International ACM
  SIGIR Conference on Research and Development in Information Retrieval (SIGIR
  2004)}, pp. 25--32. ACM, New York, NY, USA.

\bibitem[Burke(2002)]{Burke02umuai}
R.~Burke.
\newblock 2002.
\newblock Hybrid recommender systems: Survey and experiments.
\newblock \emph{User Modeling and User-Adapted Interaction}, 12: 331--370.

\bibitem[Cai and Chen(2019)]{cai2019department}
W.~Cai and L.~Chen.
\newblock 2019.
\newblock Towards a taxonomy of user feedback intents for conversational
  recommendations.
\newblock In \emph{RecSys' 19 Late-Breaking Results}, pp. 572--573.

\bibitem[Campos et~al.(2014)Campos, D{\'i}ez, and Cantador]{Campos2014}
P.~G. Campos, F.~D{\'i}ez, and I.~Cantador.
\newblock 2014.
\newblock Time-aware recommender systems: a comprehensive survey and analysis
  of existing evaluation protocols.
\newblock \emph{User Modeling and User-Adapted Interaction}, 24(1): 67--119.

\bibitem[Ca{\~n}amares and Castells(2017)]{cc17sigir}
R.~Ca{\~n}amares and P.~Castells.
\newblock 2017.
\newblock A probabilistic reformulation of memory-based collaborative
  filtering: implications on popularity biases.
\newblock In \emph{Proceedings of the $40^\textrm{th}$ Annual International ACM
  SIGIR Conference on Research and Development in Information Retrieval (SIGIR
  2017)}, pp. 215--224. ACM, New York, NY, USA.

\bibitem[Ca{\~n}amares and Castells(2018)]{cc18sigir}
R.~Ca{\~n}amares and P.~Castells.
\newblock 2018.
\newblock Should {I} follow the crowd? {A} probabilistic analysis of the
  effectiveness of popularity in recommender systems.
\newblock In \emph{Proceedings of the $41^\textrm{st}$ Annual International ACM
  SIGIR Conference on Research and Development in Information Retrieval (SIGIR
  2018)}, pp. 415--424. ACM, New York, NY, USA.

\bibitem[Ca{\~n}amares and Castells(2020)]{cc20recsys}
R.~Ca{\~n}amares and P.~Castells.
\newblock 2020.
\newblock On target item sampling in offline recommender system evaluation.
\newblock In \emph{$14^\textrm{th}$ ACM Conference on Recommender Systems
  (RecSys 2020)}, pp. 259--268. ACM, New York, NY, USA.

\bibitem[Ca{\~n}amares et~al.(2020)Ca{\~n}amares, Castells, and
  Moffat]{ccm2020irj}
R.~Ca{\~n}amares, P.~Castells, and A.~Moffat.
\newblock 2020.
\newblock Offline evaluation options for recommender systems.
\newblock \emph{Inf. Ret.}, 23(4): 387--411.

\bibitem[Castagnos et~al.(2013)Castagnos, Brun, and Boyer]{Castagnos2013}
S.~Castagnos, A.~Brun, and A.~Boyer.
\newblock 2013.
\newblock When diversity is needed... but not expected!
\newblock In \emph{Proceedings of the 3$^\textrm{rd}$ International Conference
  on Advances in Information Mining and Management}, IMMM 2013, pp. 44--50.
  Lisbon, Portugal.

\bibitem[Castells and Moffat(2022)]{castells2022}
P.~Castells and A.~Moffat.
\newblock 2022.
\newblock Offline recommender system evaluation: Challenges and new directions.
\newblock \emph{AI Magazine}, 43(2).

\bibitem[Castells et~al.(2022)Castells, Hurley, and Vargas]{chv22handbook}
P.~Castells, N.~J. Hurley, and S.~Vargas.
\newblock 2022.
\newblock Novelty and diversity in recommender systems.
\newblock In F.~Ricci, L.~Rokach, and B.~Shapira, eds., \emph{Recommender
  Systems Handbook, $3^\textrm{rd}$ edition}, pp. 603--646. Springer, New York,
  NY, USA.

\bibitem[Celma and Herrera(2008)]{ch08recsys}
O.~Celma and P.~Herrera.
\newblock 2008.
\newblock A new approach to evaluating novel recommendations.
\newblock In \emph{Proceedings of the 2$^\textrm{nd}$ ACM Conference on
  Recommender Systems}, RecSys 2008, pp. 179--186. ACM, New York, NY, USA.

\bibitem[Cen et~al.(2020)Cen, Zhang, Zou, Zhou, Yang, and Tang]{cen2020}
Y.~Cen, J.~Zhang, X.~Zou, C.~Zhou, H.~Yang, and J.~Tang.
\newblock 2020.
\newblock Controllable multi-interest framework for recommendation.
\newblock In \emph{Proceedings of the $26^\textrm{th}$ ACM SIGKDD International
  Conference on Knowledge Discovery \& Data Mining}, KDD 2020, pp. 2942--2951.
  ACM, New York, NY, USA.

\bibitem[Chaney et~al.(2018{\natexlab{a}})Chaney, Stewart, and
  Engelhardt]{chaney2018}
A.~J.~B. Chaney, B.~M. Stewart, and B.~E. Engelhardt.
\newblock 2018{\natexlab{a}}.
\newblock How algorithmic confounding in recommendation systems increases
  homogeneity and decreases utility.
\newblock In \emph{Proceedings of the 12th ACM Conference on Recommender
  Systems}, RecSys 2018, pp. 224--232. ACM, New York, NY, USA.

\bibitem[Chaney et~al.(2018{\natexlab{b}})Chaney, Stewart, and
  Engelhardt]{cse18recsys}
A.~J.~B. Chaney, B.~M. Stewart, and B.~E. Engelhardt.
\newblock 2018{\natexlab{b}}.
\newblock How algorithmic confounding in recommendation systems increases
  homogeneity and decreases utility.
\newblock In S.~Pera, M.~D. Ekstrand, X.~Amatriain, and J.~O'Donovan, eds.,
  \emph{Proceedings of the $12^\textrm{th}$ {ACM} Conference on Recommender
  Systems (RecSys 2018)}, pp. 224--232. {ACM}.

\bibitem[Chen et~al.(2015)Chen, Chen, and Wang]{Chen2015reviews}
L.~Chen, G.~Chen, and F.~Wang.
\newblock 2015.
\newblock Recommender systems based on user reviews: the state of the art.
\newblock \emph{User Model User-Adap Inter}, 25: 99--154.

\bibitem[Chen et~al.(2019)Chen, Yang, Wang, Yang, and Yuan]{Chen2019}
L.~Chen, Y.~Yang, N.~Wang, K.~Yang, and Q.~Yuan.
\newblock 2019.
\newblock How serendipity improves user satisfaction with recommendations? a
  large-scale user evaluation.
\newblock In \emph{Proceedings of the The World Wide Web Conference}, WWW 2019,
  pp. 240--250. ACM, New York, NY, USA.

\bibitem[Ciampaglia et~al.(2018)Ciampaglia, Nematzadeh, Menczer, and
  Flammini]{nematzadeh2017}
G.~L. Ciampaglia, A.~Nematzadeh, F.~Menczer, and A.~Flammini.
\newblock 2018.
\newblock How algorithmic popularity bias hinders or promotes quality.
\newblock \emph{Scientific Reports}, 8.

\bibitem[Covington et~al.(2016)Covington, Adams, and Sargin]{covington2016}
P.~Covington, J.~Adams, and E.~Sargin.
\newblock 2016.
\newblock Deep neural networks for youtube recommendations.
\newblock In \emph{Proceedings of the $10^\textrm{th}$ ACM Conference on
  Recommender Systems}, RecSys 2016, pp. 191--198. ACM, New York, NY, USA.

\bibitem[Cremonesi et~al.(2010)Cremonesi, Koren, and Turrin]{ckt10recsys}
P.~Cremonesi, Y.~Koren, and R.~Turrin.
\newblock 2010.
\newblock Performance of recommender algorithms on top-n recommendation tasks.
\newblock In \emph{Proceedings of the $4^\textrm{th}$ ACM Conf. on Recommender
  Systems (RecSys 2010)}, pp. 39--46. ACM, New York, NY, USA.

\bibitem[Dhelim et~al.(2022)Dhelim, Aung, Bouras, Ning, and
  Cambria]{Dhelim2022}
S.~Dhelim, N.~Aung, M.~A. Bouras, H.~Ning, and E.~Cambria.
\newblock 2022.
\newblock A survey on personality-aware recommendation systems.
\newblock \emph{Artif Intell Rev}, 55: 2409–2454.

\bibitem[Dong et~al.(2022)Dong, Yuan, Yao, Wang, Xu, and Zhu]{DONG2022108954}
M.~Dong, F.~Yuan, L.~Yao, X.~Wang, X.~Xu, and L.~Zhu.
\newblock 2022.
\newblock A survey for trust-aware recommender systems: A deep learning
  perspective.
\newblock \emph{Knowledge-Based Systems}, 249: 108954.

\bibitem[Ekstrand et~al.(2014)Ekstrand, Harper, Willemsen, and
  Konstan]{Ekstrand2014user}
M.~D. Ekstrand, F.~M. Harper, M.~C. Willemsen, and J.~A. Konstan.
\newblock 2014.
\newblock User perception of differences in recommender algorithms.
\newblock In \emph{Proceedings of the $8^\textrm{th}$ ACM Conference on
  Recommender Systems}, RecSys 2014, pp. 161--168. ACM, New York, NY, USA.

\bibitem[Elish et~al.(2021)Elish, Isaac, and Zemel]{elish2021}
M.~C. Elish, W.~Isaac, and R.~S. Zemel, eds.
\newblock 2021.
\newblock \emph{{ACM} Conference on Fairness, Accountability, and Transparency
  (FAccT 2021)}. {ACM}.

\bibitem[Feng et~al.(2020)Feng, Hu, Lv, Liu, Zhang, and Ou]{feng2020}
Y.~Feng, B.~Hu, F.~Lv, Q.~Liu, Z.~Zhang, and W.~Ou.
\newblock 2020.
\newblock \emph{ATBRG: Adaptive Target-Behavior Relational Graph Network for
  Effective Recommendation}, pp. 2231--2240.
\newblock SIGIR 2020. ACM, New York, NY, USA.

\bibitem[{Ferrari Dacrema} et~al.(2021){Ferrari Dacrema}, Boglio, Cremonesi,
  and Jannach]{ferraridacrema2020tois}
M.~{Ferrari Dacrema}, S.~Boglio, P.~Cremonesi, and D.~Jannach.
\newblock 2021.
\newblock A troubling analysis of reproducibility and progress in recommender
  systems research.
\newblock \emph{ACM Transactions on Information Systems}, 39(2).

\bibitem[Fleder and Hosanagar(2009)]{fh09mansci}
D.~Fleder and K.~Hosanagar.
\newblock 2009.
\newblock Blockbuster culture’s next rise or fall: The impact of recommender
  systems on sales diversity.
\newblock \emph{Management Science}, 55(5): 697--712.

\bibitem[Foltz and Dumais(1992)]{Foltz1992PersonalizedInformationDelivery}
P.~W. Foltz and S.~T. Dumais.
\newblock 1992.
\newblock Personalized information delivery: An analysis of information
  filtering methods.
\newblock \emph{Communications of the ACM}, 35(12): 51--60.

\bibitem[Garcin et~al.(2014)Garcin, Faltings, Donatsch, Alazzawi, Bruttin, and
  Huber]{gfdabh14recsys}
F.~Garcin, B.~Faltings, O.~Donatsch, A.~Alazzawi, C.~Bruttin, and A.~Huber.
\newblock 2014.
\newblock Offline and online evaluation of news recommender systems at
  swissinfo.ch.
\newblock In \emph{Proceedings of the $8^\textrm{th}$ ACM Conference on
  Recommender Systems}, RecSys 2014, pp. 169--176. ACM, New York, NY, USA.

\bibitem[Garg et~al.(2019)Garg, Gupta, Malhotra, Vig, and Shroff]{Garg:2019}
D.~Garg, P.~Gupta, P.~Malhotra, L.~Vig, and G.~Shroff.
\newblock 2019.
\newblock Sequence and time aware neighborhood for session-based
  recommendations: Stan.
\newblock In \emph{Proceedings of the $42^\textrm{nd}$ International ACM SIGIR
  Conference on Research and Development in Information Retrieval}, SIGIR 2019,
  pp. 1069--1072.

\bibitem[Ge et~al.(2010)Ge, Delgado-Battenfeld, and Jannach]{Ge2010}
M.~Ge, C.~Delgado-Battenfeld, and D.~Jannach.
\newblock 2010.
\newblock Beyond accuracy: evaluating recommender systems by coverage and
  serendipity.
\newblock In \emph{Proceedings of the 4$^\textrm{th}$ ACM Conference on
  Recommender Systems}, RecSys 2010, pp. 257--260.

\bibitem[Ge et~al.(2012)Ge, Jannach, Gedikli, and Hepp]{GeJannachEtAl2012}
M.~Ge, D.~Jannach, F.~Gedikli, and M.~Hepp.
\newblock 2012.
\newblock Effects of the placement of diverse items in recommendation lists.
\newblock In \emph{$14^\textrm{th}$ International Conference on Enterprise
  Information Systems ({ICEIS 2012})}, pp. 201--208.

\bibitem[Ghanem et~al.(2022)Ghanem, Leitner, and Jannach]{ghanem2022balancing}
N.~Ghanem, S.~Leitner, and D.~Jannach.
\newblock 2022.
\newblock Balancing consumer and business value of recommender systems: A
  simulation-based analysis.
\newblock \emph{arXiv preprint arXiv:2203.05952}.

\bibitem[Gilotte et~al.(2018)Gilotte, Calauz\`{e}nes, Nedelec, Abraham, and
  Doll\'{e}]{gcnad18wsdm}
A.~Gilotte, C.~Calauz\`{e}nes, T.~Nedelec, A.~Abraham, and S.~Doll\'{e}.
\newblock 2018.
\newblock Offline {A/B} testing for recommender systems.
\newblock In \emph{Proceedings of the $11^\textrm{th}$ ACM International
  Conference on Web Search and Data Mining (WSDM 2018)}, pp. 198--206. ACM, New
  York, NY, USA.

\bibitem[Goldberg et~al.(1992)Goldberg, Nichols, Oki, and
  Terry]{goldberg1992using}
D.~Goldberg, D.~Nichols, B.~M. Oki, and D.~Terry.
\newblock 1992.
\newblock Using collaborative filtering to weave an information tapestry.
\newblock \emph{Communications of the ACM}, 35(12): 61--70.

\bibitem[Gomez-Uribe and Hunt(2015)]{Gomez-Uribe:2015:NRS:2869770.2843948}
C.~A. Gomez-Uribe and N.~Hunt.
\newblock 2015.
\newblock The {Netflix} recommender system: Algorithms, business value, and
  innovation.
\newblock \emph{Transactions on Management Information Systems}, 6(4).

\bibitem[Gruson et~al.(2019)Gruson, Chandar, Charbuillet, {McInerney}, Hansen,
  Tardieu, and Carterette]{gccmhtc19wsdm}
A.~Gruson, P.~Chandar, C.~Charbuillet, J.~{McInerney}, S.~Hansen, D.~Tardieu,
  and B.~Carterette.
\newblock 2019.
\newblock Offline evaluation to make decisions about playlist recommendation.
\newblock In \emph{Proceedings of the $12^\textrm{th}$ ACM International
  Conference on Web Search and Data Mining (WSDM 2019)}, pp. 420--428. ACM, New
  York, NY, USA.

\bibitem[Gunawardana et~al.(2022)Gunawardana, Shani, and Yogev]{sg22handbook}
A.~Gunawardana, G.~Shani, and S.~Yogev.
\newblock 2022.
\newblock Evaluating recommender systems.
\newblock In F.~Ricci, L.~Rokach, and B.~Shapira, eds., \emph{Recommender
  Systems Handbook, $3^\textrm{rd}$ edition}, pp. 547--601. Springer, New York,
  NY, USA.

\bibitem[Guy(2022)]{g22rshb}
I.~Guy.
\newblock 2022.
\newblock Social recommender systems.
\newblock In F.~Ricci, L.~Rokach, and B.~Shapira, eds., \emph{Recommender
  Systems Handbook, $3^\textrm{rd}$ edition}, pp. 835--870. Springer, New York,
  NY, USA.

\bibitem[Harper and Konstan(2015)]{MovieLens2015}
F.~M. Harper and J.~A. Konstan.
\newblock 2015.
\newblock The movielens datasets: History and context.
\newblock \emph{ACM Transactions on Interactive Intelligent Systems}, 5(4).

\bibitem[He and Chu(2010)]{He2010}
J.~He and W.~W. Chu.
\newblock 2010.
\newblock \emph{A Social Network-Based Recommender System (SNRS)}, pp. 47--74.
\newblock Springer US.

\bibitem[He and McAuley(2016)]{Fossil2016}
R.~He and J.~McAuley.
\newblock 2016.
\newblock Fusing similarity models with markov chains for sparse sequential
  recommendation.
\newblock In \emph{Proceedings of the $16^\textrm{th}$ IEEE International
  Conference on Data Mining}, ICDM 2016, pp. 191--200.

\bibitem[He et~al.(2017)He, Liao, Zhang, Nie, Hu, and Chua]{hlznhc17www}
X.~He, L.~Liao, H.~Zhang, L.~Nie, X.~Hu, and T.-S. Chua.
\newblock 2017.
\newblock Neural collaborative filtering.
\newblock In \emph{Proceedings of the $26^\textrm{th}$ International Conference
  on World Wide Web}, WWW 2017, pp. 173--182.

\bibitem[Herlocker et~al.(2004)Herlocker, Konstan, Terveen, and
  Riedl]{herlocker2004}
J.~L. Herlocker, J.~A. Konstan, L.~G. Terveen, and J.~T. Riedl.
\newblock 2004.
\newblock Evaluating collaborative filtering recommender systems.
\newblock \emph{ACM Transactions on Information Systems}, 22(1): 5--53.

\bibitem[Hern\'{a}ndez-Lobato et~al.(2014)Hern\'{a}ndez-Lobato, Houlsby, and
  Ghahramani]{hhg14jlmr}
J.~M. Hern\'{a}ndez-Lobato, N.~Houlsby, and Z.~Ghahramani.
\newblock 2014.
\newblock Probabilistic matrix factorization with non-random missing data.
\newblock In \emph{Proceedings of the $31^\textrm{st}$ International Conference
  on Machine Learning}, ICML 2014, pp. 1512--1520. Proc. of Machine Learning
  Research, Sheffield, UK.

\bibitem[Hidasi and Karatzoglou(2018)]{hidasi2018}
B.~Hidasi and A.~Karatzoglou.
\newblock 2018.
\newblock Recurrent neural networks with top-k gains for session-based
  recommendations.
\newblock In \emph{Proceedings of the $27^\textrm{th}$ ACM International
  Conference on Information and Knowledge Management}, CIKM 2018, pp. 843--852.
  ACM, New York, NY, USA.

\bibitem[Hill et~al.(2017)Hill, Nassif, Liu, Iyer, and Vishwanathan]{hill2020}
D.~N. Hill, H.~Nassif, Y.~Liu, A.~Iyer, and S.~V.~N. Vishwanathan.
\newblock 2017.
\newblock An efficient bandit algorithm for realtime multivariate optimization.
\newblock In \emph{Proceedings of the $23^\textrm{rd}$ {ACM} {SIGKDD}
  International Conference on Knowledge Discovery and Data Mining (KDD 2017)},
  pp. 1813--1821. {ACM}.

\bibitem[Hofmann(2004)]{h04tois}
T.~Hofmann.
\newblock 2004.
\newblock Latent semantic models for collaborative filtering.
\newblock \emph{ACM Transactions on Information Systems}, 22(1): 89--115.

\bibitem[Hu and Pu(2011)]{Hu2011b}
R.~Hu and P.~Pu.
\newblock 2011.
\newblock Enhancing recommendation diversity with organization interfaces.
\newblock In \emph{Proceedings of the 16$^\textrm{th}$ International Conference
  on Intelligent User Interfaces}, IUI 2011, pp. 347--350.

\bibitem[Hu et~al.(2008)Hu, Koren, and Volinsky]{hkv08icdm}
Y.~Hu, Y.~Koren, and C.~Volinsky.
\newblock 2008.
\newblock Collaborative filtering for implicit feedback datasets.
\newblock In \emph{Proceedings of the $8^\textrm{th}$ IEEE International
  Conference on Data Mining}, ICDM 2008, pp. 15--19. IEEE Computer Society,
  Washington, DC, USA.

\bibitem[Hurley and Zhang(2011)]{hz11toit}
N.~Hurley and M.~Zhang.
\newblock 2011.
\newblock Novelty and diversity in top-{N} recommendation -- analysis and
  evaluation.
\newblock \emph{ACM Transactions on Internet Technology}, 10(4).

\bibitem[Jadidinejad et~al.(2021)Jadidinejad, Macdonald, and
  Ounis]{jmo2021tois}
A.~H. Jadidinejad, C.~Macdonald, and I.~Ounis.
\newblock sep 2021.
\newblock The simpson’s paradox in the offline evaluation of recommendation
  systems.
\newblock \emph{ACM Transactions on Information Systems}, 40(1).
\newblock ISSN 1046-8188.

\bibitem[Jannach and Adomavicius(2016)]{JannachAdomavicius2016purpose}
D.~Jannach and G.~Adomavicius.
\newblock 2016.
\newblock Recommendations with a purpose.
\newblock In \emph{Proceedings of the $10^\textrm{th}$ {ACM} Conference on
  Recommender Systems}, RecSys 2016, pp. 7--10.

\bibitem[Jannach and Adomavicius(2017)]{JannachAdomaviciusVAMS2017}
D.~Jannach and G.~Adomavicius.
\newblock 2017.
\newblock Price and profit awareness in recommender systems.
\newblock In \emph{Proceedings of the ACM RecSys 2017 Workshop on Value-Aware
  and Multi-Stakeholder Recommendation}.

\bibitem[Jannach and Bauer(2020)]{jannach2021mcnamara}
D.~Jannach and C.~Bauer.
\newblock 2020.
\newblock {Escaping the McNamara Fallacy: Towards more Impactful Recommender
  Systems Research}.
\newblock \emph{AI Magazine}, 41(4): 79--95.

\bibitem[Jannach and Hegelich(2009)]{JannachHegelich2009}
D.~Jannach and K.~Hegelich.
\newblock 2009.
\newblock A case study on the effectiveness of recommendations in the mobile
  internet.
\newblock In \emph{Proceedings of the 10th {ACM} Conference on Recommender
  Systems}, RecSys '09, pp. 205--208.

\bibitem[Jannach and Jugovac(2019)]{jannachjugovactmis2019}
D.~Jannach and M.~Jugovac.
\newblock 2019.
\newblock Measuring the business value of recommender systems.
\newblock \emph{ACM Transactions on Management Information Systems (TMIS)},
  10(4): 1--23.

\bibitem[Jannach and Ludewig(2017)]{JannachLudewig2017RecSys}
D.~Jannach and M.~Ludewig.
\newblock 2017.
\newblock When recurrent neural networks meet the neighborhood for
  session-based recommendation.
\newblock In \emph{Proceedings of the $11^\textrm{th}$ Recommender Systems
  Conference}, RecSys 2017.

\bibitem[Jannach et~al.(2010)Jannach, Zanker, Felfernig, and
  Friedrich]{JannachZankerEtAl2010}
D.~Jannach, M.~Zanker, A.~Felfernig, and G.~Friedrich.
\newblock 2010.
\newblock \emph{Recommender Systems - An Introduction}.
\newblock Cambridge University Press.

\bibitem[Jannach et~al.(2015{\natexlab{a}})Jannach, Lerche, and
  Kamehkhosh]{jannach2015}
D.~Jannach, L.~Lerche, and I.~Kamehkhosh.
\newblock 2015{\natexlab{a}}.
\newblock Beyond ``hitting the hits:'' generating coherent music playlist
  continuations with the right tracks.
\newblock In \emph{Proceedings of the $9^\textrm{th}$ ACM Conference on
  Recommender Systems}, RecSys 2015, pp. 187--194. ACM, New York, NY, USA.

\bibitem[Jannach et~al.(2015{\natexlab{b}})Jannach, Lerche, Kamehkhosh, and
  Jugovac]{jlkj15umuai}
D.~Jannach, L.~Lerche, I.~Kamehkhosh, and M.~Jugovac.
\newblock 2015{\natexlab{b}}.
\newblock What recommenders recommend: an analysis of recommendation biases and
  possible countermeasures.
\newblock \emph{User Modeling and User-Adapted Interaction}, 25(5): 427--491.

\bibitem[Jannach et~al.(2016{\natexlab{a}})Jannach, Naveed, and
  Jugovac]{JannachNaveedEtAl2016}
D.~Jannach, S.~Naveed, and M.~Jugovac.
\newblock 2016{\natexlab{a}}.
\newblock User control in recommender systems: Overview and interaction
  challenges.
\newblock In \emph{$17^\textrm{th}$ International Conference on Electronic
  Commerce and Web Technologies}, EC-Web 2016.

\bibitem[Jannach et~al.(2016{\natexlab{b}})Jannach, Resnick, Tuzhilin, and
  Zanker]{JannachResnickEtAl2016}
D.~Jannach, P.~Resnick, A.~Tuzhilin, and M.~Zanker.
\newblock 2016{\natexlab{b}}.
\newblock Recommender systems - beyond matrix completion.
\newblock \emph{Communications of the {ACM}}, 59(11): 94--102.

\bibitem[Jannach et~al.(2017)Jannach, Ludewig, and
  Lerche]{JannachLudewigLerche2017umuai}
D.~Jannach, M.~Ludewig, and L.~Lerche.
\newblock 2017.
\newblock Session-based item recommendation in e-commerce: On short-term
  intents, reminders, trends, and discounts.
\newblock \emph{User-Modeling and User-Adapted Interaction}, 27(3--5):
  351--392.

\bibitem[Jannach et~al.(2018)Jannach, Lerche, and
  Zanker]{JannachLercheZanker2018}
D.~Jannach, L.~Lerche, and M.~Zanker.
\newblock 2018.
\newblock Recommending based on implicit feedback.
\newblock In P.~Brusilovsky and D.~He, eds., \emph{Social Information Access}.
  Springer.

\bibitem[Jannach et~al.(2021)Jannach, Manzoor, Cai, and
  Chen]{jannach2021crscsur}
D.~Jannach, A.~Manzoor, W.~Cai, and L.~Chen.
\newblock 2021.
\newblock A survey on conversational recommender systems.
\newblock \emph{ACM Computing Surveys}, forthcoming.

\bibitem[Jannach et~al.(2022)Jannach, Cremonesi, and
  Quadrana]{JannachHBSessions2022}
D.~Jannach, P.~Cremonesi, and M.~Quadrana.
\newblock 2022.
\newblock Session-based recommendation.
\newblock In F.~Ricci, L.~Rokach, and B.~Shapira, eds., \emph{Recommender
  Systems Handbook, $3^\textrm{rd}$ edition}, pp. 301--224. Springer, New York,
  NY, USA.

\bibitem[Joachims(2002)]{j01kdd}
T.~Joachims.
\newblock 2002.
\newblock Optimizing search engines using clickthrough data.
\newblock In \emph{Proceedings of the $8^\textrm{th}$ ACM SIGKDD International
  Conference on Knowledge Discovery and Data Mining}, KDD 2002, pp. 133--142.
  ACM, New York, NY, USA.

\bibitem[Jugovac and Jannach(2017)]{JugovacJannachTiis2017}
M.~Jugovac and D.~Jannach.
\newblock 2017.
\newblock Interacting with recommenders - overview and research directions.
\newblock \emph{ACM Transactions on Intelligent Interactive Systems (ACM
  TiiS)}, 7(3).

\bibitem[Jugovac et~al.(2017)Jugovac, Jannach, and
  Lerche]{JugovacJannachLerche2017eswa}
M.~Jugovac, D.~Jannach, and L.~Lerche.
\newblock 2017.
\newblock Efficient optimization of multiple recommendation quality factors
  according to individual user tendencies.
\newblock \emph{Expert Systems With Applications}, 81: 321--331.

\bibitem[Kamehkhosh and Jannach(2017)]{KamehkhoshJannach2017}
I.~Kamehkhosh and D.~Jannach.
\newblock 2017.
\newblock User perception of next-track music recommendations.
\newblock In \emph{Proceedings of the 2017 Conference on User Modeling
  Adaptation and Personalization}, UMAP 2017, pp. 113--121.

\bibitem[Kang and McAuley(2018)]{SASRec2018}
W.~Kang and J.~J. McAuley.
\newblock 2018.
\newblock Self-attentive sequential recommendation.
\newblock In \emph{Proceedings of the $18^\textrm{th}$ {IEEE} International
  Conference on Data Mining}, ICDM 2018, pp. 197--206. {IEEE} Computer Society.

\bibitem[Kapoor et~al.(2015)Kapoor, Kumar, Terveen, Konstan, and
  Schrater]{Kapoor2015ILike}
K.~Kapoor, V.~Kumar, L.~Terveen, J.~A. Konstan, and P.~Schrater.
\newblock 2015.
\newblock ``i like to explore sometimes'': Adapting to dynamic user novelty
  preferences.
\newblock In \emph{Proceedings of the $9^\textrm{th}$ ACM Conference on
  Recommender Systems}, RecSys 2015, pp. 19--26.

\bibitem[Kirshenbaum et~al.(2012)Kirshenbaum, Forman, and
  Dugan]{KirshenbaumForbes}
E.~Kirshenbaum, G.~Forman, and M.~Dugan.
\newblock 2012.
\newblock A live comparison of methods for personalized article recommendation
  at forbes.com.
\newblock In \emph{Machine Learning and Knowledge Discovery in Databases}, pp.
  51--66.

\bibitem[Knijnenburg and Willemsen(2015)]{kw15rshb}
B.~P. Knijnenburg and M.~C. Willemsen.
\newblock 2015.
\newblock Evaluating recommender systems with user experiments.
\newblock In F.~Ricci, L.~Rokach, and B.~Shapira, eds., \emph{Recommender
  Systems Handbook}, pp. 309--352. Springer, New York, NY, USA.

\bibitem[Knijnenburg et~al.(2012)Knijnenburg, Willemsen, Gantner, Soncu, and
  Newell]{umuai2012knijnenburg}
B.~P. Knijnenburg, M.~C. Willemsen, Z.~Gantner, H.~Soncu, and C.~Newell.
\newblock 2012.
\newblock Explaining the user experience of recommender systems.
\newblock \emph{User Modeling and User-Adapted Interaction}, 22: 441--504.

\bibitem[Koren(2008)]{k08sigkdd}
Y.~Koren.
\newblock 2008.
\newblock Factorization meets the neighborhood: a multifaceted collaborative
  filtering model.
\newblock In \emph{Proceedings of the $14^\textrm{th}$ ACM SIGKDD International
  Conference on Knowledge Discovery and Data Mining (KDD 2008)}, pp. 426--434.
  ACM, New York, NY, USA.

\bibitem[Koren(2009)]{k09kdd}
Y.~Koren.
\newblock 2009.
\newblock Collaborative filtering with temporal dynamics.
\newblock In \emph{Proceedings of the $15^\textrm{th}$ ACM SIGKDD International
  Conference on Knowledge Discovery and Data Mining (KDD 2009)}, pp. 447--456.
  ACM, New York, NY, USA.

\bibitem[Koren et~al.(2009)Koren, Bell, and Volinsky]{kbv09computer}
Y.~Koren, R.~M. Bell, and C.~Volinsky.
\newblock 2009.
\newblock Matrix factorization techniques for recommender systems.
\newblock \emph{Computer}, 42(8): 30--37.

\bibitem[Kouki et~al.(2020)Kouki, Fountalis, Vasiloglou, Cui, Liberty, and
  Al~Jadda]{kouki2020}
P.~Kouki, I.~Fountalis, N.~Vasiloglou, X.~Cui, E.~Liberty, and K.~Al~Jadda.
\newblock 2020.
\newblock {From the lab to production: {A} case study of session-based
  recommendations in the home-improvement domain}.
\newblock In \emph{Proceedings of the $14^\textrm{th}$ ACM Conference on
  Recommender Systems}, RecSys 2020, pp. 140--149.

\bibitem[Krichene and Rendle(2020)]{kr20kdd}
W.~Krichene and S.~Rendle.
\newblock 2020.
\newblock On sampled metrics for item recommendation.
\newblock In \emph{Proceedings of the $26^\textrm{th}$ ACM SIGKDD International
  Conference on Knowledge Discovery and Data Mining (KDD 2020)}, pp.
  1748--1757. ACM, New York, NY, USA.

\bibitem[Lathia et~al.(2010)Lathia, Hailes, Capra, and Amatriain]{lhca10sigir}
N.~Lathia, S.~Hailes, L.~Capra, and X.~Amatriain.
\newblock 2010.
\newblock Temporal diversity in recommender systems.
\newblock In \emph{Proceedings of the 33$^\textrm{rd}$ Annual International ACM
  SIGIR Conference on Research and Development in Information Retrieval}, SIGIR
  2010, pp. 210--217. ACM, New York, NY, USA.

\bibitem[Lathia(2010)]{l10phd}
N.~K. Lathia.
\newblock 2010.
\newblock \emph{Evaluating collaborative filtering over time}.
\newblock PhD thesis, University College London, {UK}.

\bibitem[Latifi et~al.(2020)Latifi, Mauro, and Jannach]{latifi2020sessionaware}
S.~Latifi, N.~Mauro, and D.~Jannach, 2020.
\newblock Session-aware recommendation: A surprising quest for the
  state-of-the-art.

\bibitem[Li et~al.(2016)Li, Karatzoglou, and Gentile]{Li2016}
S.~Li, A.~Karatzoglou, and C.~Gentile.
\newblock 2016.
\newblock Collaborative filtering bandits.
\newblock In \emph{Proceedings of the $39^\textrm{th}$ International ACM SIGIR
  Conference on Research and Development in Information Retrieval}, SIGIR 2016,
  pp. 539--548. ACM, New York, NY, USA.

\bibitem[Liang et~al.(2018)Liang, Krishnan, Hoffman, and Jebara]{liang2018}
D.~Liang, R.~G. Krishnan, M.~D. Hoffman, and T.~Jebara.
\newblock 2018.
\newblock Variational autoencoders for collaborative filtering.
\newblock In \emph{Proceedings of the World Wide Web Conference}, WWW 2018, pp.
  689--698. International World Wide Web Conferences Steering Committee,
  Republic and Canton of Geneva, CHE.

\bibitem[Linden et~al.(2003)Linden, Smith, and York]{linden2003}
G.~Linden, B.~Smith, and J.~York.
\newblock 2003.
\newblock Amazon.com recommendations: Item-to-item collaborative filtering.
\newblock \emph{{IEEE} Internet Computing}, 7(1): 76--80.

\bibitem[Liu et~al.(2007)Liu, Bennett, Elkan, Smyth, and Tikk]{Liu2007}
B.~Liu, J.~Bennett, C.~Elkan, P.~Smyth, and D.~Tikk.
\newblock 2007.
\newblock Kdd cup and workshop 2007.
\newblock In \emph{Proceedings of the $13^\textrm{th}$ ACM SIGKDD International
  Conference on Knowledge Discovery and Data Mining}, KDD 2007. ACM.

\bibitem[Liu et~al.(2020)Liu, Cheng, Dong, He, Pan, and Ming]{lcdhpm20sigir}
D.~Liu, P.~Cheng, Z.~Dong, X.~He, W.~Pan, and Z.~Ming.
\newblock 2020.
\newblock A general knowledge distillation framework for counterfactual
  recommendation via uniform data.
\newblock In \emph{Proceedings of the $43^\textrm{rd}$ International ACM SIGIR
  Conference on Research and Development in Information Retrieval}, SIGIR 2020,
  pp. 831--840. ACM, New York, NY, USA.

\bibitem[Liu(2009)]{l09now}
T.-Y. Liu.
\newblock 2009.
\newblock Learning to rank for information retrieval.
\newblock \emph{Foundations and Trends® in Information Retrieval}, 3(3):
  225--331.

\bibitem[Ludewig and Jannach(2019)]{ludewigjannach2019radio}
M.~Ludewig and D.~Jannach.
\newblock 2019.
\newblock User-centric evaluation of session-based recommendations for an
  automated radio station.
\newblock In \emph{Proceedings of the $13^\textrm{th}$ ACM Conference on
  Recommender Systems}, RecSys 2019, pp. 516--520.

\bibitem[Ludewig et~al.(2021)Ludewig, Latifi, Mauro, and
  Jannach]{ludewiglatifiumuai2020}
M.~Ludewig, S.~Latifi, N.~Mauro, and D.~Jannach.
\newblock 2021.
\newblock Empirical analysis of session-based recommendation algorithms.
\newblock \emph{User Modeling and User-Adapted Interaction}, 31: 149--181.

\bibitem[Marlin and Zemel(2009)]{mz09recsys}
B.~M. Marlin and R.~S. Zemel.
\newblock 2009.
\newblock Collaborative prediction and ranking with non-random missing data.
\newblock In \emph{Proceedings of the $3^\textrm{rd}$ ACM Conference on
  Recommender Systems (RecSys 2009)}, pp. 5--12. ACM, New York, NY, USA.

\bibitem[Masthoff and Delić(2022)]{Masthoff2022}
J.~Masthoff and A.~Delić.
\newblock 2022.
\newblock Group recommender systems: Beyond preference aggregation.
\newblock In F.~Ricci, L.~Rokach, and B.~Shapira, eds., \emph{Recommender
  Systems Handbook, $3^\textrm{rd}$ edition}, pp. 381--420. Springer, New York,
  NY, USA.

\bibitem[McNee et~al.(2006)McNee, Riedl, and Konstan]{mrk06chi}
S.~M. McNee, J.~Riedl, and J.~A. Konstan.
\newblock 2006.
\newblock Being accurate is not enough: How accuracy metrics have hurt
  recommender systems.
\newblock In \emph{Proceedings of ACM CHI 2006 Conference on Human Factors in
  Computing Systems}, CHI 2006, pp. 1097--1101. ACM, New York, NY, USA.

\bibitem[Mehrotra et~al.(2018{\natexlab{a}})Mehrotra, McInerney, Bouchard,
  Lalmas, and Diaz]{Mehrotra2018}
R.~Mehrotra, J.~McInerney, H.~Bouchard, M.~Lalmas, and F.~Diaz.
\newblock 2018{\natexlab{a}}.
\newblock Towards a fair marketplace: Counterfactual evaluation of the
  trade-off between relevance, fairness \& satisfaction in recommendation
  systems.
\newblock In \emph{Proceedings of the 27$^\textrm{th}$ ACM International
  Conference on Information and Knowledge Management}, CIKM 2018, pp.
  2243--2251. ACM, New York, NY, USA.

\bibitem[Mehrotra et~al.(2018{\natexlab{b}})Mehrotra, McInerney, Bouchard,
  Lalmas, and Diaz]{mmbld18cikm}
R.~Mehrotra, J.~McInerney, H.~Bouchard, M.~Lalmas, and F.~Diaz.
\newblock 2018{\natexlab{b}}.
\newblock Towards a fair marketplace: Counterfactual evaluation of the
  trade-off between relevance, fairness \& satisfaction in recommendation
  systems.
\newblock In \emph{Proceedings of the 27$^\textrm{th}$ ACM International
  Conference on Information and Knowledge Management}, CIKM 2018, pp.
  2243--2251. ACM, New York, NY, USA.

\bibitem[Mena-Maldonado et~al.(2021)Mena-Maldonado, Cañamares, Castells, Ren,
  and Sanderson]{mccrs21tois}
E.~Mena-Maldonado, R.~Cañamares, P.~Castells, Y.~Ren, and M.~Sanderson.
\newblock 2021.
\newblock Popularity bias in false-positive metrics for recommender systems
  evaluation.
\newblock \emph{ACM Transactions on Information Systems}.

\bibitem[Merton(1968)]{Merton56}
R.~K. Merton.
\newblock 1968.
\newblock The matthew effect in science.
\newblock \emph{Science}, 159(3810): 56--63.

\bibitem[Mikolov et~al.(2013)Mikolov, Sutskever, Chen, Corrado, and
  Dean]{mikolov2013}
T.~Mikolov, I.~Sutskever, K.~Chen, G.~Corrado, and J.~Dean.
\newblock 2013.
\newblock Distributed representations of words and phrases and their
  compositionality.
\newblock In \emph{Proceedings of the $26^\textrm{th}$ International Conference
  on Neural Information Processing Systems}, NIPS 2013, pp. 3111--3119. Curran
  Associates Inc., Red Hook, NY, USA.

\bibitem[Moffat and Zobel(2008)]{Moffat2008}
A.~Moffat and J.~Zobel.
\newblock December 2008.
\newblock Rank-biased precision for measurement of retrieval effectiveness.
\newblock \emph{ACM Transactions on Information Systems}, 27(1): 2:1--2:27.
\newblock ISSN 1046-8188.

\bibitem[Musto et~al.(2022)Musto, de~Gemmis, Lops, Narducci, and
  Semeraro]{Musto2022}
C.~Musto, M.~de~Gemmis, P.~Lops, F.~Narducci, and G.~Semeraro.
\newblock 2022.
\newblock Semantics and content-based recommendations.
\newblock In F.~Ricci, L.~Rokach, and B.~Shapira, eds., \emph{Recommender
  Systems Handbook, $3^\textrm{rd}$ edition}, pp. 251--298. Springer, New York,
  NY, USA.

\bibitem[Nikolakopoulos et~al.(2022)Nikolakopoulos, Ning, Desrosiers, and
  Karypis]{ndk22handbook}
A.~N. Nikolakopoulos, X.~Ning, C.~Desrosiers, and G.~Karypis.
\newblock 2022.
\newblock Trust your neighbors: A comprehensive survey of neighborhood-based
  methods for recommender systems.
\newblock In F.~Ricci, L.~Rokach, and B.~Shapira, eds., \emph{Recommender
  Systems Handbook, $3^\textrm{rd}$ edition}, pp. 39--89. Springer, New York,
  NY, USA.

\bibitem[Ning and Karypis(2011)]{nk11icdm}
X.~Ning and G.~Karypis.
\newblock 2011.
\newblock Slim: Sparse linear methods for top-n recommender systems.
\newblock In \emph{Proceedings of the $11^\textrm{th}$ IEEE International
  Conference on Data Mining}, ICDM 2011, pp. 497--506. IEEE Computer Society,
  Washington, DC, USA.

\bibitem[Oh et~al.(2011)Oh, Park, Yu, Song, and Park]{Oh2011}
J.~Oh, S.~Park, H.~Yu, M.~Song, and S.-T. Park.
\newblock 2011.
\newblock Novel recommendation based on personal popularity tendency.
\newblock In \emph{Proceedings of the $11^\textrm{th}$ IEEE Conference on Data
  Mining}, ICDM 2011, pp. 507--516.

\bibitem[Okura et~al.(2017)Okura, Tagami, Ono, and Tajima]{otot17kdd}
S.~Okura, Y.~Tagami, S.~Ono, and A.~Tajima.
\newblock 2017.
\newblock Embedding-based news recommendation for millions of users.
\newblock In \emph{Proceedings of the $23^\textrm{rd}$ ACM SIGKDD International
  Conference on Knowledge Discovery and Data Mining}, KDD 2017, pp. 1933--1942.
  ACM, New York, NY, USA.

\bibitem[Palomares et~al.(2021)Palomares, Porcel, Pizzato, Guy, and
  Herrera{-}Viedma]{Reciprocal2021}
I.~Palomares, C.~Porcel, L.~Pizzato, I.~Guy, and E.~Herrera{-}Viedma.
\newblock 2021.
\newblock Reciprocal recommender systems: Analysis of state-of-art literature,
  challenges and opportunities towards social recommendation.
\newblock \emph{Inf. Fusion}, 69: 103--127.

\bibitem[Patil and Taillie(1982)]{pt82jasa}
G.~P. Patil and C.~Taillie.
\newblock 1982.
\newblock Diversity as a concept and its measurement.
\newblock \emph{Journal of the American Statistical Association}, 77(379):
  548--561.

\bibitem[Patro et~al.(2020)Patro, Biswas, Ganguly, Gummadi, and
  Chakraborty]{Patro2020}
G.~K. Patro, A.~Biswas, N.~Ganguly, K.~P. Gummadi, and A.~Chakraborty.
\newblock 2020.
\newblock Fairrec: Two-sided fairness for personalized recommendations in
  two-sided platforms.
\newblock In \emph{Proceedings of the Web Conference}, WWW 2020, pp.
  1194--1204. ACM / IW3C2.

\bibitem[Pazzani and Billsus(2007)]{pb07aw}
M.~J. Pazzani and D.~Billsus.
\newblock 2007.
\newblock \emph{Content-Based Recommendation Systems}, pp. 325--341.
\newblock Springer-Verlag, Berlin, Heidelberg.

\bibitem[Pennington et~al.(2014)Pennington, Socher, and
  Manning]{pennington2014}
J.~Pennington, R.~Socher, and C.~Manning.
\newblock Oct. 2014.
\newblock {G}lo{V}e: Global vectors for word representation.
\newblock In \emph{Proceedings of the 2014 Conference on Empirical Methods in
  Natural Language Processing ({EMNLP})}, pp. 1532--1543. ACL, Doha, Qatar.

\bibitem[Pu et~al.(2011)Pu, Chen, and Hu]{Pu:2011:UEF:2043932.2043962}
P.~Pu, L.~Chen, and R.~Hu.
\newblock 2011.
\newblock A user-centric evaluation framework for recommender systems.
\newblock In \emph{Proceedings of the $5^\textrm{th}$ ACM Conference on
  Recommender Systems}, RecSys 2011, pp. 157--164.

\bibitem[Quadrana et~al.(2018)Quadrana, Cremonesi, and
  Jannach]{QuadranaetalCSUR2018}
M.~Quadrana, P.~Cremonesi, and D.~Jannach.
\newblock 2018.
\newblock Sequence-aware recommender systems.
\newblock \emph{ACM Computing Surveys}, 51(4).

\bibitem[Radlinski and Craswell(2017)]{RadlinskiConvSearch2017}
F.~Radlinski and N.~Craswell.
\newblock 2017.
\newblock A theoretical framework for conversational search.
\newblock In \emph{Proceedings of the 2017 Conference on Conference Human
  Information Interaction and Retrieval}, CHIIR 2017, pp. 117--126.

\bibitem[Rendle(2010)]{rendle2010}
S.~Rendle.
\newblock 2010.
\newblock Factorization machines.
\newblock In \emph{Proceedings of the $10^\textrm{th}$ IEEE International
  Conference on Data Mining}, ICDM 2010, pp. 995--1000.

\bibitem[Rendle(2012)]{rendle2012}
S.~Rendle.
\newblock May 2012.
\newblock Factorization machines with libfm.
\newblock \emph{ACM Transactions on Intelligent Systems and Technology}, 3(3).

\bibitem[Rendle et~al.(2009)Rendle, Freudenthaler, Gantner, and
  Schmidt{-}Thieme]{rfgs09uai}
S.~Rendle, C.~Freudenthaler, Z.~Gantner, and L.~Schmidt{-}Thieme.
\newblock 2009.
\newblock {BPR:} bayesian personalized ranking from implicit feedback.
\newblock In J.~A. Bilmes and A.~Y. Ng, eds., \emph{Proceedings of the
  $25^\textrm{th}$ Conference on Uncertainty in Artificial Intelligence (UAI
  2009)}, pp. 452--461. {AUAI} Press.

\bibitem[Rendle et~al.(2010)Rendle, Freudenthaler, and
  Schmidt-Thieme]{rendle10FPMC}
S.~Rendle, C.~Freudenthaler, and L.~Schmidt-Thieme.
\newblock 2010.
\newblock Factorizing personalized markov chains for next-basket
  recommendation.
\newblock In \emph{Proceedings of the $19^\textrm{th}$ International Conference
  on World Wide Web}, WWW 2010, pp. 811--820.

\bibitem[Rendle et~al.(2020)Rendle, Krichene, Zhang, and
  Anderson]{rkza20recsys}
S.~Rendle, W.~Krichene, L.~Zhang, and J.~R. Anderson.
\newblock 2020.
\newblock Neural collaborative filtering vs. matrix factorization revisited.
\newblock In \emph{Proceedings of the $14^\textrm{th}$ {ACM} Conference on
  Recommender Systems (RecSys 2020)}, pp. 240--248. {ACM}.

\bibitem[Resnick et~al.(1994)Resnick, Iacovou, Suchak, Bergstrom, and
  Riedl]{ResnickGrouplens1994}
P.~Resnick, N.~Iacovou, M.~Suchak, P.~Bergstrom, and J.~Riedl.
\newblock 1994.
\newblock Grouplens: An open architecture for collaborative filtering of
  netnews.
\newblock In \emph{Proceedings of the ACM Conference on Computer Supported
  Cooperative Work}, CSCW 1994, pp. 175--186.

\bibitem[Ricci et~al.(2022)Ricci, Rokach, and Shapira]{RSHandbook2022}
F.~Ricci, L.~Rokach, and B.~Shapira, eds.
\newblock 2022.
\newblock \emph{Recommender Systems Handbook, $3^\textrm{rd}$ edition}, 3rd.
\newblock Springer, New York, NY, USA.

\bibitem[Rossetti et~al.(2016)Rossetti, Stella, and Zanker]{rsz16recsys}
M.~Rossetti, F.~Stella, and M.~Zanker.
\newblock 2016.
\newblock Contrasting offline and online results when evaluating recommendation
  algorithms.
\newblock In \emph{Proceedings of the $10^\textrm{th}$ ACM Conf. on Recommender
  Systems}, RecSys 2016, pp. 31--34. ACM, New York, NY, USA.

\bibitem[Salakhutdinov and Mnih(2007)]{sm07nips}
R.~Salakhutdinov and A.~Mnih.
\newblock 2007.
\newblock Probabilistic matrix factorization.
\newblock In J.~C. Platt, D.~Koller, Y.~Singer, and S.~T. Roweis, eds.,
  \emph{Proceedings of the Twenty-First Annual Conference on Neural Information
  Processing Systems (NIPS 2007)}, pp. 1257--1264. Curran Associates, Inc.

\bibitem[Sanderson(2010)]{s10now}
M.~Sanderson.
\newblock 2010.
\newblock Test collection based evaluation of information retrieval systems.
\newblock \emph{Foundations and Trends in Information Retrieval}, 4(4):
  247--375.

\bibitem[{Sanderson} and {Croft}(2012)]{sc2012pie}
M.~{Sanderson} and W.~B. {Croft}.
\newblock 2012.
\newblock The history of information retrieval research.
\newblock \emph{Proceedings of the IEEE}, 100(Special Centennial Issue):
  1444--1451.

\bibitem[Sanz{-}Cruzado and Castells(2018)]{sc18recsys}
J.~Sanz{-}Cruzado and P.~Castells.
\newblock 2018.
\newblock Enhancing structural diversity in social networks by recommending
  weak ties.
\newblock In S.~Pera, M.~D. Ekstrand, X.~Amatriain, and J.~O'Donovan, eds.,
  \emph{Proceedings of the $12^\textrm{th}$ {ACM} Conference on Recommender
  Systems (RecSys 2018)}, pp. 233--241. {ACM}.

\bibitem[Sarwar et~al.(2001)Sarwar, Karypis, Konstan, and Riedl]{sarwar2010}
B.~Sarwar, G.~Karypis, J.~Konstan, and J.~Riedl.
\newblock 2001.
\newblock Item-based collaborative filtering recommendation algorithms.
\newblock In \emph{Proceedings of the $10^\textrm{th}$ International Conference
  on World Wide Web}, WWW 2001, pp. 285--295. ACM, New York, NY, USA.

\bibitem[Schafer et~al.(1999)Schafer, Konstan, and Riedl]{Shafer1999ecommerce}
J.~B. Schafer, J.~Konstan, and J.~Riedl.
\newblock 1999.
\newblock Recommender systems in e-commerce.
\newblock In \emph{Proceedings of the $1^\textrm{st}$ ACM Conference on
  Electronic Commerce}, EC 1999, pp. 158--166.

\bibitem[Schnabel et~al.(2016)Schnabel, Swaminathan, Singh, Chandak, and
  Joachims]{ssscj16icml}
T.~Schnabel, A.~Swaminathan, A.~Singh, N.~Chandak, and T.~Joachims.
\newblock June 2016.
\newblock Recommendations as treatments: Debiasing learning and evaluation.
\newblock In \emph{Proceedings of the $3^\textrm{rd}$ International Conference
  on Machine Learning (ICML 2016)}, pp. 1670--1679. Proceedings of Machine
  Learning Research, Sheffield, UK.

\bibitem[Semerci et~al.(2019)Semerci, Gruson, Edwards, Lacker, Gibson, and
  Radosavljevic]{10.1145/3298689.3346977}
O.~Semerci, A.~Gruson, C.~Edwards, B.~Lacker, C.~Gibson, and V.~Radosavljevic.
\newblock 2019.
\newblock Homepage personalization at spotify.
\newblock In \emph{Proceedings of the $13^\textrm{th}$ ACM Conference on
  Recommender Systems}, RecSys 2019, p.~527.

\bibitem[Shi et~al.(2012)Shi, Karatzoglou, Baltrunas, Larson, Oliver, and
  Hanjalic]{sbloh12recsys}
Y.~Shi, A.~Karatzoglou, L.~Baltrunas, M.~A. Larson, N.~Oliver, and A.~Hanjalic.
\newblock 2012.
\newblock Climf: learning to maximize reciprocal rank with collaborative
  less-is-more filtering.
\newblock In P.~Cunningham, N.~J. Hurley, I.~Guy, and S.~S. Anand, eds.,
  \emph{Proceedings of the $6^\textrm{th}$ {ACM} Conference on Recommender
  Systems}, pp. 139--146. {ACM}.

\bibitem[Smyth and McClave(2001)]{smyth2001}
B.~Smyth and P.~McClave.
\newblock 2001.
\newblock Similarity vs. diversity.
\newblock In \emph{Proceedings of the 4$^\textrm{th}$ International Conference
  on Case-Based Reasoning}, ICCBR 2001, pp. 347--361. Springer-Verlag, London,
  UK, UK.

\bibitem[Song et~al.(2018)Song, Yang, Cao, and Xu]{sycx19cikm}
B.~Song, X.~Yang, Y.~Cao, and C.~Xu.
\newblock 2018.
\newblock Neural collaborative ranking.
\newblock In \emph{Proceedings of the $27^\textrm{th}$ ACM International
  Conference on Information and Knowledge Management}, CIKM 2018, pp.
  1353--1362. ACM, New York, NY, USA.

\bibitem[Steck(2010)]{s10kdd}
H.~Steck.
\newblock 2010.
\newblock Training and testing of recommender systems on data missing not at
  random.
\newblock In \emph{Proceedings of the $16^\textrm{th}$ ACM SIGKDD International
  Conference on Knowledge Discovery and Data Mining (KDD 2010)}, pp. 713--722.
  ACM, New York, NY, USA.

\bibitem[Steck(2011)]{s11recsys}
H.~Steck.
\newblock 2011.
\newblock Item popularity and recommendation accuracy.
\newblock In \emph{Proceedings of the $5^\textrm{th}$ ACM Conference on
  Recommender Systems (RecSys 2011)}, pp. 125--132. ACM, New York, NY, USA.

\bibitem[Steck(2013)]{s13recsys}
H.~Steck.
\newblock 2013.
\newblock Evaluation of recommendations: rating prediction and ranking.
\newblock In \emph{Proceedings of the $7^\textrm{th}$ ACM Conference on
  Recommender Systems (RecSys 2013)}, pp. 213--220. ACM, New York, NY, USA.

\bibitem[Steck(2018)]{steck2018calibration}
H.~Steck.
\newblock 2018.
\newblock Calibrated recommendations.
\newblock In \emph{ACM RecSys 2018}, pp. 154--162.

\bibitem[Steck(2019)]{steck2019}
H.~Steck.
\newblock 2019.
\newblock Embarrassingly shallow autoencoders for sparse data.
\newblock In \emph{Proceedings of the World Wide Web Conference}, WWW 2019, pp.
  3251--3257. ACM, New York, NY, USA.

\bibitem[Steck et~al.(2021)Steck, Baltrunas, Elahi, Liang, Raimond, and
  Basilico]{steck2021}
H.~Steck, L.~Baltrunas, E.~Elahi, D.~Liang, Y.~Raimond, and J.~Basilico.
\newblock 2021.
\newblock Deep learning for recommender systems: {A} netflix case study.
\newblock \emph{{AI} Magazine}, 42(3): 7--18.

\bibitem[Stray(2020)]{stray2020}
J.~Stray.
\newblock 2020.
\newblock Aligning ai optimization to community well-being.
\newblock \emph{International Journal of Community Well-Being}, 3: 443--463.

\bibitem[Sun et~al.(2019{\natexlab{a}})Sun, Liu, Wu, Pei, Lin, Ou, and
  Jiang]{BERT4REC2019}
F.~Sun, J.~Liu, J.~Wu, C.~Pei, X.~Lin, W.~Ou, and P.~Jiang.
\newblock 2019{\natexlab{a}}.
\newblock Bert4rec: Sequential recommendation with bidirectional encoder
  representations from transformer.
\newblock In \emph{Proceedings of the $28^\textrm{th}$ ACM International
  Conference on Information and Knowledge Management}, CIKM 2019, pp.
  1441--1450. ACM, New York, NY, USA.

\bibitem[Sun et~al.(2019{\natexlab{b}})Sun, Liu, Wu, Pei, Lin, Ou, and
  Jiang]{slwploj19cikm}
F.~Sun, J.~Liu, J.~Wu, C.~Pei, X.~Lin, W.~Ou, and P.~Jiang.
\newblock 2019{\natexlab{b}}.
\newblock Bert4rec: Sequential recommendation with bidirectional encoder
  representations from transformer.
\newblock In \emph{Proceedings of the $28^\textrm{th}$ ACM International
  Conference on Information and Knowledge Management}, CIKM 2019, pp.
  1441--1450. ACM, New York, NY, USA.
\newblock \url{https://doi.org/10.1145/3357384.3357895}.

\bibitem[Sun et~al.(2019{\natexlab{c}})Sun, Guo, Yang, Fang, Guo, Zhang, and
  Burke]{SUN2019100879}
Z.~Sun, Q.~Guo, J.~Yang, H.~Fang, G.~Guo, J.~Zhang, and R.~Burke.
\newblock 2019{\natexlab{c}}.
\newblock Research commentary on recommendations with side information: A
  survey and research directions.
\newblock \emph{Electronic Commerce Research and Applications}, 37: 100879.

\bibitem[Sutton and Barto(2021)]{sb21rlintro}
R.~S. Sutton and A.~G. Barto.
\newblock 2021.
\newblock \emph{Reinforcement learning - an introduction, second edition}.
\newblock Adaptive computation and machine learning. {MIT} Press.

\bibitem[Swaminathan et~al.(2017)Swaminathan, Krishnamurthy, Agarwal,
  Dud\'{i}k, Langford, Jose, and Zitouni]{skadljz17nips}
A.~Swaminathan, A.~Krishnamurthy, A.~Agarwal, M.~Dud\'{i}k, J.~Langford,
  D.~Jose, and I.~Zitouni.
\newblock 2017.
\newblock Off-policy evaluation for slate recommendation.
\newblock In \emph{Proceedings of the $31^\textrm{st}$ Conference on Neural
  Information Processing Systems (NIPS 2017)}, pp. 3635--3645. Curran
  Associates, Inc., Red Hook, NY, USA.

\bibitem[Tak\'{a}cs and Tikk(2012)]{takacs2012}
G.~Tak\'{a}cs and D.~Tikk.
\newblock 2012.
\newblock Alternating least squares for personalized ranking.
\newblock In \emph{Proceedings of the $6^\textrm{th}$ ACM Conference on
  Recommender Systems (RecSys 2012)}, pp. 83--90. ACM, New York, NY, USA.

\bibitem[Tang and Wang(2018)]{caser2018}
J.~Tang and K.~Wang.
\newblock 2018.
\newblock Personalized top-n sequential recommendation via convolutional
  sequence embedding.
\newblock In \emph{Proceedings of the $11^\textrm{th}$ ACM International
  Conference on Web Search and Data Mining}, WSDM 2018, pp. 565--573. ACM, New
  York, NY, USA.

\bibitem[Trattner and Jannach(2019)]{TrattnerJannach2019}
C.~Trattner and D.~Jannach.
\newblock 2019.
\newblock Learning to recommend similar items from human judgements.
\newblock \emph{User Modeling and User-Adapted Interaction}, 30: 1--49.

\bibitem[Vapnik(1998)]{vapnik1998}
V.~N. Vapnik.
\newblock 1998.
\newblock \emph{Statistical Learning Theory}.
\newblock Wiley-Interscience.

\bibitem[Vargas and Castells(2011)]{vc11recsys}
S.~Vargas and P.~Castells.
\newblock 2011.
\newblock Rank and relevance in novelty and diversity metrics for recommender
  systems.
\newblock In \emph{Proceedings of the 5$^\textrm{th}$ ACM Conference on
  Recommender Systems}, RecSys 2011, pp. 109--116. ACM, New York, NY, USA.

\bibitem[Vargas and Castells(2014)]{Vargas2014}
S.~Vargas and P.~Castells.
\newblock 2014.
\newblock Improving sales diversity by recommending users to items.
\newblock In \emph{Proceedings of the 8$^\textrm{th}$ ACM Conference on
  Recommender Systems}, RecSys 2014, pp. 145--152. ACM, New York, NY, USA.

\bibitem[Vargas et~al.(2011)Vargas, Castells, and Vallet]{vcv11sigir}
S.~Vargas, P.~Castells, and D.~Vallet.
\newblock 2011.
\newblock Intent-oriented diversity in recommender systems.
\newblock In \emph{Proceedings of the 34$^\textrm{th}$ Annual International ACM
  SIGIR Conference on Research and Development in Information Retrieval}, SIGIR
  2011, pp. 1211--1212. ACM, New York, NY, USA.

\bibitem[Veloso et~al.(2014)Veloso, Ribeiro, Lacerda, Moura, Hata, and
  Ziviani]{vrlmhz14tist}
A.~Veloso, M.~Ribeiro, A.~Lacerda, E.~Moura, I.~Hata, and N.~Ziviani.
\newblock Dec. 2014.
\newblock Multi-objective pareto-efficient approaches for recommender systems.
\newblock \emph{Special Issue on Novelty and Diversity in Recommender Systems,
  ACM Transactions on Information Systems and Technology}, 5(4).

\bibitem[Wasilewski and Hurley(2016)]{wh16recsys}
J.~Wasilewski and N.~Hurley.
\newblock 2016.
\newblock Intent-aware diversification using a constrained plsa.
\newblock In \emph{Proceedings of the 10$^\textrm{th}$ ACM Conference on
  Recommender Systems}, RecSys 2016, pp. 39--42. ACM, New York, NY, USA.

\bibitem[Whittaker et~al.(2021)Whittaker, Looney, Reed, and
  Votta]{whittaker2021}
J.~Whittaker, S.~Looney, A.~Reed, and F.~Votta.
\newblock 2021.
\newblock Recommender systems and the amplification of extremist content.
\newblock \emph{Internet Policy Review}, 10(2): 1--29.

\bibitem[Wu et~al.(2017)Wu, Ahmed, Beutel, Smola, and Jing]{wu2017}
C.-Y. Wu, A.~Ahmed, A.~Beutel, A.~J. Smola, and H.~Jing.
\newblock 2017.
\newblock Recurrent recommender networks.
\newblock In \emph{Proceedings of the $10^\textrm{th}$ ACM International
  Conference on Web Search and Data Mining}, WSDM 2017, pp. 495--503. ACM, New
  York, NY, USA.

\bibitem[Wu and Grbovic(2020)]{wg20sigir}
L.~Wu and M.~Grbovic.
\newblock 2020.
\newblock How airbnb tells you will enjoy sunset sailing in barcelona?
  recommendation in a two-sided travel marketplace.
\newblock In \emph{Proceedings of the $43^\textrm{rd}$ International ACM SIGIR
  Conference on Research and Development in Information Retrieval}, SIGIR 2020,
  pp. 2387--2396. ACM, New York, NY, USA.

\bibitem[Wu et~al.(2022)Wu, He, Wang, Zhang, and Wang]{Wu2022}
L.~Wu, X.~He, X.~Wang, K.~Zhang, and M.~Wang.
\newblock 2022.
\newblock A survey on accuracy-oriented neural recommendation: From
  collaborative filtering to information-rich recommendation.
\newblock \emph{IEEE Transactions on Knowledge and Data Engineering}, pp. 1--1.
\newblock \doi{10.1109/TKDE.2022.3145690}.

\bibitem[Xiao and Benbasat(2007)]{Xiao:2007:EPR:2017327.2017335}
B.~Xiao and I.~Benbasat.
\newblock 2007.
\newblock E-commerce product recommendation agents: Use, characteristics, and
  impact.
\newblock \emph{MIS Quarterly}, 31(1): 137--209.

\bibitem[Xin et~al.(2020)Xin, Karatzoglou, Arapakis, and Jose]{xin2020}
X.~Xin, A.~Karatzoglou, I.~Arapakis, and J.~M. Jose.
\newblock 2020.
\newblock \emph{Self-Supervised Reinforcement Learning for Recommender
  Systems}, pp. 931--940.
\newblock SIGIR 2020. ACM, New York, NY, USA.

\bibitem[Xin et~al.(2022{\natexlab{a}})Xin, Karatzoglou, Arapakis, and
  Jose]{xin2022}
X.~Xin, A.~Karatzoglou, I.~Arapakis, and J.~M. Jose.
\newblock 2022{\natexlab{a}}.
\newblock Supervised advantage actor-critic for recommender systems.
\newblock In \emph{Proceedings of the $15^\textrm{th}$ ACM International
  Conference on Web Search and Data Mining}, WSDM 2022, pp. 1186--1196. ACM,
  New York, NY, USA.

\bibitem[Xin et~al.(2022{\natexlab{b}})Xin, Pimentel, Karatzoglou, Ren,
  Christakopoulou, and Ren]{xin2022b}
X.~Xin, T.~Pimentel, A.~Karatzoglou, P.~Ren, K.~Christakopoulou, and Z.~Ren.
\newblock 2022{\natexlab{b}}.
\newblock Rethinking reinforcement learning for recommendation: A prompt
  perspective.
\newblock In \emph{Proceedings of the $45^\textrm{th}$ International ACM SIGIR
  Conference on Research and Development in Information Retrieval}, SIGIR 2022.

\bibitem[Yao and Huang(2017)]{Yao2017}
S.~Yao and B.~Huang.
\newblock 2017.
\newblock Beyond parity: Fairness objectives for collaborative filtering.
\newblock In I.~Guyon, U.~V. Luxburg, S.~Bengio, H.~Wallach, R.~Fergus,
  S.~Vishwanathan, and R.~Garnett, eds., \emph{Advances in Neural Information
  Processing Systems}, volume~30. Curran Associates, Inc.

\bibitem[Yao and Harper(2018)]{Yao2018judging}
Y.~Yao and F.~M. Harper.
\newblock 2018.
\newblock Judging similarity: A user-centric study of related item
  recommendations.
\newblock In \emph{Proceedings of the $12^\textrm{th}$ ACM Conference on
  Recommender Systems}, RecSys 2018, pp. 288--296.

\bibitem[Yuan et~al.(2019)Yuan, Karatzoglou, Arapakis, Jose, and He]{yuan2019}
F.~Yuan, A.~Karatzoglou, I.~Arapakis, J.~M. Jose, and X.~He.
\newblock 2019.
\newblock A simple convolutional generative network for next item
  recommendation.
\newblock In \emph{Proceedings of the $12^\textrm{th}$ ACM International
  Conference on Web Search and Data Mining}, WSDM 2019, pp. 582--590. ACM, New
  York, NY, USA.

\bibitem[Zhai et~al.(2003)Zhai, Cohen, and Lafferty]{zcl03sigir}
C.~X. Zhai, W.~W. Cohen, and J.~Lafferty.
\newblock 2003.
\newblock Beyond independent relevance: Methods and evaluation metrics for
  subtopic retrieval.
\newblock In \emph{Proceedings of the 26$^\textrm{th}$ Annual International ACM
  SIGIR Conference on Research and Development in Information Retrieval}, SIGIR
  2003, pp. 10--17. ACM, New York, NY, USA.

\bibitem[Zhang et~al.(2020)Zhang, Adomavicius, Gupta, and
  Ketter]{Gedas2020Longitudinal}
J.~Zhang, G.~Adomavicius, A.~Gupta, and W.~Ketter.
\newblock 2020.
\newblock Consumption and performance: Understanding longitudinal dynamics of
  recommender systems via an agent-based simulation framework.
\newblock \emph{Information Systems Research}, 31(1).

\bibitem[Zhang et~al.(2022)Zhang, Tay, Yao, Sun, and Zhang]{zhang2022}
S.~Zhang, Y.~Tay, L.~Yao, A.~Sun, and C.~Zhang.
\newblock 2022.
\newblock Deep learning for recommender systems.
\newblock In F.~Ricci, L.~Rokach, and B.~Shapira, eds., \emph{Recommender
  Systems Handbook, $3^\textrm{rd}$ edition}, pp. 173--210. Springer US, New
  York, NY.

\bibitem[Zhang et~al.(2021)Zhang, Feng, He, Wei, Song, Ling, and
  Zhang]{zhang2021}
Y.~Zhang, F.~Feng, X.~He, T.~Wei, C.~Song, G.~Ling, and Y.~Zhang.
\newblock 2021.
\newblock Causal intervention for leveraging popularity bias in recommendation.
\newblock In \emph{Proceedings of the 44th International ACM SIGIR Conference
  on Research and Development in Information Retrieval}, SIGIR 2021, pp.
  11--20. ACM, New York, NY, USA.
\newblock ISBN 9781450380379.

\bibitem[Ziegler et~al.(2005)Ziegler, McNee, Konstan, and Lausen]{Ziegler2005}
C.-N. Ziegler, S.~M. McNee, J.~A. Konstan, and G.~Lausen.
\newblock 2005.
\newblock Improving recommendation lists through topic diversification.
\newblock In \emph{Proceedings of the 14$^\textrm{th}$ International Conference
  on World Wide Web}, WWW 2005, pp. 22--32. ACM, New York, NY, USA.

\bibitem[Zuckerberg(2018)]{zuckerberg2018}
M.~Zuckerberg, 2018.
\newblock A blueprint for content governance and enforcement.
\newblock Retrieved from the Facebook Newsroom website: https://www.
  facebook.com/notes/mark-zuckerberg/ablueprint-for-content-governance-andenforcement/10156443129621634.

\end{thebibliography}

\end{document}